\documentclass{article}
\usepackage[a4paper, total={5.5 in, 10in}]{geometry}
\usepackage[utf8]{inputenc}
\usepackage{physics}
\usepackage{amsthm}
\usepackage{amsmath}
\usepackage{amssymb}
\usepackage{tcolorbox}
\usepackage{hyperref}
\usepackage{tcolorbox}
\usepackage{simpler-wick}
\usepackage{xcolor}
\usepackage{slashed}
\usepackage{graphics}
\usepackage{authblk}
\usepackage{etoolbox}

\usepackage[numbers,sort&compress]{natbib}

\newtheorem{exercise}{Exercise}[section]

\newcommand{\hint}[1]{\par\noindent\textbf{Hint:} #1}
\newcommand{\bonus}[1]{\par\noindent\textbf{Bonus:} #1}
\AtEndEnvironment{exercise}{\par\noindent\rule{\textwidth}{0.4pt}\vspace{1em}}
\numberwithin{equation}{section}

\title{\textbf{Lectures on the Spinor and Twistor Formalism in 3D Conformal Field Theory}}
\author{\Large Dhruva K.S.}
\date{}

\affil[1]{Indian Institute of Science Education and Research, Dr Homi Bhabha Road, Pashan, Pune, India}

\affil[1]{\texttt{k.s.dhruva@students.iiserpune.ac.in}}

\begin{document}
\setcounter{section}{-1}

\maketitle

\vspace{-1em}
{\centering\Large\bfseries Abstract\par}
\vspace{0.5em}
\noindent
These notes are based on my lectures given at $\text{ST}^4$ 2025 held at IISER Bhopal. We study the application of spinor and twistor methods to three dimensional conformal field theories in these notes. They are divided into three parts dealing with spinor helicity, twistors and super-twistors respectively. In the first part, we introduce the off-shell spinor helicity formalism and apply it in several contexts including double copy relations, connection to four dimensional scattering amplitudes, correlators in Chern-Simons matter theories and the holography of chiral higher spin theory. The second part of the notes introduces the twistor space formalism. After discussing the geometry of twistor space, we derive the Penrose transform for conserved currents, scalars with arbitrary scaling dimension as well as generic non-conserved operators. We also explicitly show how the spinor and twistor approaches are related. We discuss how correlators of these operators and conserved currents in particular drastically simplify in twistor space unveiling their hidden simplicity. We also extend our construction to super-conformal field theories and develop a manifest super-twistor space formalism and derive the supersymmetric Penrose transform. We find that the supersymmetric correlators are simple and natural generalizations of their non-supersymmetric counterparts.
The notes are made to be self-contained and also include over $50$ exercises that illustrate the formalism.

\vspace{1em}
\hrule
\vspace{1.5em}
\tableofcontents 

\section{Prologue}
A cornerstone of the modern amplitudes program is the use of a pair of spinors to describe both the four-dimensional null momenta and helicity of massless particles. This leads to dramatic simplifications in the expressions of scattering amplitudes, unveils hidden structures, and provides new ways to bootstrap them through the use of recursion relations\footnote{See the $\text{ST}^4$ $2025$ lectures by Amit Suthar on Amplitudes and references therein such as \cite{Elvang:2013cua} for more details.}. Given these successes in four-dimensional flat space, a natural question to ask is whether a similar story exists in other space-times.

To start with, one could ask about the other two maximally symmetric spacetimes: Anti-de Sitter (AdS) and de Sitter (dS), where the natural scattering like observables are boundary correlators. They enjoy the same symmetries as a Lorentzian/Euclidean conformal field theory (CFT) in one lower dimension, respectively. This is most fruitfully realized by the AdS/CFT correspondence \cite{Maldacena:1997re,Gubser:1998bc,Witten:1998qj}, which broadly speaking is the equivalence between a \( d+1 \) dimensional quantum gravitational theory in asymptotically AdS spacetime and a \( d \) dimensional non-gravitational scale invariant quantum field theory\footnote{See Gopal Yadav's $\text{ST}^4$ lectures to appear on the AdS$/$CFT correspondence and references therein for more details.} on its ``boundary".

Focusing on a four-dimensional bulk, the question that we ask is whether the simplicity of four-dimensional flat space (tree level) scattering amplitudes of massless particles carries over to boundary conformal correlators in \( \text{AdS}_4 \)? These are correlators of symmetric, traceless, conserved currents which---through AdS/CFT---describe spin-\( s \) massless gauge bosons. The first thing to do in this pursuit is to find out the correct kinematic variables to describe these correlators. In flat space, these are the spinor helicity variables that exponentially simplify traditional momentum space methods. In \( \text{AdS}_4 \), it is not yet fully clear what the correct set of kinematic variables are.

Traditionally, CFT has been analyzed in position space \cite{Belavin:1984vu,Rattazzi:2008pe,Poland:2022qrs,Hartman:2022zik} but in more recent years has garnered interest befitting a momentum space formulation \cite{Maldacena:2011nz,McFadden:2011kk,Coriano:2013jba,Bzowski:2013sza,Ghosh:2014kba,Bzowski:2015pba,Bzowski:2017poo,Bzowski:2018fql,Farrow:2018yni,Isono:2019ihz,Bautista:2019qxj,Gillioz:2019lgs,Baumann:2019oyu,Lipstein:2019mpu,Baumann:2020dch,Jain:2020rmw,Jain:2020puw,Jain:2021wyn,Jain:2021qcl,Jain:2021vrv,Baumann:2021fxj,Jain:2021gwa,Jain:2021whr,Gillioz:2022yze,Caloro:2022zuy,Marotta:2022jrp,Bzowski:2023jwt,S:2024zqp,Gupta:2024yib,Jain:2024bza,Marotta:2024sce,Coriano:2024ssu,Gillioz:2025yfb}. This has a variety of applications such as cosmology \cite{Maldacena:2011nz,McFadden:2011kk,Ghosh:2014kba,Arkani-Hamed:2015bza,Arkani-Hamed:2018kmz,Baumann:2022jpr}, the study of AdS amplitudes \cite{Raju:2010by,Raju:2011mp,Raju:2012zs,Albayrak:2018tam,Albayrak:2019yve,Gadde:2022ghy,Armstrong:2022mfr,Albayrak:2023jzl,Mei:2023jkb,Chowdhury:2023khl,Albayrak:2023kfk,Chowdhury:2024wwe,Chopping:2024oiu,Chowdhury:2024snc,Chowdhury:2024dcy}, double copy relations \cite{Farrow:2018yni,Lipstein:2019mpu,Jain:2021qcl}, connection to flat space S-matrices \cite{Maldacena:2011nz,Raju:2012zr}, the study of anomalies \cite{Coriano:2023hts,Coriano:2023cvf,Coriano:2023gxa,Hartman:2023qdn,Hartman:2023ccw,Hartman:2024xkw} etc... The tools we discuss in these notes are natural extensions of the momentum space approach which should help us towards achieving a unified description of aspects of Physics in Minkowski, AdS and dS spacetimes. 

One natural direction to take is the following: on-shell spinor helicity variables in 4 dimensional flat space induce \emph{off-shell} spinor helicity variables in 3 dimensional flat space \cite{Maldacena:2011nz,Baumann:2020dch,Baumann:2021fxj,Jain:2021vrv}.  These variables are appropriate to describe the \( \text{CFT}_3 \) correlators/\( \text{AdS}_4 \) amplitudes of interest to us and are subject of the first part of these notes. 

The second part of these notes also take a leaf out of the scattering amplitudes toolkit: The use of twistor variables. Twistors were  first introduced by Penrose \cite{Penrose:1967wn} and first applied in the context of the modern approach to scattering amplitudes in \cite{Nair:1988bq,Witten:2003nn,Mason:2009sa,Arkani-Hamed:2009hub}. In the context of 3d CFT correlators, twistors were first developed in \cite{Baumann:2024ttn}. An earlier work \cite{Jain:2023idr} also employed Grassmann twistors for 3d super-conformal field theories. For more work over the past year on twistors and super-twistors, please see \cite{Bala:2025gmz,Bala:2025jbh,Bala:2025qxr,Rost:2025uyj,Mazumdar:2025egx}. Wightman functions of conserved currents are extremely simple when cast in twistor variables. We also extend our methods to accommodate arbitrary operators in twistor space, which from the \( \text{AdS}_4 \) bulk perspective is the ability to describe particles with any mass and spin. This presents an inviting framework for the study of AdS$_4$ boundary correlators of gluons, gravitons etc.. as well as a potential stage for the spinning conformal bootstrap.

The third and final part of the notes deals with 3d super-conformal field theories in super-twistor space. Similar to the non-supersymmetric case, in contrast to the position (super) space approach \cite{Osborn:1998qu,Park:1999pd,Park:1999cw,Nizami:2013tpa,Buchbinder:2015qsa,Buchbinder:2015wia,Kuzenko:2016cmf,Buchbinder:2021gwu,Buchbinder:2021izb,Buchbinder:2021kjk,Buchbinder:2021qlb,Jain:2022izp,Buchbinder:2023fqv,Buchbinder:2023ndg}, this reformulation unveils features of super-correlators that are obscure otherwise \cite{Jain:2023idr,Bala:2025jbh,Bala:2025qxr}. After discussing the geometry of super-twistor space, we shall derive the supersymmetric Penrose transform for super-currents and discuss the construction of their correlation functions. 

These notes are structured as follows:
\subsection*{\textbf{A Brief Outline}}
$\bullet$ In Part \ref{part:SH}, we introduce an off-shell spinor helicity formalism for the study of 3d CFT correlators. This loosely follows the review portions given in \cite{Jain:2024bza,Bala:2025gmz}. We study its application to double copy relations \cite{Jain:2021qcl}, the flat space limit \cite{Maldacena:2011nz,Raju:2012zr}, correlators in Chern-Simons matter theories \cite{Jain:2022ajd} and the holography of chiral higher spin theory \cite{Jain:2024bza}. We conclude this part with a discussion of Lorentzian 3d CFT based on the discussion in \cite{Bala:2025gmz}.\\
$\bullet$ In part \ref{part:Twistors}, we introduce a twistor space framework for 3d CFT. The flow of the material closely follows \cite{Bala:2025qxr} and also \cite{Baumann:2024ttn,Bala:2025gmz} in several places. We discuss the geometry of twistor space, the Penrose transform, the Witten transform, the conformal invariants and the construction of Wightman functions of conserved currents. We also do so for parity odd correlators and those involving scalars.\\
$\bullet$ The subject of part \ref{part:SuperTwistors} are super-conformal field theories formulated in super-twistor space. The flow and contents of this part are heavily based on \cite{Bala:2025qxr} where we discuss the geometry of super-twistor space, derive the super-Penrose transform, discuss its relation to the super-Witten transform and the construction of correlators of super-currents as well as super-scalars. We also follow \cite{Bala:2025jbh} in some parts. Finally, we discuss contact terms and parity odd super-correlators in this context which is also based on material in \cite{Bala:2025qxr}.\\
$\bullet$ Finally, we summarize the contents of the three parts in part \ref{part:Conclusion}.\\
$\bullet$ There is an accompanying appendix \ref{app:CFTreview} to these notes on a brief review of the basics of 3d CFT for a reader not very  familiar with the subject or one who wants a quick refresher.\\

\subsection*{Prerequisites}
The prerequisites for reading these notes are basic Quantum field theory, a little group and representation theory and optionally some knowledge of conformal field theory and scattering amplitudes. Familiarity with the basics of AdS/CFT is also useful. Wolfram Mathematica is also invaluable for doing the calculations including those in the $55$ exercises in the notes. 

Questions, comments or suggestions about the notes or exercises are welcome.

\part{Spinor Helicity}\label{part:SH}

\section{Spinor helicity variables in three dimensions}\label{sec:SH3d}
The aim of this section is to derive and develop off-shell spinor helicity variables to describe three dimensional quantum field theories. We begin by discussing spinors in both the Lorentzian and Euclidean contexts. We setup the conformal generators and the conformal Ward identities in this language and discuss the general form of two, three point correlation functions of conserved currents in Euclidean space with its Lorentzian counterpart being deferred to section \ref{sec:LorentzianCFT}. Finally, we briefly discuss the construction of four and higher point functions. In particular, we derive the spinor helicity conformal partial waves which serve as the building blocks of correlators and discuss some simple examples involving conserved currents. The main references for this section are \cite{Bala:2025gmz,Jain:2024bza}.

\subsection{The Lorentzian Spacetime $\mathbb{R}^{2,1}$}\label{subsec:LorentzSH}
Consider a three dimensional momentum vector $p_\mu$. Since the Lorentz group in three dimensions is $SO(2,1)\cong SL(2,\mathbb{R})/\mathbb{Z}_2$, we can view it is a symmetric representation of the latter group by trading it for a $2\times 2$ matrix as follows:
\begin{align}\label{momentummatrix1}
    p_{a}^{b}=(\sigma^\mu)^b_a p_\mu.
\end{align}
$(\sigma^\mu)^a_b$ are $2\times 2$ matrices which obey the Clifford algebra,
\begin{align}
    \{\sigma_\mu,\sigma_\nu\}=\eta_{\mu\nu},
\end{align}
where $\eta_{\mu\nu}=\text{Diag}(-1,1,1)$ is the standard flat metric on $\mathbb{R}^{2,1}$.
\begin{exercise}\label{ex:5sec1}
Prove the homomorphism between the groups $SL(2,\mathbb{R})$ and $SO(2,1)$.
\hint{The explicit form of the sigma matrices we work with are given by,
\begin{align}\label{Paulixyz}
    (\sigma^x)^a_b=\begin{pmatrix}
        0&&1\\
        1&&0
    \end{pmatrix},(\sigma^t)^a_b=\begin{pmatrix}
        0&&-1\\
        1&&0
    \end{pmatrix},(\sigma^z)^a_b=\begin{pmatrix}
        1&&0\\
        0&&-1
    \end{pmatrix}.
\end{align}
Indices are raised and lowered as usual using the Levi-Civita symbol,
\begin{align}\label{LeviCivita2dLorentz}
    \epsilon_{ab}=\epsilon^{ab}=\begin{pmatrix}
        0&&1\\
        -1&&0
    \end{pmatrix},
\end{align}
via $p_{ab}=\epsilon_{cb}p_{a}^{c}, p^{ab}=\epsilon^{ac}p_{c}^{b}$. It is also easy to check that $p^{ab}$ and $p_{ab}$ are both symmetric in their indices.
Define the momentum matrix as in \eqref{momentummatrix1} with $p_\mu=(p_t,p_x,p_z)$. Show that $\text{Det}(p)=(p_t)^2-(p_x)^2-(p_z)^2=\frac{1}{2}p_{ab}p^{ab}=-p^2$ and that it is preserved under an $SL(2,\mathbb{R})$ transformation $p\to g^T p g$ where $g\in SL(2,\mathbb{R})$.}
\end{exercise}
We can express \eqref{momentummatrix1} using a pair of fundamental $SL(2,\mathbb{R})$ spinors $\lambda$ and $\Bar{\lambda}$ as follows:
\begin{align}\label{3dSHLorentzian}
    p_a^b=\frac{(\lambda_a\Bar{\lambda}^b+\lambda^b\Bar{\lambda}_a)}{2}.
\end{align}
Writing out the momentum matrix explicitly we obtain,
\begin{align}\label{explicitpabR21}
    p_a^b=\begin{pmatrix}
        p_z&&p_x-p_t\\
        p_x+p_t&&-p_z
    \end{pmatrix}.
\end{align}
For $(p_t,p_x,p_z)$ to represent a valid three momentum vector we require that,
\begin{align}
    p_t,p_x,p_z\in \mathbb{R}\implies  (p_a^b)^*=p_a^b.
\end{align}
This reality condition can be satisfied if we take the spinors in the RHS of \eqref{3dSHLorentzian} be purely real viz,
\begin{align}\label{LorentzianReality}
    \lambda_a^*=\lambda_a,\Bar{\lambda}_a^*=\Bar{\lambda}_a:~~\textbf{Lorentzian Reality Conditions}.
\end{align}
\begin{exercise}
    Show that the Lorentzian reality conditions on the spinors $\lambda$ and $\Bar{\lambda}$ is \eqref{LorentzianReality} for space-like momenta. What happens for time-like momenta? 
    \hint{Solve for three of the four combined components of $\lambda$ and $\Bar{\lambda}$ using \eqref{3dSHLorentzian}. Should we use a different decomposition of the momentum vector in terms of spinors for time-like momenta?}
\end{exercise}
The description \eqref{3dSHLorentzian} has a redundancy. Rescaling the two spinors by opposite amounts leaves the momentum invariant. This is called the \textit{stabilizer} or in physics parlance, the little group:
\begin{align}\label{LorentzLittleGroup}
    \lambda\to \frac{1}{r}\lambda,\Bar{\lambda}\to r\Bar{\lambda}, r\in\mathbb{R}\implies p_a^b\to p_a^b.
\end{align}
An interesting alternative way to arrive at this formalism is through a dimensional reduction of on-shell four dimensional spinor helicity variables.  This is illustrated by the following exercise:
\begin{exercise}\label{ex:dimredLorentz}
Our starting point is Klein space $\mathbb{R}^{2,2}$, the space-time with two spatial and two temporal directions. Consider a null momentum $p_\mu=(p_t,p_w,p_x,p_z)$. The four dimensional Klein group $SO(2,2)$ is homomorphic to $SL(2,\mathbb{R})\times SL(2,\mathbb{R})$. A vector representation of $SO(2,2)$ corresponds to the $(\frac{1}{2},\frac{1}{2})$ representation of $SL(2,\mathbb{R})\times SL(2,\mathbb{R})$. A null vector in particular can be written as a direct product of a $(\frac{1}{2},0)$ representation $\lambda_a$ and $(0,\frac{1}{2})$ representation $\tilde{\lambda}^{\Dot{a}}$.
    \begin{align}\label{4dKleinmomentum}
        p^\mu\to p_{a}^{\Dot{a}}=\begin{pmatrix}
            p_w+p_z&&p_x-p_t\\
            p_x+p_t&&p_w-p_z
        \end{pmatrix}=\lambda_a \tilde{\lambda}^{\Dot{a}},
    \end{align}
    where $-p_t^2-p_w^2+p_x^2+p_z^2=0$. 
Show that the reality condition $\lambda_a^*=\lambda_a,\tilde{\lambda}^{\Dot{a}*}=\tilde{\lambda}^{\Dot{a}}$ ensures that the four components of the momentum vector are real.
There is a redundancy in the description \eqref{4dKleinmomentum} viz
    \begin{align}\label{KleinMomentumLittleGroup}
        \lambda_a\to \frac{1}{r}\lambda_a,\tilde{\lambda}^{\Dot{a}}\to r\tilde{\lambda}^{\Dot{a}},r\in\mathbb{R}\implies p_a^{\Dot{a}}\to p_a^{\Dot{a}}.
    \end{align}
This is the manifestation of the fact that the compact (reduced) little group $SO(2)\cong U(1)$ of massless particles in $\mathbb{R}^{3,1}$ corresponds to the non compact little group $\mathbb{R}-\{0\}$ in Klein space.

Introduce a special vector that breaks the $SL(2,\mathbb{R})\times SL(2,\mathbb{R})$ symmetry down to a single $SL(2,\mathbb{R})$:
\begin{align}
    \epsilon_{a\Dot{a}}=\begin{pmatrix}
        0&&1\\
        -1&&0
    \end{pmatrix}.
\end{align}
Show that we can obtain the general three dimensional off-shell momentum \eqref{explicitpabR21} via,
\begin{align}
    p_{a}^{b}=-\epsilon^{bc}\frac{1}{2}(p_{(a}^{\Dot{a}}\epsilon_{c)\Dot{a}}),
\end{align}
where $\epsilon^{bc}$ is the usual Levi-Civita symbol \eqref{LeviCivita2dLorentz}. Also prove that the $SL(2,\mathbb{R})$ spinor $\Bar{\lambda}_a$ is given by,
\begin{align}\label{lambdatildelambdabarrelation}
    \Bar{\lambda}_a=\epsilon_{\Dot{a}a}\tilde{\lambda}^{\Dot{a}}.
\end{align}
The little group redundancy \eqref{LorentzLittleGroup} is also inherited from the same in its four dimensional counterpart \eqref{4dKleinmomentum}.
\end{exercise}
Thus, we can naturally view a three dimensional off-shell space-like momentum as a dimensional reduction of a null Klein space momentum vector. 

Lets now consider the other three dimensional flat space, $\mathbb{R}^3$, and compare and contrast spinors with their $\mathbb{R}^{2,1}$ counterparts.
\subsection{The Euclidean Space $\mathbb{R}^3$}\label{subsec:EuclidSH}
Consider a general Euclidean three-momentum $p_\mu=(p_x,p_y,p_z)$. Since the rotation group in 3d is $SO(3)$, we can make use of its homomorphism with $SU(2)$ to represent it as a $2\times 2$ matrix as follows:
\begin{align}\label{momentummatrixEuclid1}
    p_\mu\to p_{a}^{b}=(\sigma^\mu)^b_a p_\mu=\begin{pmatrix}
        p_z&&p_x-ip_y\\
        p_x+ip_y&&-p_z
    \end{pmatrix}.
\end{align}
$(\sigma^\mu)_b^a$ are the usual Pauli matrices whose explicit forms are,
\begin{align}\label{paulixyzEuclid}
     (\sigma^x)^a_b=\begin{pmatrix}
        0&&1\\
        1&&0
    \end{pmatrix},(\sigma^y)^a_b=\begin{pmatrix}
        0&&-i\\
        i&&0
    \end{pmatrix},(\sigma^z)^a_b=\begin{pmatrix}
        1&&0\\
        0&&-1
        \end{pmatrix}
\end{align}
They obey the Euclidean Clifford algebra,
\begin{align}
    \{\sigma_\mu,\sigma_\nu\}=\delta_{\mu\nu}.
\end{align}
The fundamental equation of 3d spinor helicity variables is \eqref{momentummatrixEuclid1} expressed as a bi-spinor by employing a pair of $SU(2)$ fundamental spinors $\lambda$ and $\Bar{\lambda}$ as follows:
\begin{align}\label{3dEuclidSH}
    p_{a}^{b}=\frac{(\lambda_a\Bar{\lambda}^b+\Bar{\lambda}_a\lambda^b)}{2},
\end{align}
where the presence of the two terms in \eqref{3dEuclidSH} ensures tracelessness as well as the fact that the matrix is full rank.
There is a redundancy in this description however, since $p_{ab}$ which contains three independent components, is represented in terms of two 2-component (complex) spinors. To reduce the degrees of freedom, we must first impose reality conditions. For physical momenta we require $p_x,p_y,p_z\in\mathbb{R}$. This is satisfied provided we take the matrix in \eqref{momentummatrixEuclid1} to be Hermitian,
\begin{align}
    (p_a^b)^\dagger=p_b^a.
\end{align}
Using the relation to the spinor variables \eqref{3dEuclidSH}, we see that this is satisfied if,
\begin{align}\label{EuclideanReality}
\lambda_a^\dagger=\Bar{\lambda}^a,\Bar{\lambda}_a^\dagger=\lambda^a~~  (\textbf{Euclidean Reality Conditions}).
\end{align}
Therefore, $\lambda$ and $\Bar{\lambda}$ are Hermitian conjugates of each other. Further, one can see in \eqref{3dEuclidSH} that there is a $U(1)$ redundancy,
\begin{align}\label{EuclideanSHU1redundency}
    \lambda\to e^{-\frac{i\theta}{2}}\lambda,\Bar{\lambda}\to e^{+\frac{i\theta}{2}}\Bar{\lambda},
\end{align}
which together with the reality condition \eqref{EuclideanReality} ensures that the degrees of freedom on both sides of \eqref{3dEuclidSH} match. Contrast this with the Lorentzian reality conditions \eqref{LorentzianReality} where the two spinors were real and independent and the little group was $\mathbb{R}$ \eqref{LorentzLittleGroup} rather than $U(1)$ \eqref{EuclideanSHU1redundency}.
\begin{exercise}\label{ex:dimredEuclid}
    Consider a null momentum $p_\mu=(p_t,p_x,p_y,p_z)$ in Minkowski space $\mathbb{R}^{3,1}$. Recall that the four dimensional Lorentz group $SO(3,1)$ is isomorphic to $SL(2,\mathbb{C})/\mathbb{Z}_2\cong \frac{SU(2)\times SU(2)^*}{\mathbb{Z}_2}$. A vector representation of $SO(3,1)$ corresponds to the $(\frac{1}{2},\frac{1}{2})$ representation of $\frac{SU(2)\times SU(2)^*}{\mathbb{Z}_2}$. A null vector in particular can be written as a direct product of a $(\frac{1}{2},0)$ representation $\lambda_a$ and $(0,\frac{1}{2})$ representation $\tilde{\lambda}^{\Dot{a}}$.
    \begin{align}\label{4dMinkowskimomentum}
        p^\mu\to p_{a}^{\Dot{a}}=\begin{pmatrix}
            p_t+p_z&&p_x-ip_y\\
            p_x+i p_y&&p_t-p_z
        \end{pmatrix}=\lambda_a \tilde{\lambda}^{\Dot{a}},
    \end{align}
    where $-p_t^2+p_x^2+p_y^2+p_z^2=0$.
Show that the reality condition $\lambda_a^\dagger=\tilde{\lambda}^{\Dot{a}}$ ensures that the four components of the momentum vector are real. Introduce a special vector that breaks the $SL(2,\mathbb{C})$ symmetry down to a $SU(2)$ subgroup,
\begin{align}
    \epsilon_{a\Dot{a}}=\begin{pmatrix}
        0&&1\\
        -1&&0
    \end{pmatrix}.
\end{align}
Show that we can obtain the general three dimensional off-shell momentum \eqref{momentummatrixEuclid1} via,
\begin{align}
    p_{a}^{b}=-\epsilon^{bc}\frac{1}{2}(p_{(a}^{\Dot{a}}\epsilon_{c)\Dot{a}}),
\end{align}
where $\epsilon^{bc}$ is the usual $SU(2)$ Levi-Civita symbol \eqref{LeviCivita2dLorentz}. Also prove that the $SU(2)$ spinor $\Bar{\lambda}_a$ is given by,
\begin{align}
    \Bar{\lambda}_a=\epsilon_{\Dot{a}a}\tilde{\lambda}^{\Dot{a}}.
\end{align}
The little group redundancy \eqref{EuclideanSHU1redundency} is also inherited from the same in its four dimensional counterpart \eqref{4dMinkowskimomentum}.
\end{exercise}
\subsection{The Helicity basis}\label{subsec:HelicityBasis}
Now that we have been acquainted with the spinors in both Lorentzian and Euclidean signature, lets pay justice to the second part of the formalism: Helicity. As we discussed at the beginning of this section, our immediate focus is on correlation functions of conserved currents in 3d CFT which describe scattering of massless gauge bosons in $\text{AdS}_4$. Lets now define these operators more precisely. A symmetric traceless conserved current $J_s^{a_1\cdots a_{2s}}(p)$ contains $2s$ spinor indices\footnote{These indices are those of the $SU(2)$ fundamental representation in the Euclidean case and the $SL(2,\mathbb{R})$ fundamental representation in the Lorentzian setting.} and is traceless with respect to any of them. In conformal field theories, it is a well known fact that such a  spin-s current in three dimensions has a scaling dimension $\Delta=s+1$. For a brief review of the basics of conformal field theory with a focus in $d=3$, please check out appendix \ref{app:CFTreview}. Now, the helicity basis for these currents is defined with the help of our spinors $\lambda$ and $\Bar{\lambda}$ as follows:
\begin{align}\label{3dpolarizationspinors}
    \zeta_{-}^a=\frac{\lambda^a}{\sqrt{p}},\zeta_{+}^a=\frac{\Bar{\lambda}^a}{\sqrt{p}},
\end{align}
where $p=-\frac{\lambda\cdot\Bar{\lambda}}{2}$ is the magnitude of the three-momentum. We assign $\lambda$ and $\Bar{\lambda}$ a helicity of $-\frac{1}{2}$ and $+\frac{1}{2}$ respectively which is consistent with their little group ``charge"  \eqref{LorentzLittleGroup},\eqref{EuclideanSHU1redundency} which we can also think of as inherited from their four dimensional counterparts in light of exercises \eqref{ex:dimredLorentz} and \eqref{ex:dimredEuclid}. The positive and negative helicity components of the current which have helicity $+s$ and $-s$ respectively are then defined as,
\begin{align}\label{Jspm}
    J_s^{\pm}(\lambda,\Bar{\lambda})=\zeta_{\pm a_1}\cdots \zeta_{\pm a_{2s}}J_s^{a_1\cdots a_{2s}}(p).
\end{align}
The beauty of the helicity basis is that a complicated quantity with $2s$ indices is traded for just two components. As the below exercise illustrates, the remaining components are proportional to the divergence of the current which is zero since it is conserved.
\begin{exercise}
    (a)Prove that a symmetric traceless integer spin conserved current in three dimensions has only two independent components regardless of spin. (b) Show that the mixed helicity components constructed similar to \eqref{Jspm} using \eqref{3dpolarizationspinors} are all proportional to the divergence of the current.
    \hint{The constraints on the current are,
    \begin{align}
        \epsilon_{a_i a_j}J_s^{\cdots a_i a_j\cdots}=0,~~ p_{a_i a_j}J_s^{\cdots a_i a_j\cdots}=0.
    \end{align}}
    Note that $p_{ab}$ \eqref{3dSHLorentzian}, \eqref{3dEuclidSH} can be written in terms of the polarization spinors \eqref{3dpolarizationspinors} as $p^{ab}=p\zeta_{-}^{(a}\zeta_{+}^{b)}$.\\
    (c)Prove that $J_s^{a_1\cdots a_{2s}}(p)$ can be expressed in terms of its components $J_s^{\pm}$ as, follows:
    \begin{align}\label{Jsinhelicitybasis}
        J_s^{a_1\cdots a_{2s}}=\bigg(\frac{-1}{2}\bigg)^{s}\bigg(\zeta^{a_1}_{-}\cdots \zeta^{a_{2s}}_{-}J_s^{+}+\zeta^{a_1}_{+}\cdots \zeta^{a_{2s}}_{+}J_s^{-}\bigg)
    \end{align}
    \bonus{Does the result of exercise $(a)$ remind you of a familiar result? Understand this in light of the AdS/CFT correspondence.}
\end{exercise}
For future convenience, we define the rescaled currents with a hat,
\begin{align}\label{Jsrescaled}
    \hat{J}_s^{\pm}(\lambda,\Bar{\lambda})=\frac{J_s^{\pm}(\lambda,\Bar{\lambda})}{p^{s-1}}.
\end{align}
Our objective is to constrain and bootstrap correlators of these currents. Given the fact that we are interested in conformal field theories, let us now see the implication of this symmetry on these objects.

\subsection{Conformal Ward identities}\label{subsec:ConformalWardId}
The three dimensional Lorentzian and Euclidean conformal groups are respectively $SO(3,2)$ and $SO(4,1)$. The correlation functions of our currents are constrained by these symmetries. In order to see what these are, one possible approach to take is to start from the usual position space generators and Fourier transform them and then convert to spinor helicity variables which is done in appendix \ref{app:CFTreview}. Here, we use dimensional analysis and little group invariance to construct the generators acting on the rescaled currents \eqref{Jsrescaled}.
\subsection*{The Poincare generators}
These ones are obvious. We already have the momentum generator in spinor helicity variables \eqref{3dSHLorentzian}, \eqref{3dEuclidSH},  which gives,
\begin{align}\label{PGenSH}
    P_{ab}=\lambda_{(a}\Bar{\lambda}_{b)}.
\end{align}
Lets now construct the Rotation generators $M_{\mu\nu}$. In three dimensions, it is convenient to dualize $M_{\mu\nu}$ and to consider $M_{\rho}=\epsilon_{\mu\nu\rho}M^{\mu\nu}$ instead. We can then contract the index $\rho$ with the Pauli matrices \eqref{Paulixyz}/\eqref{paulixyzEuclid} to obtain a symmetric quantity $M_{ab}$. As far as the counting goes, $M_{ab}$ has three independent components which is the correct number in three dimensions (two rotations and one Lorentz boost or simply three rotations). To figure out $M_{ab}$ in terms of $\lambda$ and $\Bar{\lambda}$ we first note that it should be dimensionless as well as a first order differential operator as is appropriate for a rotation generator. Further, it should be invariant under little group scaling \eqref{LorentzLittleGroup}, \eqref{EuclideanSHU1redundency}. An ansatz satisfying all these constraints is,
\begin{align}\label{MGenSH}
    M_{ab}=\frac{1}{2}\big(\lambda_{(a}\frac{\partial}{\partial \lambda^{b)}}+\Bar{\lambda}_{(a}\frac{\partial}{\partial \Bar{\lambda}^{b)}}\big).
\end{align}
There is no spin matrix contribution to this generator since the currents \eqref{Jsrescaled} have no free indices.
\begin{exercise}
 The Poincare algebra in the usual vector notation is given by,
 \begin{align}
     &[P_\mu,P_\nu]=0,\notag\\
     &[M_{\mu\nu},P_{\rho}]=\eta_{\mu\rho}P_{\nu}-\eta_{\nu\rho}P_{\mu},\notag\\
     &[M_{\mu\nu},M_{\rho\sigma}]=\eta_{\mu\rho}M_{\nu\sigma}-\eta_{\nu\rho}M_{\mu\sigma}-\eta_{\mu\sigma}M_{\nu\rho}+\eta_{\nu\sigma}M_{\mu\rho}.
 \end{align}
    Show that upon converting to the spinor notation (by contracting every vector index with a Pauli matrix) that  \eqref{PGenSH} and \eqref{MGenSH} form a valid representation of the Poincare algebra.
\end{exercise}
\subsection*{Scale and special conformal generators}
The theories we will consider are scale invariant. Lets consider our rescaled currents \eqref{Jsrescaled}. The conserved currents $J_s$ in position space have scaling dimension $\Delta=s+1$. A Fourier transform brings this down to $s+1-3=s-2$. Rescaling by $\frac{1}{p^{s-1}}$ further brings it down to $s-2-(s-1)=-1$. Given the fact the both $\lambda$ and $\Bar{\lambda}$ have scaling dimension $\frac{1}{2}$ as is clear from \eqref{PGenSH} we can guess the following form for the Dilatation generator that generates scale transformations:
\begin{align}\label{DGenSH}
    D=\frac{1}{2}\bigg(\lambda^a\frac{\partial}{\partial\lambda^a}+\Bar{\lambda}^a\frac{\partial}{\partial\Bar{\lambda}^a}+2\bigg).
\end{align}
The extra factor of $+1$ is a consequence of the fact that the rescaled current has scaling dimension $-1$. 

Finally, conformal field theories are invariant under what are known as special conformal transformations. In position space, we can think of these transformations as an inversion followed by a translation followed by another inversion. This tells us that in three dimensions, it has three independent parameters. This generator has scaling dimension $-1$ and also has to be invariant under little group scaling. Further, one can show by a Fourier transform that this operator must be a second order differential operator. This leads us to a natural choice,
\begin{align}\label{KGenSH}
    K_{ab}=\frac{\partial^2}{\partial\lambda^{(a}\partial\Bar{\lambda}^{b)}}.
\end{align}
\begin{exercise}
   The Conformal algebra is given by,
\begin{align}\label{confalgebra}
&[M_{\mu\nu},M_{\rho\sigma}]=i(\delta_{\mu\rho}M_{\nu\sigma}-\delta_{\nu\rho}M_{\mu\sigma}-\delta_{\mu\sigma}M_{\nu\rho}+\delta_{\nu\sigma}M_{\mu\rho}),\notag\\
&[M_{\mu\nu},P_{\alpha}]=i(\delta_{\mu\alpha}P_\nu-\delta_{\nu\alpha}P_\mu),~~[M_{\mu\nu},K_{\alpha}]=i(\delta_{\mu\alpha}K_\nu-\delta_{\nu\alpha}K_\mu),\notag\\
&[P_{\mu},K_{\nu}]=2i(\delta_{\mu\nu}D-M_{\mu\nu}),~~[D,P_{\mu}]=iP_\mu,~~[D,K_\mu]=-iK_\mu,
    \end{align}
    with the remaining commutators vanishing. After converting this to spinor notation or vice versa, show that our generators \eqref{PGenSH},\eqref{MGenSH}, \eqref{DGenSH} and \eqref{KGenSH} obey this algebra.
    \bonus{Show that this algebra is isomorphic to the Lorentz algebra in $5$ dimensions $\mathfrak{so}(4,1)$ or $\mathfrak{so(3,2)}$. Therefore, the Euclidean and Lorentzian conformal groups in $3$ dimensions are $SO(4,1)$ and $SO(3,2)$ respectively. This is the starting point for the \textit{embedding} space approach to CFTs where one treats conformal transformations as Lorentz transformations in a  spacetime with two extra dimensions. Although we wont discuss the embedding space approach much, we will rediscover it when we deal with twistor variables in the second part of the notes.}
\end{exercise}
\subsection*{The Conformal Ward identities}
Consider a generic $n-$point correlation function of our rescaled currents;
\begin{align}\label{Jsnpoint}
    \langle \hat{J}_{s_1}^{\pm}(\lambda_1,\Bar{\lambda}_1)\cdots \hat{J}_{s_n}^{\pm}(\lambda_n,\Bar{\lambda}_n)\rangle.
\end{align}
Using \eqref{PGenSH}, translation invariance reads,
\begin{align}\label{SHPID}
    \sum_{i=1}^{n}\lambda_{i(a}\Bar{\lambda}_{ib)} \langle \hat{J}_{s_1}^{\pm}(\lambda_1,\Bar{\lambda}_1)\cdots \hat{J}_{s_n}^{\pm}(\lambda_n,\Bar{\lambda}_n\rangle=0\implies  \langle \hat{J}_{s_1}^{\pm}(\lambda_1,\Bar{\lambda}_1)\cdots \hat{J}_{s_n}^{\pm}(\lambda_n,\Bar{\lambda}_n\rangle=\delta^3(p_1+\cdots+p_n)\Gamma_n.
\end{align}
Lorentz invariance \eqref{MGenSH} tells us that the function $\Gamma_n$ can only depend on the invariants formed using the 2d Levi Civita symbol \eqref{LeviCivita2dLorentz} $\epsilon_{ab}$. These include the following classes of angle brackets:
\begin{align}\label{SHMID}
    \langle i j\rangle=\lambda_{ia}\lambda_j^b~,\langle \Bar{i}\Bar{j}\rangle=\Bar{\lambda}_{ia}\Bar{\lambda}_j^a,\langle i \Bar{j}\rangle=\lambda_{ia}\Bar{\lambda}_j^a, p_i=-\frac{1}{2}\lambda_{ia}\Bar{\lambda}_i^a.
\end{align}
Note in particular that in contrast to on-shell spinor helicity variables in $4d$, we can form invariants by contracting barred and unbarred spinors. 
Scale invariance that is implemented by \eqref{DGenSH} is satisfied if $\Gamma_n$ has scaling dimension $3-n$:
\begin{align}\label{SHDID}
    \sum_{i=1}^{n}\frac{1}{2}\big(\lambda_{i}^a\frac{\partial}{\partial\lambda_i^a}+\Bar{\lambda}_i^a\frac{\partial}{\partial\Bar{\lambda}_i^a}+2(\frac{3-n}{n})\big)\Gamma_n=0
\end{align}
For example, three point functions with the momentum conserving delta function stripped are dimensionless.
Finally the constraints due to special conformal transformations \eqref{KGenSH} read,
\begin{align}\label{SHKID}
    \sum_{i=1}^{n}\frac{\partial^2}{\partial\lambda_{i}^{(a}\partial\Bar{\lambda}_{i}^{b)}}\Gamma_n=\frac{1}{2}\bigg(\sum_{i=1}^{n}\frac{s_i(s_i+1)}{p_i^{s_i+1}}\zeta_{i\pm a}\zeta_{i \pm b}p_i^{a_1 a_2}\zeta_{i\pm}^{a_3}\cdots \zeta_{i\pm}^{a_{2s}}\bigg)\langle\langle \cdots \hat{J}_{s_i a_1\cdots a_{2s}}(\lambda_i,\Bar{\lambda}_i)\cdots\rangle\rangle,
\end{align}
where the double bracketed notation means that we have stripped out the momentum conserving delta function.
Note that rather than the action being zero, it leads to the Ward-Takahashi identity of the currents involved. The derivation of this result is sketched in appendix \ref{app:CFTreview}. One way to simplify life is to work with Wightman functions in Lorentzian signature in which the currents are identically conserved. The Euclidean results can then by obtained by a Wick rotation. We will indeed take this approach in a later section. Before we proceed, there is another important constraint we must impose on correlators of our conserved currents. The little group covariance. Given the fact that the currents \eqref{Jspm} are contracted with $2s$ $\lambda$s for the negative helicity component and $2s$ $\Bar{\lambda}$s for the positive helicity one, we require the following recaling property of our correlators:
\begin{align}\label{SHHelicityID}
    \langle \cdots \hat{J}_{s_i}^{\pm}(\frac{1}{r}\lambda_i,r\Bar{\lambda}_i)\cdots\rangle=r^{\pm 2 s_i}\langle \cdots \hat{J}_{s_i}^{\pm}(\lambda_i,\Bar{\lambda}_i)\cdots\rangle.
\end{align}
Armed with the conformal Ward identities and the helicity identity, lets discuss the general form of two and three point correlators of currents. So far our discussion is general and applies to both Euclidean and Lorentzian signature. Now, we specialize to Euclidean space to discuss correlation functions. Lorentzian CFTs are discussed in section \ref{sec:LorentzianCFT}.
\subsection{General form of Euclidean Two and Three point current correlators}\label{subsec:TwoThreePointSH}
\subsection*{Two Points}
Before we proceed, lets get familiar with two point kinematics with the following exercise:
\begin{exercise}
    Two point momentum conservation is the statement that $p_1^\mu+p_2^\mu=0$. Using spinor helicity variables this reads,
    \begin{align}
        \lambda_{1(a}\Bar{\lambda}_{1b)}+\lambda_{2(a}\Bar{\lambda}_{2b)}=0.
    \end{align}
    Assuming that $\langle 12\rangle\ne 0$ show that the above equation implies,
    \begin{align}
        &\langle 1 \Bar{2}\rangle=\langle \Bar{1} 2\rangle=0\notag\\
        &\langle 12\rangle\langle \Bar{1}\Bar{2}\rangle=4 p_1^2=4p_2^2=\langle 1 \Bar{1}\rangle^2=\langle 2 \Bar{2}\rangle^2.
    \end{align}
\end{exercise}
Translation invariance yields a momentum conserving delta function. It is also easy to see that the little group property \eqref{SHHelicityID} can only be satisfied if $s_1=s_2=s$. Further, it requires that the answer is of the form $\sum_{m=0}^{2s}c_m(p_1)\langle 1 2\rangle^{-m}\langle\Bar{1}\Bar{2}\rangle^{2s-m}$. However, in light of the above exercise, this sum collapses to but one term. using dimensional analysis \eqref{SHDID} and the fact that the only other independent invariant that is little group invariant is $p_i$ yields the result,
\begin{align}\label{JsJspp}
    \langle \hat{J}_{s}^{+}(\lambda_1,\Bar{\lambda}_1)J_{s}^{+}(\lambda_2,\Bar{\lambda}_2)\rangle=(c_{\text{even}}+i c_{\text{odd}})\frac{\langle \Bar{1} \Bar{2}\rangle^{2s}}{p_1^{2s-1}}\delta^3(p_1+p_2).
\end{align}
We have included a complex coefficient $c_{even}+i c_{odd}$ in front. Lets look at the negative helicity two point function. Using the fact that the positive and negative helicity currents are related by a Hermitian conjugation\footnote{This follows from the definition of the currents \eqref{Jspm} and the Euclidean reality conditions \eqref{EuclideanReality}. In the Lorentzian case, the positive and negative helicity currents will turn out to be CPT conjugates of each other as we shall see in a later section.} we obtain,
\begin{align}\label{JsJsmm}
    \langle \hat{J}_{s}^{-}(\lambda_1,\Bar{\lambda}_1)J_{s}^{-}(\lambda_2,\Bar{\lambda}_2)\rangle=(-1)^{2s}(c_{\text{even}}-i c_{\text{odd}})\frac{\langle 1 2\rangle^{2s}}{p_1^{2s-1}}\delta^3(p_1+p_2).
\end{align}
The $(-1)^{2s}$ factor is due to the fact that Hermitian conjugation takes $\langle ij\rangle$ to $\langle \Bar{j}\Bar{i}\rangle$ and vice versa. 
The reader might be wondering why there are two coefficients rather than just one. Doesnt conformal invariance fix two point functions up to a single overall constant? Yes and No. It fixes the position space correlator at separated points up to an overall constant. However, there could exist contact term contributions that have support only at $x_1=x_2$ that could potentially be conformally invariant. Indeed, this is the case for conserved currents in three dimensions. Lets go to position space for a bit and consider the two point function of a spin-1 current to begin with. The usual parity even two point function is given by,
\begin{align}
    \langle J_{\mu}(x_1)J_{\nu}(x_2)\rangle\supset \bigg(\delta_{\mu\nu}-2\frac{x_{12\mu}x_{12\nu}}{x_{12}^2}\bigg)\frac{1}{x_{12}^4}.
\end{align}
We have $x_{12}=|x_1-x_2|$ and $x_{12\mu}=x_{1\mu}-x_{2\mu}$ above. It is easy to see that current conservation with respect to both currents is satisfied. In three dimensions however, there is another possibility of a conserved quantity which uses the parity odd 3d Levi-Civita symbol $\epsilon_{\mu\nu\rho}$:
\begin{align}
    \langle J_{\mu}(x_1)J_{\nu}(x_2)\rangle\supset \epsilon_{\mu\nu\rho}\partial_{1\rho}\delta^3(x_1-x_2).
\end{align}
This object is automatically conserved with respect to the two currents. It also scales nicely thanks to the scaling property of the delta function. Invariance under special conformal transformations is a bit harder to see but can be shown with a little work. Therefore, the general two point function is given by,
\begin{align}\label{JJposspacetwopoint}
    \langle J_{\mu}(x_1)J_{\nu}(x_2)\rangle=c_{\text{even}} \bigg(\delta_{\mu\nu}-2\frac{x_{12\mu}x_{12\nu}}{x_{12}^2}\bigg)\frac{1}{x_{12}^4}+c_{\text{odd}}\epsilon_{\mu\nu\rho}\partial_{1\rho}\delta^3(x_1-x_2).
\end{align}
Performing a Fourier transform yields,
\begin{align}\label{JJmomentumspace2point}
    \langle J_\mu(p_1)J_{\nu}(p_2)\rangle=\bigg(c_{even}(\delta_{\mu\nu}-\frac{p_{1\mu}p_{1\nu}}{p_1^2})p_1+c_{\text{odd}}\epsilon_{\mu\nu\rho}p_1^\rho\bigg)\delta^3(p_1+p_2).
\end{align}
\begin{exercise}
    Show that upon converting the above expression to spinor variables and contracting with the polarization spinors \eqref{3dpolarizationspinors}, we reproduce the results \eqref{JsJspp} and \eqref{JsJsmm} for $s=1$. Generalize the position space result to currents of arbitrary spin and show that they give rise to the general results in \eqref{JsJspp} and \eqref{JsJsmm}. Are such parity odd contributions to two point functions possible in $d>3$?
    \bonus{How do such parity odd contributions arise in the holographic context from the AdS$_4$ perspective?}
\end{exercise}
Therefore, the existence of two independent coefficients in the two point function of a current reflects the existence of a possible parity odd contribution in three dimensions.  There is another very interesting fact about this parity odd contribution. It can in fact be obtained from its parity even counterpart from what is known as the \textit{epsilon} transformation \cite{Jain:2021gwa}. Consider the momentum space parity even two point function \eqref{JJmomentumspace2point}. Lets define the following operation:
\begin{align}\label{epsilonTransformspin1}
   (\epsilon\cdot J)_{\mu}(p)=\frac{\epsilon_{\mu\nu\rho}}{p}p^\nu J^\rho(p).
\end{align}
Performing this operation on the parity even two point function structure in \eqref{JJmomentumspace2point} yields,
\begin{align}
    \frac{\epsilon_{\mu\rho\sigma}}{p_1}p_1^\sigma\big(\delta^\rho_\nu-\frac{p_1^\rho p_{1\nu}}{p_1^2}\big)p_1\delta^3(p_1+p_2)=\epsilon_{\mu\nu \rho}p_1^\rho \delta^3(p_1+p_2),
\end{align}
which is precisely the parity odd structure in \eqref{JJmomentumspace2point}! In fact, for general integer spin conserved currents, the epsilon transform is given by,
\begin{align}\label{epsilonTransformspins}
    (\epsilon\cdot J_s)_{\mu_1\cdots \mu_{s}}(p)=p^\nu\frac{\epsilon_{\nu\rho (\mu_1}}{p}J^{\rho}_{s\mu_2\cdots \mu_s)}.
\end{align}
This is a conformally invariant transformation that preserves the spin and the scaling dimension of the current. However, it brings out the additional parity odd epsilon symbol thereby changing the discrete parity label when this current is inserted in a correlator. The easiest way to check its conformal invariance is to use spinor helicity variables. Lets take the epsilon transform formula \eqref{epsilonTransformspins}, convert vector indices to spinor ones and contract with the polarization spinors \eqref{3dpolarizationspinors}\footnote{It is useful to note that $\epsilon^{\mu\nu\rho}=\frac{1}{2i}\text{Tr}(\sigma^\mu\sigma^\nu \sigma^\rho)$.}. The resulting positive and negative helicity components are given by,
\begin{align}
    (\epsilon\cdot J_s)^{\pm}=\pm i J_s^{\pm}.
\end{align}
Therefore, the epsilon transform becomes a simple multiplicative operator in spinor helicity variables. This is also consistent with our results for the SH variables two point functions \eqref{JsJspp} and \eqref{JsJsmm}. 
\subsection*{Three point functions}
Lets now move on to the three point case. There are eight possible helicity configurations out of which half of them can be obtained via Hermitian conjugation from the other half. Using little group scaling \eqref{SHHelicityID}, we can write down the general form of these correlators (suppressing the momentum conserving delta function):
\begin{align}\label{3ptJs1s2s3genSH}
    &\langle J_{s_1}^{+}J_{s_2}^{+}J_{s_3}^{+}\rangle=\langle \Bar{1}\Bar{2}\rangle^{s_1+s_2-s_3}\langle\Bar{2}\Bar{3}\rangle^{s_2+s_3-s_1}\langle \Bar{3}\Bar{1}\rangle^{s_3+s_1-s_2}f_{+++}(p_1,p_2,p_3),\notag\\
    &\langle J_{s_1}^{+}J_{s_2}^{+}J_{s_3}^{-}\rangle=\langle \Bar{1}\Bar{2}\rangle^{s_1+s_2-s_3}\langle\Bar{2}3\rangle^{s_2+s_3-s_1}\langle 3\Bar{1}\rangle^{s_3+s_1-s_2}f_{++-}(p_1,p_2,p_3),\notag\\
    &\langle J_{s_1}^{+}J_{s_2}^{-}J_{s_3}^{+}\rangle=\langle \Bar{1}2\rangle^{s_1+s_2-s_3}\langle 2\Bar{3}\rangle^{s_2+s_3-s_1}\langle \Bar{3}\Bar{1}\rangle^{s_3+s_1-s_2}f_{+-+}(p_1,p_2,p_3),\notag\\
    &\langle J_{s_1}^{-}J_{s_2}^{+}J_{s_3}^{+}\rangle=\langle 1\Bar{2}\rangle^{s_1+s_2-s_3}\langle\Bar{2}\Bar{3}\rangle^{s_2+s_3-s_1}\langle \Bar{3}1\rangle^{s_3+s_1-s_2}f_{-++}(p_1,p_2,p_3).
\end{align}
The remaining four helicities are obtained by Hermitian conjugation from the above ones. Further, one has to impose scale and conformal invariance that will restrict the functional from of the functions appearing in the above equation. We shall not do this in detail here but only quote the general structure. In fact, we shall do so for the general momentum space correlator which contains all the above spinor helicity expressions as its components. It is important to separate out two distinct cases based on the spin-triangle inequality $s_i+s_j\ge s_k, i\ne j\ne k\in\{1,2,3\}$. 
\subsection*{Inside the triangle}
A general three point function of conserved currents obeying $s_i+s_j\ge s_k, i\ne j\ne k\in\{1,2,3\}$ consists of three different consistent contributions \cite{Giombi:2011rz}:
\small
\begin{align}\label{genspins1s2s3momspace}
    &\langle J_{s_1}(p_1,\zeta_1)J_{s_2}(p_2,\zeta_2)J_{s_3}(p_3,\zeta_3)\rangle\notag\\&= c_{s_1s_2s_3}^{h,\text{even}}\langle J_{s_1}(p_1,\zeta_1)J_{s_2}(p_2,\zeta_2)J_{s_3}(p_3,\zeta_3)\rangle_{h,\text{even}}+c_{s_1s_2s_3}^{h,\text{odd}}\langle J_{s_1}(p_1,\zeta_1)J_{s_2}(p_2,\zeta_2)J_{s_3}(p_3,\zeta_3)\rangle_{\text{h,odd}}\notag\\&+c_{s_1s_2s_3}^{nh}\langle J_{s_1}(p_1,\zeta_1)J_{s_2}(p_2,\zeta_2)J_{s_3}(p_3,\zeta_3)\rangle_{\text{nh,even}}.
\end{align}
\normalsize
Given that the form of the SCT Ward identity \eqref{SHKID} is a second order in-homogeneous differential equation, we can separate the solution into a homogeneous and particular solution. The first two terms in the above equation are homogeneous solutions in that they are identically conserved and lead to zero on the RHS of \eqref{SHKID} in every helicity. The third solution with the label $nh$ indicates the non-homogeneity of the SCT Ward identity and saturates the current Ward-Takahashi identity of the correlator. There is also a fourth solution which is a parity odd non-homogeneous correlator but we do not discuss it here. See \cite{Jain:2021vrv} for more details. We have also contracted the spinor indices of the currents with polarization spinors $\zeta_i$ which we are arbitrary for the moment. When going to spinor helicity variables, we will of course take them to be of the form \eqref{3dpolarizationspinors}. The homogeneous correlators in particular are quite easy to solve in general as we illustrate in the below exercise. 
\begin{exercise}
    Lets start with the $(+++)$ helicity configuration. We have seen that Translation and Lorentz invariance yield the form \eqref{3ptJs1s2s3genSH} for the correlator. After acting with the dilatation and SCT generators on the correlator, we obtain differential equations for $f_{+++}$. Show that they are solved by,
    \begin{align}
        f_{+++}(p_1,p_2,p_3)=\frac{1}{(p_1+p_2+p_3)^{s_1+s_2+s_3}}.
    \end{align}
    There also exist three other solutions obtained from the above one by flipping $p_i\to -p_i$. However, in the Euclidean context, these solutions were shown to be inconsistent with the Fourier transform of the known position space result known as the OPE consistency criteria \cite{Maldacena:2011nz,Mata:2012bx,Jain:2021whr}. Also, one can also show that the homogeneous correlator vanishes in all mixed helicity configurations. Whatever possible answers there exist are inconsistent with the OPE. However, when dealing with Wightman functions in the Lorentzian case in section \ref{sec:LorentzianCFT},  all these solutions can and will occur as we shall discuss subsequently. 
\end{exercise}
Therefore, the Euclidean homogeneous correlators exist only in the $(+++)$ and the $(---)$ helicities. What about the non-homogeneous even correlator? It is an interesting fact that it is non-zero only in the mixed helicity configurations although much harder to prove. For some explicit examples and physical arguments supporting the statements to follow, please see \cite{Jain:2024bza}.  To summarize the structures, we have the following statements about the Euclidean current correlators:

\underline{\textbf{Statement}~1:} For correlators obeying the spin triangle inequality, the homogeneous parts, $ \langle J_{s_1}J_{s_2}J_{s_3}\rangle_{h}$ are only nonzero in the $(- - -)$ and $(+ + +)$ helicity configurations.
\begin{align}\label{statement1}
    &\langle J_{s_1}^{-}J_{s_2}^{-}J_{s_3}^{-}\rangle_h\ne 0\,,\qquad \langle J_{s_1}^{+}J_{s_2}^{+}J_{s_3}^{+}\rangle_h\ne 0\,,\notag\\&\langle J_{s_1}^{h_1}J_{s_2}^{h_2}J_{s_3}^{h_3}\rangle_h=0~\text{for all other helicities}.
\end{align}
 As for the odd correlators, we have the following fact: In spinor-helicity variables, they are equal to the even homogeneous correlators up to factors of $\pm i$ that depend on the helicity. This is a consequence of the epsilon transform \eqref{epsilonTransformspins} which for three point functions is given by,
 \begin{align}\label{hoddeptofheven}
     \langle (\epsilon\cdot J_{s_1})J_{s_2}J_{s_3}\rangle_{\text{h,even}}\sim \langle J_{s_1}J_{s_2}J_{s_3}\rangle_{\text{h,odd}}
 \end{align}
\underline{\textbf{Statement}~2:} The nonhomogeneous correlators inside the triangle, $ \langle J_{s_1}J_{s_2}J_{s_3}\rangle_{nh}$ are only nonzero in the mixed helicity configurations, i.e. they are zero in the $(- - -)$ and $(+ + +)$ helicity configurations. This is true up to contact terms contributions.
\begin{align}\label{statement2}
    &\langle J_{s_1}^{-}J_{s_2}^{-}J_{s_3}^{-}\rangle_{nh}=\langle J_{s_1}^{+}J_{s_2}^{+}J_{s_3}^{+}\rangle_{nh}=0\,,~~~\notag\\
    &\langle J_{s_1}^{h_1}J_{s_2}^{h_2}J_{s_3}^{h_3}\rangle_{nh}\ne 0~\text{for all other (net nonzero) helicity configurations}.
\end{align}
\underline{\textbf{Statement}~3:} There is another interesting statement. The zero-helicity sector (sum of the helicities of the operators is zero) non-homogeneous correlators are identically zero. This statement is also of course true up to contact term contributions. Thus, we can find contact terms to remove any potential contributions.
\begin{align}\label{statement3}
    \langle J_{s_1}^{h_1}J_{s_2}^{h_2}J_{s_3}^{h_3}\rangle=0~~\text{if the net helicity is zero}.
\end{align}
We now proceed to further understand the meaning of the homogeneous and non-homogeneous correlators.
\subsection*{An expression in terms of Free theory correlators}
Two obvious conformal field theories in three dimensions are the free massless bosonic and fermionic theories. It turns out that the homogeneous and non-homogeneous correlators have simple expressions in terms of the corresponding correlators in these theories. First of all, let us note that these free theories possess what is known as \textit{higher spin symmetry} which includes conformal symmetry as a sub-group. Just as conformal transformations are generated by the stress tensor (its co-dimension one integrals contracted with the conformal killing vectors give rise to the conformal generators), the higher spin symmetries are generated by higher spin currents $J_s^{\mu_1\cdots \mu_s}$. In the free (complex) bosonic theory for example they schematically take the form,
\begin{align}
    J_s^{\mu_1\cdots \mu_s}=\Bar{\phi}\partial^{\mu_1}\cdots\partial^{\mu_s}\phi+\cdots.
\end{align}
Tracelessness with respect to any pair of indices and conservation can uniquely fix the expression up to a single overall normalization factor.
\begin{exercise}
    Derive the exact form of these currents in the free $U(1)$ massless complex Klein Gordon theory (For real scalar fields, there are only even spin conserved currents). What happens to these symmetries when we turn on a mass? What about if we add a $\phi^6$ classically marginal coupling? 
    \bonus{Lets add some flavour and promote the bosonic fields from $U(1)$ to a $U(N)$ fundamental representation ($O(N)$ in the real case). Lets add in some self-interactions via the relevant $\phi^4$ coupling. Lets work at large $N$ and flow to the famous Wilson Fisher fixed point. How is the conservation of the higher spin currents ($s>2$) affected?}
\end{exercise}
Similar currents also exist in the free Fermionic theory. Lets denote these correlators with subscripts,
\begin{align}
    \langle J_{s_1}J_{s_2}J_{s_3}\rangle_{FB/FF},
\end{align}
where $FB$ and $FF$ stand for free boson and free fermion respectively. The homogeneous and non-homogeneous correlators are then given by,
\begin{align}\label{handnhinfreetheory}
    &\langle J_{s_1}J_{s_2}J_{s_3}\rangle_{h}=\langle J_{s_1}J_{s_2}J_{s_3}\rangle_{FF}-\langle J_{s_1}J_{s_2}J_{s_3}\rangle_{FB}\notag\\
    &\langle J_{s_1}J_{s_2}J_{s_3}\rangle_{nh}=\langle J_{s_1}J_{s_2}J_{s_3}\rangle_{FF}+\langle J_{s_1}J_{s_2}J_{s_3}\rangle_{FB}.
\end{align}
Simply put, the difference between the fermionic and bosonic correlators cancels out the Ward Takahashi identities due to current conservation whereas the sum adds them up. Therefore, we can write down the three point function of higher spin currents in any CFT purely in terms of free theory correlators in SH variables as follows:
\begin{align}
    \langle J_{s_1}J_{s_2}J_{s_3}\rangle=(c_{s_1s_2s_3}^{h}+c_{s_1s_2s_3}^{nh})\langle J_{s_1}J_{s_2}J_{s_3}\rangle_{FF}+(-c_{s_1s_2s_3}^{h}+c_{s_1s_2s_3}^{nh})\langle J_{s_1}J_{s_2}J_{s_3}\rangle_{FB}.
\end{align}
Here, the coefficient $c_{s_1s_2s_3}^{h}=c_{s_1s_2s_3}^{h,\text{even}}\pm i c_{s_1s_2s_3}^{h,\text{odd}}$ is a complex coefficient whereas $c_{s_1s_2s_3}^{nh}$ is a real quantity proportional to the two point function that occurs in the Ward-Takahashi identity of this correlator.
\subsection*{Perspective from the Bulk}
Lets also provide some intuition to these correlators by turning our eye towards the $\text{(A)dS}_4$ bulk where these quantities have natural interpretations, see \cite{Jain:2021qcl} for a detailed discussion.
Consider Gluons in the bulk in the Poincare patch (which covers all of AdS in the Euclidean case). The Yang Mills action (schematically) takes the form,
\begin{align}
    \int d^3 x \frac{dz}{z^4}~F^2=\int d^3 x dz\big( \partial A\partial A+g_{YM}AA\partial A+g_{YM}^2AAAA\big).
\end{align}
The term with three gauge fields will contribute to the three point function. In fact, this three point function is the non-homogeneous three point function we discussed above. This is due to the fact that the coefficient of the non-homogenenous correlator is fixed by the Ward-Takahashi identity of the correlator which is indeed proportional to the two point function whose coefficient from the bulk perspective also arises from the same term.
How about the homogeneous contributions? For this we need to consider the following higher derivative interactions:
\begin{align}
    \frac{1}{\Lambda^2}\int d^3 x \frac{dz}{z^4}\bigg(F^3+\tilde{F}F^2\bigg),
\end{align}
where $\tilde{F}$ is the Hodge dual of $F$ and $\Lambda$ is a parameter with dimension $1$.  These respectively contribute a parity even and parity odd contribution to the three point function which from the CFT perspective are the homogeneous even and odd correlator respectively. One can generalize this to higher spins. For instance for the graviton three point function, we have the Einstein Hilbert action, the $\text{Weyl}^3$ term and its parity odd counterpart. Since vertices in AdS and flat space are in $1-1$ correspondence, we can then view the number of independent structures for a spin s conserved current three point function as equal to the number of independent three point flat space scattering amplitudes of spin s massless gauge Bosons which is indeed $3$. One can similarly generalize this argument to three point interactions involving distinct particles/ correlators of non-identical currents.
\subsection*{Outside the triangle}
Finally, lets look at the case where the spin triangle inequality $s_i+s_j\ge s_k, i\ne j\ne k\in\{1,2,3\}$ is violated: Here, it turns out that the parity odd structure is inconsistent with current conservation and therefore does not exist \cite{Jain:2021whr}. There is also no consistent parity even homogeneous solution. Rather, there are two parity even solutions which are both non-homogeneous. In terms of free theory correlators we have,
\begin{align}
    \langle J_{s_1}J_{s_2}J_{s_3}\rangle=c_{s_1s_2s_3}^{nh,1} \langle J_{s_1}J_{s_2}J_{s_3}\rangle_{FF-FB}+c_{s_1s_2s_3}^{nh,2} \langle J_{s_1}J_{s_2}J_{s_3}\rangle_{FF+FB}.
\end{align}
Similar to the independent correlators inside the triangle having support only in particular helicities, we have statements for this correlators too. For example, consider $s_1$ to be the largest spin in the correlator. The $FF-FB$ correlator then is non zero only in the $(+--)$ and $(---)$ configs and their conjugate helicities. On the other hand, the $FF+FB$ correlator is non-zero only in $(+-+)$, $(++-)$ and its conjugate helicities. More details can be found in \cite{Jain:2024bza}. Another important point to note is that both solutions outside the triangle have a non zero Ward Takahashi identity. This is because the free bosonic and free fermionic Ward Takahashi identities are distinct outside the triangle and do not cancel out in the $FF-FB$ combination \cite{Jain:2021whr}. Thus we see that the situation inside and outside the triangle are quite distinct. This concludes our discussion of the generalities of Euclidean three point correlators of conserved currents.
\subsection{Four and higher point functions}\label{subsec:FourHigherPointSH}
At the level of four and higher point functions, conformal invariance is no longer strong enough to fix the functional form of correlation functions. In position space for instance, four point functions are fixed up to functions of what are called conformal cross ratios. For example consider a correlator of identical scalar operators. It takes the form,
\begin{align}
    \langle O_{\Delta}(x_!)\cdots O_{\Delta}(x_4)\rangle=\frac{1}{x_{12}^{2\Delta}x_{34}^{2\Delta}}F(u,v),
\end{align}
where the cross ratios $u,v$ are given by,
\begin{align}
    u=\frac{x_{12}^2x_{34}^2}{x_{13}^2 x_{24}^2}~,v=\frac{x_{14}^2x_{23}^2}{x_{13}^2 x_{24}^2}.
\end{align}
$n$ point functions take a similar form but with a unknown function of $n(n-3)/2$ cross ratios\footnote{There is a subtlety in this formula. For instance in $d=1$, there is only one independent cross ratio at the level of four points. In $d=2$ at the level of five points we have $4$ rather than $5$ independent cross ratios. One has to account for these and related constraints in $d$ dimensions in writing down the precise formula which is left as an exercise to the reader. The above formula $n(n-3)/2$ is only an upper bound on the possible independent cross ratios.}. In momentum space, the general expression for $n$ point scalar correlators was discovered in  \cite{Bzowski_2020,Bzowski_2021,Caloro_2023} and the result takes the form of a simplex integral. For spinning correlators, the analogous expressions remain unknown. However, given the fact that we are working with a conformal field theory which in particular has a convergent operator product expansion in position space\footnote{One can perform the OPE between two operators as long as they are well separated from any other operator. For a detailed discussion of the OPE please refer to the standard and excellent resources \cite{Simmons-Duffin:2016gjk,Rychkov:2016iqz}.},
\begin{align}
    \mathcal{O}_i(x_i)\mathcal{O}_j(x_j)=\sum_{k}f_{ijk}(x_{ij},\partial_{j})\mathcal{O}_k(x_j),
\end{align}
The $f_{ijk}$ encode the three point function coefficient. We can see this by multiplying the above equation with a third operator. The only term which is picked out on the RHS is when the very same operator appears in the OPE since only two point functions of identical operators are non-zero in CFT up to contact term contributions. We have also suppressed the possible vector or more generally spinor indices of all operators involved. Given this formula, one can in principle construct all higher point correlators from lower point ones. For example, performing an OPE between the first two and last two operators in a four point function we obtain,
\begin{align}
    \langle \mathcal{O}_1(x_1)\mathcal{O}_2(x_2)\mathcal{O}_3(x_3)\mathcal{O}_4(x_4)\rangle=\sum_{k,l}f_{12 k}(x_{12},\partial_2)f_{34 l}(x_{34}\partial_4)\langle \mathcal{O}_{k}(x_2)\mathcal{O}_l(x_4)\rangle.
\end{align}
However, in $d>2$ it is famously known that every four and higher point correlator receives contributions from infinitely many (primary) operators, see section 10.3 of \cite{Simmons-Duffin:2016gjk} for instance. Therefore, this sum is always an infinite one and requires detailed knowledge of the spectrum of operators and their three point function coefficients. However, in the traditional conformal bootstrap programme, we also supplement the OPE with what is known as crossing symmetry. Basically, it states that performing the OPE say between $1,2$ and $3,4$ should yield the same result as performing it between $1,3$  and $2,4$ in a common domain of convergence. In momentum space, the issues of convergence of OPE become more complicated since the Fourier transform instructs as to integrate over all positions, including coincident ones. However, what we can still do is look at the contribution of a single exchanged operator which is called a conformal block. 

Consider a general four point function,
\begin{align}
    \langle \mathcal{O}_1(x_1)\mathcal{O}_2(x_2)\mathcal{O}_3(x_3)\mathcal{O}_4(x_4)\rangle.
\end{align}
Lets insert a complete set of states in the middle. In a CFT, the operator spectrum of the theory serves this purpose thanks to the operator state correspondence \cite{Simmons-Duffin:2016gjk,Rychkov:2016iqz}:
\begin{align}
        \sum_{\alpha\in\mathcal{O},P^\mu \mathcal{O},\cdots}\langle \mathcal{O}_1(x_1)\mathcal{O}_2(x_2)|\alpha\rangle\langle \alpha|\mathcal{O}_3(x_3)\mathcal{O}_4(x_4)\rangle,
\end{align}
where $\alpha$ runs over the spectrum of primary operators and their descendants in the theory. Now, what is this complete set? Well, the projector has to be dimensionless and conformally invariant. This heavily restricts the form it can take and leads us to an ansatz,
\begin{align}
    |\alpha\rangle\langle \alpha|=\int d^3 x\mathcal{O}_{\mu_1\cdots \mu_{s}}(x)|0\rangle\langle 0|\mathcal{\tilde{O}}^{\mu_1\cdots \mu_s}(x)
\end{align}
$\mathcal{\tilde{O}}$ is called the shadow of $\mathcal{O}$. It has scaling dimension $3-\Delta$ and the same spin as $\mathcal{O}$ to ensure that the integrand is invariant under Lorentz and scale transformations. For scalar operators, we have,
\begin{align}
    \tilde{O}(x)=\int \frac{d^3 y}{|x-y|^{2(3-\Delta)}}O(y).
\end{align}
The contribution to the correlator due to the exchange of a single operator with scaling dimension $\Delta$ and spin $s$ is called a conformal block. This is equivalent to (with a caveat \cite{Simmons-Duffin:2012juh}),
\begin{align}\label{CPW1}
    G_{\Delta,s}=\int d^3 x   \langle \mathcal{O}_1(x_1)\mathcal{O}_2(x_2)\mathcal{O}^{\mu_1\cdots \mu_s}(x)\rangle\langle \mathcal{\tilde{O}}_{\mu_1\cdots \mu_s}(x)\mathcal{O}_3(x_3)\mathcal{O}_4(x_4)\rangle.
\end{align}
The caveat is that the above expression is democratic in the operator and its shadow so one has to do some more work to extract the contribution just due to the operator. This is through a Monodromy projection \cite{Simmons-Duffin:2012juh} which roughly speaking yields the conformal block with short distance behavior consistent with the exchange of the operator and not its shadow. We shall however, not attempt this in these lectures and instead shall be content with looking at the conformal partial wave, which is what \eqref{CPW1} is called. Our interest is in the spinor helicity version of \eqref{CPW1}. The first step is to convert the above CPW to momentum space. Lets first define the shadow of a general spinning operator. By conformal invariance and dimensional analysis we see that it takes the form \cite{Dolan:2011dv},
\begin{align}\label{genShadow}
    \tilde{O}_{\mu_1\cdots \mu_s}(x)\propto \int d^3 y \frac{1}{|x-y|^{2(3-\Delta)}}\mathcal{I}_{\mu_1\cdots \mu_s;\nu_1\cdots \nu_s}(x-y)\mathcal{O}^{\nu_1\cdots \nu_s}(y).
\end{align}
$\mathcal{I}$ is called the Inversion multi-tensor. It is made out of the following object  simply called an inversion tensor\footnote{This is the Jacobian of the transformation $x^\mu\to\frac{x^\mu}{x^2}$ which is a conformal transformation, although one disconnected from the identity transformation.}:
\begin{align}
    I_{\mu\nu}(x)=(\delta_{\mu\nu}-\frac{2 x_\mu x_{\nu}}{x^2}).
\end{align}
Recall that this is the same tensor that appeared in the spin-1 conformal two point function \eqref{JJposspacetwopoint}. The Inversion multi-tensors are symmetric traceless products of $s$ copies of the Inversion tensor. Our focus is on conserved currents. For these, the shadow transform \eqref{genShadow} drastically simplifies! The result after converting to momentum space is given by (ignoring overall proportionality constants),
\begin{align}\label{shadowformulamomspace}
\tilde{J}_s^{\mu_1\cdots \mu_s}(p)=\frac{1}{p^{2s-1}}\Pi^{\mu_1\cdots \mu_s;\nu_1\cdots \nu_s}(p)J_{s \nu_1\cdots \nu_s}(p).
\end{align}
$\Pi^{\mu_1\cdots \mu_s}$ is a transverse traceless projector. It is traceless with respect to any pair of indices and transverse with respect to every index which is good for a conserved current. For example for a spin-1 current we have the two indexed object (there is no tracelessness for a spin-1 object)
\begin{align}
    \pi^{\mu\nu}(p)=\delta^{\mu\nu}-\frac{p^\mu p^\nu}{p^2}.
\end{align}
For spin-2 we have,
\begin{align}
    \Pi^{\mu\nu;\rho\sigma}(p)=\pi^{\mu\nu}(p)\pi^{\rho\sigma}(p)-(\pi^{\mu\rho}(p)\pi^{\nu\sigma}(p)+\pi^{\mu\sigma}(p)\pi^{\mu\rho}(p)),
\end{align}
and so on for higher spins.
Contracting with the polarizations to go the helicity basis we obtain the super simple formulae for the shadow currents,
\begin{align}\label{shadowcurrent}
    \tilde{J}_s^{\pm}(p)=\frac{1}{p^{2s-1}}J_s^{\pm}(p).
\end{align}
\begin{exercise}
    Derive \eqref{shadowformulamomspace} from \eqref{genShadow}. Also show \eqref{shadowcurrent} by contracting with the polarization spinors \eqref{3dpolarizationspinors}.
\end{exercise}
Lets now use this. First, we need to get the CPW \eqref{CPW1} in momentum space. By a simple Fourier transform we get,
\begin{align}\label{CPWmomspace}
    G_{\Delta,s}=\int d^3 p \langle \mathcal{O}_1(p_1)\mathcal{O}_2(p_2)\mathcal{O}^{\mu_1\cdots \mu_s}(p)\rangle\langle \mathcal{\tilde{O}}_{\mu_1\cdots \mu_s}(-p)\mathcal{O}(p_3)\mathcal{O}(p_4)\rangle.
\end{align}
Specializing to exchanges of conserved currents and using the helicity basis \eqref{Jsinhelicitybasis} we can use the formula \eqref{shadowcurrent} to obtain,
\begin{align}
    J^{\mu_1\cdots \mu_s}(p)\rangle\langle \tilde{J}_{\mu_1\cdots \mu_s}(-p)=\frac{1}{p^{2s-1}}(J_s^{-}(p)\rangle\langle \tilde{J}_s^{+}(-p)+J_s^{+}(p)\rangle\langle \tilde{J}_s^{-}(-p)).
\end{align}
Given $p^{ab}=\lambda^{(a}\Bar{\lambda}^{b)}$, we can obtain the spinor decomposition of $-p^{ab}$ simply by taking $\Bar{\lambda}\to -\Bar{\lambda}$. Using the formula for the shadow current and using the above in the CPW, we obtain\footnote{A word of caution. We have been ignoring overall numerical constants in this analysis},
\begin{align}
    G_{\Delta,s}=\int \frac{d^3 p}{p^{2s-1}}&\bigg(\langle \mathcal{O}_1(p_1)\mathcal{O}_2(p_2)J_s^{-}(p)\rangle\langle J_s^{+}(-p)\mathcal{O}_3(p_3)\mathcal{O}_4(p_4)\rangle\notag\\
    &+\langle \mathcal{O}_1(p_1)\mathcal{O}_2(p_2)J_s^{+}(p)\rangle\langle J_s^{-}(-p)\mathcal{O}_3(p_3)\mathcal{O}_4(p_4)\rangle\bigg)
\end{align}
However, recall that three point functions in momentum space come with a momentum conserving delta function and therefore the above integral is trivial! The result is,
\begin{align}
    G_{\Delta,s}=\delta^3(p_1+p_2+p_3+p_4)\frac{1}{|p_1+p_2|^{2s-1}}&\bigg(\langle \mathcal{O}_1(p_1)\mathcal{O}_2(p_2)J_s^{-}(-p_1-p_2)\rangle\langle J_s^{+}(p_1+p_2)\mathcal{O}_3(p_3)\mathcal{O}_4(p_4)\rangle\notag\\
    &+\langle \mathcal{O}_1(p_1)\mathcal{O}_2(p_2)J_s^{+}(-p_1-p_2)\rangle\langle J_s^{-}(p_1+p_2)\mathcal{O}_3(p_3)\mathcal{O}_4(p_4)\rangle\bigg)
\end{align}
The external operators $\mathcal{O}_i(p_i)$ can be any operators including conserved currents but is not limited to the same. They can be scalars or even generic non conserved currents. Also, this is in the s channel with $(12)$ and $(34)$ paired together. One can also do it in the $t$ and $u$ channels. Lets do one simple example of a s channel exchange.
\subsection*{``Gluon" exchange}
Consider a correlator of four spin-1 non-Abelian conserved currents which are dual to Gluons in AdS$_4$. The first two operators have negative helicity whereas the third and fourth have positive helicity which is the analog of the flat space MHV configuration. 
\begin{align}
    \langle J^{A_1-}J^{A_2-}J^{A_3 +}J^{A_4 +}\rangle.
\end{align}
Lets look at the s channel CPW due to the exchange of another ``gluon". We shall also focus on the non-homogeneous contribution to both three point functions which from the bulk perspective is the Yang-Mills three point function $(s=|p_1+p_2|$ below and we are suppressing the momentum conserving delta function)
\small
\begin{align}\label{gluonCPW1}
    &G_{1}^{A_1A_2A_3A_4}=  \langle J^{A_1-}J^{A_2-}J^{A+}\rangle\frac{1}{s}\langle J^{A-}J^{A_3 +}J^{A_4 +}\rangle\notag\\
    &=f^{A_1A_2A}f^{A_3A_A}\Bigg(\frac{\langle 12\rangle^3 \langle 3 4\rangle^3}{p_1p_2p_3p_4s^3}(s+p_1-p_2)(s-p_1+p_2)(s+p_3-p_4)(s-p_3+p_4)\notag\\&~~~~~~~~~~~~~~~~~~~~~~\big(1-\frac{3s}{p_1+p_2+s})(1-\frac{3s}{p_3+p_4+s}\big)\frac{1}{\langle 2 \lambda\rangle\langle 1\lambda\rangle\langle \Bar{3}\Bar{\lambda}\rangle\langle\Bar{4}\Bar{\lambda}\rangle}\Bigg).
\end{align}
\normalsize
We used the fact that the Gluon three point function is given by \cite{Jain:2024bza},
\begin{align}
    \langle J^{A_1 -}(p_1)J^{A_2-}(p_2)J^{A_3 +}(p_3)\rangle=\frac{\langle 12\rangle^4}{\langle 12\rangle\langle 2 3\rangle\langle 3 1\rangle}\frac{(p_1-p_2+p_3)(-p_1+p_2+p_3)}{p_1 p_2 p_3}(1-\frac{3p_3}{p_1+p_2+p_3})
\end{align}
In \eqref{gluonCPW1}, the spinors $\lambda,\Bar{\lambda}$ are the ones corresponding to the exchanged momentum and read,
\begin{align}
    \lambda_{(a}\Bar{\lambda}_{b)}=-p_{1ab}-p_{2ab}.
\end{align}
Using this we can simplify the $\lambda$ dependent brackets in the denominator. One can show using momentum conservation that we get,
\begin{align}
    \langle 2 \lambda\rangle\langle 1\lambda\rangle\langle \Bar{3}\Bar{\lambda}\rangle\langle\Bar{4}\Bar{\lambda}\rangle=((p_1+p_2+s)\langle 1\Bar{3}\rangle+\langle 12\rangle\langle \Bar{2}\Bar{3}\rangle)((p_1+p_2+s)\langle 2\Bar{4}\rangle+\langle 12\langle \Bar{3}\Bar{4}\rangle).
\end{align}
Thus we see that even this simple CPW with a spin-1 exchange is not so simple. For a simpler example, lets look at a scalar exchange conformal partial wave.
\subsection*{Scalar exchange for Abelian currents}
Lets look at the simpler case of the exchange of a $\Delta=1$ scalar for a correlator of four Abelian currents. The result is quite simple really and reads,
\begin{align}
    G_{\Delta=1}=\frac{\langle 12\rangle^2\langle \Bar{3}\Bar{4}\rangle^2}{s^3(p_1+p_2+s)^2(p_3+p_4+s)^2}.
\end{align}
Although we have many tools at our disposal, the study of spinning four point functions in momentum space and spinor helicity variables has not had much progress and remains an important area to make more progress in. The bulk AdS$_4$ perspective offers an excellent avenue to analyze holographic CFT correlators. See \cite{Armstrong:2020woi} for a computation of the four point Yang-Mills AdS amplitude for example. This concludes our discussion of Euclidean momentum space and spinor helicity variable correlators in CFT$_3$ for this section.

In the next few sections, we shall see several applications of what we have learnt so far. In particular, we discuss double copy relations between three point correlators and a flat space limit from the AdS$_4$ perspective. We also apply momentum space techniques to Chern-Simons matter theories which are an interesting class of 3d conformal field theories. We show that spinor helicity variables provide a pathway to identify the holographic dual of chiral higher spin theory in AdS$_4$.

\section{Double Copy Relations}\label{sec:DoubleCopy}
The traditional double copy is a famous observation between scattering amplitudes in gauge theory and gravity, see \cite{Adamo:2022dcm} for a review. Roughly speaking, the idea is that the ``square" of gauge theory amplitudes are amplitudes in gravity. For instance consider the three point MHV Gluon and Graviton scattering amplitudes in Yang-Mills and Einstein gravity respectively, We have,
\begin{align}
    &A^{--+}_{3,YM}=\frac{\langle 12\rangle^4}{\langle 1 2\rangle\langle 2 3\rangle 3 1\rangle},~~ A^{--+}_{3,GR}=\bigg(\frac{\langle 12\rangle^4}{\langle 1 2\rangle\langle 2 3\rangle 3 1\rangle^2}\bigg)\notag\\
    &\implies (A_{3,YM}^{--+})^2=A^{--+}_{3,GR}.
\end{align}
As for the amplitudes arising due to higher derivative $F^3$ and $R^3$ three point amplitudes, there exists a similar relation,
\begin{align}
    A_{3,F^3}^{---}=\langle 12\rangle\langle 23\rangle\langle 3 1\rangle,~A_{3,R^3}=(A_{3,F^3}^{---})^2.
\end{align}
The question we ask is whether similar relations hold for AdS$_4$ amplitudes and therefore CFT$_3$ correlators? Lets work with our spinor helicity variables. At the level of three points, recall from the previous section that there are three independent solutions: homogeneous even, homogeneous odd and a non-homogeneous contribution. For the homogeneous part we derived a general expression. In the $(---)$ helicity we have (supressing the overall coefficient),
\begin{align}
    \langle J_{s_1}^{-}J_{s_2}^{-}J_{s_3}^{-}\rangle_h=\frac{\langle 1 2\rangle^{s_1+s_2-s_3}\langle 2 3\rangle^{s_3+s_1-s_2}\langle 3 1\rangle^{s_3+s_1-s_2}}{E^{s_1+s_2+s_3}}.
\end{align}
This immediately leads to an infinite number of three point double copy relations \cite{Jain:2021qcl}:
\begin{align}
    \langle J_{s_1}^{-}J_{s_2}^{-}J_{s_3}^{-}\rangle_h=\langle J_{s_1'}^{-}J_{s_2'}^{-}J_{s_3'}^{-}\rangle_h \langle J_{s_1''}^{-}J_{s_2''}^{-}J_{s_3''}^{-}\rangle_h,~s_i'+s_i''=s_i.
\end{align}
A special case of this includes,
\begin{align}
    \langle TTT\rangle_{h}=\langle JJJ\rangle_h^2,
\end{align}
which from the bulk perspective implies that the $\text{AdS}_4$ graviton three point amplitude due to the $R^3$ vertex is the square of the corresponding gauge theory gluon amplitude due to the $F^3$ vertex. 

For the non-homogeneous correlators there do not exist such nice double copy relations such as between the Einstein gravity three point correlator and the Yang Mills three point correlator. The situation at higher points is also not yet as well understood and is an important area of research, see \cite{Lipstein:2019mpu,Albayrak:2020fyp,Lipstein:2023pih} and references therein.

\section{Connection to 4d S-Matrices}\label{sec:flatlimit}
There is another interesting aspect about correlators in momentum space and spinor helicity variables. It turns out that if we take a particular limit, they turn into one-higher dimensional scattering amplitudes \cite{Maldacena:2011nz,Raju:2012zr}! Let us illustrate this with the example of Einstein gravity three point function. In the $(--+)$ helicity we have the following complicated result \cite{Jain:2024bza}:
\tiny
\begin{align}
    &\langle T^{-}T^{-}T^{+}\rangle\notag\\&=\frac{\langle 12\rangle^{6}}{96 \langle 2 3\rangle^2 \langle 3 1\rangle^2 p_1^3 p_2^3 p_3^2 E^2}((p_1-p_2)^2-p_3^2)^2\bigg((p_1+p_2)^2(5p_1^2+2p_1p_2+5p_2^2)+(p_1+p_2)(9 p_1^2+2 p_1 p_2+9 p_2^2)p_3+(9 p_1^2+16 p_1 p_2+9 p_2^2)p_3^2+5(p_1+p_2)p_3^3\bigg),
\end{align}
\normalsize
where $E=p_1+p_2+p_3$. Lets now do the following. First replace all $p_3=E-p_1-p_2$. Then take the limit $E\to 0$. We obtain,
\begin{align}
    \lim_{E\to 0}\langle T^{-}T^{-}T^{+}\rangle=\frac{1}{E^2}\frac{\langle 12\rangle^6}{\langle 23\rangle^2\langle 3 1\rangle^2}+\cdots.
\end{align}
The coefficient of the most singular term is nothing but the flat space scattering amplitude of three gravitons in Einstein's theory of gravity! This is a generic feature of three point correlators in CFT$_3$. The coefficient of the leading pole in the limit $E\to 0$ is the corresponding flat space scattering amplitude. For more work in this direction we refer the reader to the literature with references to be provided soon. 
\begin{exercise}
    Compute the four point Witten diagram in AdS $\phi^4$ theory for conformally coupled scalars. Take the flat space limit and show that we obtain the familiar constant result. Why is $E\to 0$ the flat space limit? 
\end{exercise}

\section{Correlators in higher spin theories}\label{sec:CSmattercorrelators}
Lets now study the applications of spinor helicity variables to Chern-Simons matter theories at large $N$ which also exhibit slightly broken higher spin symmetry \cite{Giombi:2011kc,Aharony:2011jz,Maldacena:2011jn,Maldacena:2012sf}. In terms of the computation of correlators in these theories, there has been a lot of work \cite{Giombi:2011rz,Maldacena:2011jn,Maldacena:2012sf,Aharony:2012nh,Chowdhury:2017vel,Sezgin:2017jgm,Skvortsov:2018uru,Yacoby:2018yvy,Chowdhury:2018uyv,Aharony:2018pjn,Jain:2020rmw,Jain:2020puw,Jain:2021gwa,Jain:2021whr,Caron-Huot:2021kjy,Jain:2021vrv,Jain:2021wyn,Jain:2021qcl,Gerasimenko:2021sxj,Scalea:2023dpw, Bedhotiya:2015uga,Turiaci:2018nua,Li:2019twz,Silva:2021ece,Kalloor:2019xjb,Jain:2022ajd,Jain:2023juk,Kukolj:2024yyo}. We focus on Chern-Simons coupled to fermions which is called the Quasi-Fermionic theory.

The (Euclidean) Action of this theory is given by,
\begin{align}
    S=\frac{k}{4\pi}\int d^3 x~ \epsilon^{\mu\nu\rho}(A^a_\mu \partial_\nu A^a_\rho+\frac{2}{3} f^{abc}A_{\mu}^a A_{\nu}^b A_{\rho}^c)+\int d^3 x \Bar{\psi}(\slashed{\partial}-\slashed{A})\psi.
\end{align}
The gauge field $A_{\mu}^a$ is in the adjoint representation of the gauge group $U(N)$ and the fermions are in the fundamental representation. This theory is clearly conformally invariant since the level $k$ is quantized due to invariance under large gauge transformations. We focus on large $N$ and large 
$k$ keeping the t'Hooft coupling $\lambda=\frac{N}{k}$ fixed. The singlet spectrum of this theory consists of a parity odd $\Delta=2+\order{\frac{1}{N}}$ scalar $O_2$, conserved spin-$1$ and spin-$2$ currents $J$, $T$ and weakly nonconserved currents $J_s,s=3,4,\cdots$ with $\Delta_s=s+1+\order{\frac{1}{N}}$. Recall our discussion in the earlier section about the higher spin symmetry of the free bosonic and free fermionic theories. We see that turning on a Chern-Simons gauge field breaks the higher spin symmetry at $\order{\frac{1}{N}}$ since that is the order at which the higher spin currents acquire an anomalous dimension.

This theory is dual to the $SU(N_b)$ critical bosonic theory. The latter is the obtained via a RG flow to the Wilson-Fisher fixed point starting with the below action:
\begin{align}
    S_{CB,CS}=\int d^3 x\bigg(D_\mu \bar{\phi}D_\mu \phi+\frac{\lambda_4}{N_b}(\bar{\phi}\phi)^2+i\epsilon^{\mu\nu\rho}\frac{\kappa_b}{4\pi}\Tr{A_\mu\partial_\nu A_\rho-\frac{2i}{3}A_\mu A_\nu A_\rho}\bigg).
\end{align}
We work at large $N$ i.e. $N_b\to\infty$ as well as $\lambda_4\to \infty$, $\kappa_b\to \infty$. We do this while keeping fixed $\frac{\lambda_4}{N_b}$ and $\lambda_b=\frac{N_b}{\kappa_b}$.  The spectrum of this theory includes a parity even $\Delta=2+\order{\frac{1}{N_b}}$ scalar $O_2$, conserved spin-$1$ and spin-$2$ currents $J$ and $T$ and weakly nonconserved currents $J_s,s=3,4,\cdots$ with $\Delta=s+1+\order{\frac{1}{N_b}}$. The non-conservation starts at $\order{\frac{1}{N}}$:
\begin{align}
    \partial\cdot J_s=\order{\frac{1}{N}}.
\end{align}
These theories are also conjectured to be dual to Vasiliev's parity violating higher spin theory in $\text{AdS}_4$ \cite{Giombi:2011kc}. Lets now discuss correlators in these theories. We have already found the general form of two and three point correlators in generic conformal field theories up to coefficients, see \eqref{JsJspp}, \eqref{JsJsmm} and \eqref{genspins1s2s3momspace}.  These coefficients can be calculated explicitly since we have the action. At two points we obtain,
 \begin{equation}
     \langle J_{s}J_{s}\rangle_\text{QF}=\tilde{N}\langle J_{s}J_{s}\rangle_{\text{even}}+\tilde{N}\tilde{\lambda}\langle J_{s}J_{s}\rangle_{\text{odd}},
 \end{equation}
where the even and odd correlators are precisely what we discussed earlier. Note in particular that the parity odd contact term is produced as a one-loop contribution to the two point function thanks to the CS gauge field! The values of these constants in terms of the parameters of the theory are,
\begin{equation}
		\lambda_f=\frac{N_f}{\kappa_f}~,\tilde{N}=2N_f\frac{\sin\pi\lambda_f}{\pi\lambda_f},\quad\tilde{\lambda}=\tan\left(\frac{\pi\lambda_f}{2}\right)
	\end{equation}
Using the epsilon transform we obtain,
\begin{align}
     \langle J_{s}J_{s}\rangle_\text{QF}=\tilde{N}\langle J_{s}J_{s}\rangle_{\text{FF}}+\tilde{N}\tilde{\lambda}\langle \epsilon\cdot J_{s}J_{s}\rangle_{\text{FF}}.
\end{align}
Converting to spinor helicity variables yields,
 \begin{align}
     \langle J_s^{-}J_s^{-}\rangle_{QF}=\frac{N e^{i\pi\lambda_f}}{\pi\lambda_f}\langle J_{s}^{-}J_{s}^{-}\rangle_{FF}.
 \end{align}
 Note that the answer is simply a phase factor times the free fermionic result (which is the same as the even correlator).
 
Moving on to three points we find,
 \begin{align}\label{3ptansmz}
	 \langle J_{s_1}J_{s_2}J_{s_3}\rangle_{\text{QF}}=\frac{\tilde{N}}{1+\tilde{\lambda}^2}\bigg[ \langle J_{s_1}J_{s_2}J_{s_3}\rangle_\text{FF}+\tilde{\lambda}\langle J_{s_1}J_{s_2}J_{s_3}\rangle_\text{odd}+\tilde{\lambda}^2\langle J_{s_1}J_{s_2}J_{s_3}\rangle_\text{FB} \bigg].
	\end{align}
 When $\tilde{\lambda}=0$ (for which $\tilde{N}=N$ as well), the answer coincides with the free fermionic result. As we turn on the coupling, an odd contribution is generated due to the CS gauge field. However, as we go to strong coupling, the result is that of the free bosonic theory! This suggests a nice anyonic interpretation for CS matter theory correlators! This is made more manifest in spinor helicity variables\footnote{We used the fact that the odd correlator is the epsilon transform of the homogeneous correlator.}. \begin{align}\label{3ptsphcs}
		    \langle J_{s_1}J_{s_2}J_{s_3}\rangle_\text{QF}=&\frac{\tilde{N}}{2}\bigg[ \langle J_{s_1}J_{s_2}J_{s_3}\rangle_{\text{FF+FB}}+e^{-i\pi\lambda_f}\langle J_{s_1}J_{s_2}J_{s_3}\rangle_{\text{FF-FB}} \bigg]\nonumber\\
		    =&\tilde{N}e^{-\frac{i\pi\lambda_f}{2}}\bigg[ \cos{\frac{\pi\lambda_f}{2}}\langle J_{s_1}J_{s_2}J_{s_3}\rangle_\text{FF}+i\sin{\frac{\pi\lambda_f}{2}}\langle J_{s_1}J_{s_2}J_{s_3}\rangle_\text{FB} \bigg]
		\end{align}
	For, $\lambda_f=0$, we get the FF correlator and for $\lambda_f=1$, we get the FB correlator. This suggests a nice anyonic picture. Up to three points, the functional form of CFT correlators are fixed up to constants by symmetries. What is interesting about this theory is that we can actually go even further beyond and compute the correlators exactly at large $N$. In particular, the existence of the higher spin currents $J_s,s>2$ and their associated higher spin charges $Q_s\sim \int d\Sigma J_s$ lead to what are called slightly broken higher spin equations. In particular, lets focus on the spin-4 current $J_4^{\mu_1\mu_2\mu_3 \mu_4}$. Its associated charge is ($n_{\sigma}$ is the normal to the constant time slice on which this charge is defined),
    \begin{align}
        Q_4^{\mu\nu\rho}=\int d\Sigma~n_{\sigma}J_4^{\mu\nu\rho\sigma}(x).
    \end{align}
    The associated Ward-Takahashi identity reads,
    \begin{align}\label{HigherSpinWardId}
        \langle [Q_4^{\mu\nu\rho},J_{s_1}\cdots J_{s_n}]\rangle=\int d^3 x\langle \partial_{\sigma}J_4^{\mu\nu\rho\sigma}(x)J_{s_1}\cdots J_{s_n}]\rangle
    \end{align}
    \begin{exercise}
        Derive a general current conservation Ward-Takahashi identity from the path integral.
    \end{exercise}
    If the current is conserved, the RHS vanishes (throwing away the boundary term) and we obtain the higher spin charge Ward identities. However, in the CS-matter theory, the current is non-conserved at $\order{\frac{1}{N}}$. For spin-4, the divergence takes the form \cite{Jain:2022ajd},
    \begin{align}
        \partial_\sigma J^\sigma_{\mu\nu\rho}&=r_0\partial_{\mu}O(x)T_{\nu\rho}(x)+\epsilon_{\mu a b}\big( a_0 J_a(x)\partial_{\rho}\partial_{\nu}J_{b}(x)+b_1 J_a(x)\partial_{b}\partial_{\nu}J_{\rho}(x)+e_0 J_{\nu}(x)\partial_{a}\partial_{\rho}J_{b}(x)\big).
    \end{align}
    The constants $r_0,a_0,b_1,c_0$ all equal $\frac{1}{N}$ times a function of $\tilde{\lambda}$. Exact expressions can be found in \cite{Jain:2022ajd}. On the LHS of the Ward-Takahashi identity, we can figure out the algebra of $Q_4$ with any of the spin-s currents explicitly too. Further, given the fact that the RHS of \eqref{HigherSpinWardId} has a $\frac{1}{N}$ factor, only the disconnected part of the correlator appearing inside the integrand will contribute. This is called large-N factorization. This is to ensure that each of the two correlators the integrand each gives a factor of $N$ thus ensuring the expression is of $\order{N}$. This equation can in principle be used to solve for the CS+matter theory correlators. However, what is interesting is that we can actually avoid explicitly solving this equation. It turns out that writing an ansatz inspired by the two and three point functions, we can actually map this interacting theory equation to the free bosonic and free fermionic equations at each order \cite{Jain:2022ajd}. We find for instance,
    \begin{align}
        \langle J_{s_1}J_{s_2}J_{s_3}J_{s_4}\rangle_{\text{QF}}&=\frac{\tilde{N}}{(1+\tilde\lambda^2)}\bigg[\langle J_{s_1}J_{s_2}J_{s_3}J_{s_4}\rangle_\text{FF}+\tilde\lambda \langle\epsilon \cdot J_{s_1}J_{s_2}J_{s_3}J_{s_4}\rangle_\text{FF-CB}+\tilde\lambda^2\langle J_{s_1}J_{s_2}J_{s_3}J_{s_4}\rangle_\text{CB}\bigg]
    \end{align}
Here, $CB$ stands for the correlator in the critical bosonic theory which is obtained by starting with the free bosonic theory, deforming with a $\phi^4$ interaction and flowing to a IR fixed point. The good part is that at large $N$, all correlators of the CB theory are determined quite easily given its FB counterparts. In SH variables we obtain,
\begin{align}
    \langle J_{s_1}J_{s_2}J_{s_3}J_{s_4}\rangle_{\text{QF}}&=\tilde{N}e^{-\frac{i\pi \lambda_f}{2}}\big[\cos{\frac{\pi\lambda_f}{2}}\langle J_{s_1}J_{s_2}J_{s_3}J_{s_4}\rangle_{\text{FF}}+i \sin{\frac{\pi\lambda_f}{2}}\langle J_{s_1}J_{s_2}J_{s_3}J_{s_4}\rangle_{\text{CB}}\big],
\end{align}
    where we have suppressed the helicities of the operators since this is schematically the form in any helicity. One can even go further and constrain five and higher point functions using the slightly broken higher spin equatons, see  \cite{Jain:2022ajd}. However, it is important to note that the above solutions are one solution to the equations. There could be other solutions. Only explicit computations like in \cite{Kalloor:2019xjb} and the recent paper \cite{Kukolj:2024yyo} will confirm or rule out these solutions.

\section{Holography of Chiral higher spin theory}\label{sec:chiralhigherspin}

In our final application of spinor helicity, we shall turn our eyes towards holography. This section is based on \cite{Jain:2024bza}. See also \cite{Aharony:2024nqs} where the authors independently derive the same conclusions. The theory of interest to us is Chiral higher spin theory in AdS$_4$ \cite{Metsaev:2018xip,Skvortsov:2018uru,Skvortsov:2020wtf,Sharapov:2022awp}. It is a local, non-unitary, possibly UV finite theory of quantum gravity. It can be thought of as a higher spin version of self dual Yang Mills and self dual Gravity. Its spectrum contains massless particles of all positive integer spins but it only features a certain subset of all possible interactions. In particular, the (anti-)chiral theory contains vertices with only net (negative)positive helicities. For example, in the anti-chiral theory we have,
            \begin{align}
                A_{3,\text{graviton}}^{---},A_{3,\text{graviton}}^{--+}\ne 0~,~A_{3,\text{graviton}}^{+++}=A_{3,\text{graviton}}^{++-}=0.
            \end{align}
            Given this specific feature of the vertices, the question we ask is: What is the dual CFT?
First of all, the CFT must clearly be non-unitary due to the fact that there is a $(---)$ but no $(+++)$ vertex in the bulk which violates unitarity. Second, the CFT must possess a higher spin spectrum of currents dual to the massless gauge bosons in the bulk. Finally, by the usual rules of AdS/CFT, the CFT must be a large $N$ theory which from the bulk should represent a classical AdS geometry with fluctuations on top that represent the propagation of the higher spin particles. One clear candidate are the Chern-Simons matter theories. They are large N theories and possess an infinite tower of higher spin currents. However, they are unitary. What we shall show is that there actually exists a closed subsector of Chern-Simons matter theories that is dual to (anti-)Chiral higher-spin gravity in $\text{AdS}_4$. Clearly, the spinor helicity formalism is most well suited to address this question since chiral higher spin theory is also naturally formulated in the helicity basis.  Lets set up the notation first. We normalize our operators such that,
\begin{align}\label{rescaling}
    &\Tilde{J}_s^{\pm}=e^{\mp i \theta}J_s^{\pm}\notag\\
    &\Tilde{O}_2=\cos{\theta}~O_2,
\end{align}
where $\theta=\frac{\pi \lambda}{2}$ and $\lambda=\frac{N}{k}$ is the 't Hooft coupling in the Chern-Simons matter theory. Further, we normalize our currents such that two points are $\order{1}$ in contrast to the previous section. We work in Euclidean signature where  $J_s^{+}$ and $J_s^{-}$ are Hermitian conjugates of each other which the rescaling respects for $\theta\in\mathbb{R}$. However, for complex values of $\theta$, this is no longer true and $J_s^{+}$ and $J_s^{-}$ become independent quantities. 

We now state two limits that, as we shall show, project us into the anti-chiral or chiral sectors.
\begin{align}\label{anti-chirallimit}
    \textbf{anti-chiral limit:~} (\Tilde{N}\to \infty, \Tilde{\lambda}\to -i\equiv\theta\to -i\infty)~\textbf{with~}\frac{e^{i\theta}}{\sqrt{\Tilde{N}}}=g_{ac}, 
\end{align}
and,
\begin{align}\label{chirallimit}
    \textbf{chiral limit:~} (\Tilde{N}\to \infty, \Tilde{\lambda}\to i\equiv\theta\to i\infty)~\textbf{with~}\frac{e^{-i\theta}}{\sqrt{\Tilde{N}}}=g_{c}.
\end{align}
\subsection{Two points}
The two-point functions of the spin-$s$ currents in the QF theory are given by,
\begin{align}
    \langle J_s J_s\rangle_{QF}=\langle J_s J_s\rangle_{even}+\Tilde{\lambda}\langle J_s J_s\rangle_{odd}\,,
\end{align}
where the odd piece is given in terms of the even one through an epsilon transformation as we discussed earlier.

In spinor-helicity variables, the expression for the correlator in the two nonzero helicity configurations is given by(note that this and all other expressions in spinor-helicity variables are after the rescaling \eqref{rescaling}):
\begin{align}\label{2pointspins}
    \langle \Tilde{J}_s^{-}\Tilde{J}_s^{-}\rangle_{QF}=(1
    -i\Tilde{\lambda})\langle J_s^{-}J_s^{-}\rangle_{even}\,,\qquad\langle \Tilde{J}_s^{+}\Tilde{J}_s^{+}\rangle_{QF}=(1+i\Tilde{\lambda})\langle J_s^{+}J_s^{+}\rangle_{even}.
\end{align}
Let us now take the limit $\Tilde{\lambda}\to -i$ in \eqref{2pointspins}. In this limit, only the $(- -)$ configuration survives. We obtain,
\begin{align}
    \lim_{\Tilde{\lambda}\to -i}\bigg(\langle \Tilde{J}_s^{-}\Tilde{J}_s^{-}\rangle_{QF}=2\langle J_s^{-}J_s^{-}\rangle_{even}\,,\qquad\langle \Tilde{J}_s^{+}\Tilde{J}_s^{+}\rangle_{QF}=0\bigg).
\end{align}
Similarly, as $\Tilde{\lambda}\to i$, we obtain only the $(+ +)$ helicity configuration. 

Clearly, these limits work as desired at the level of two points, projecting us onto a negative helicity or positive helicity subspace.

\subsection{Three points}
We will show that \eqref{anti-chirallimit} and \eqref{chirallimit} do indeed produce (anti-)chiral three-point functions. Lets consider a correlator of a stress tensor with two $U(1)$ currents which from the bulk perspective represents a cubic interaction between a graviton and two photons. More examples can be found in \cite{Jain:2024bza}.
This correlator is given by,
\begin{align}\label{TJJcorr}
    \langle TJJ\rangle_{QF}=\frac{1}{\sqrt{\Tilde{N}}(1+\Tilde{\lambda}^2)}\big(\langle TJJ\rangle_{FF}+\Tilde{\lambda}~\epsilon\cdot\langle TJJ\rangle_{FF-FB}+\Tilde{\lambda}^2\langle TJJ\rangle_{FB}\big).
\end{align}
Using our knowledge about correlators inside the triangle,  we obtain the expression of this correlator in the eight helicity configurations:
\begin{align}
    &\langle \Tilde{T}^{-}\Tilde{J}^{-}\Tilde{J}^{-}\rangle=\frac{e^{i\theta}}{\sqrt{\Tilde{N}}}\langle T^{-}J^{-}J^{-}\rangle_h\,, \qquad&\langle \Tilde{T}^{+}\Tilde{J}^{+}\Tilde{J}^{+}\rangle&=\frac{e^{-i\theta}}{\sqrt{\Tilde{N}}}\langle T^{+}J^{+}J^{+}\rangle_h\notag\,, \\
    &\langle \Tilde{T}^{-}\Tilde{J}^{-}\Tilde{J}^{+}\rangle=\frac{e^{i\theta}}{\sqrt{\Tilde{N}}}\langle T^{-}J^{-}J^{+}\rangle_{nh}\,, \qquad&\langle \Tilde{T}^{+}\Tilde{J}^{+}\Tilde{J}^{-}\rangle&=\frac{e^{-i\theta}}{\sqrt{\Tilde{N}}}\langle T^{+}J^{+}J^{-}\rangle_{nh}\notag\,, \\
    &\langle \Tilde{T}^{-}\Tilde{J}^{+}\Tilde{J}^{-}\rangle=\frac{e^{i\theta}}{\sqrt{\Tilde{N}}}\langle T^{-}J^{+}J^{-}\rangle_{nh}\,, \qquad&\langle \Tilde{T}^{+}\Tilde{J}^{-}\Tilde{J}^{+}\rangle&=\frac{e^{-i\theta}}{\sqrt{\Tilde{N}}}\langle T^{+}J^{-}J^{+}\rangle_{nh}\notag\,, \\
    &\langle \Tilde{T}^{-}\Tilde{J}^{+}\Tilde{J}^{+}\rangle=0\,,&\langle \Tilde{T}^{+}\Tilde{J}^{-}\Tilde{J}^{-}\rangle&=0\,.
\end{align}
Let us now take the anti-chiral limit defined in \eqref{anti-chirallimit}. In this limit we obtain,
\begin{align}\label{TJJantichiral}
    &\langle \Tilde{T}^{-}\Tilde{J}^{-}\Tilde{J}^{-}\rangle=g_{ac}\langle T^{-}J^{-}J^{-}\rangle_h\,, \qquad&\langle \Tilde{T}^{+}\Tilde{J}^{+}\Tilde{J}^{+}\rangle&=0\notag\,, \\
    &\langle \Tilde{T}^{-}\Tilde{J}^{-}\Tilde{J}^{+}\rangle=g_{ac}\langle T^{-}J^{-}J^{+}\rangle_{nh}\,, \qquad&\langle \Tilde{T}^{+}\Tilde{J}^{+}\Tilde{J}^{-}\rangle&=0\notag\,, \\
    &\langle \Tilde{T}^{-}\Tilde{J}^{+}\Tilde{J}^{-}\rangle=g_{ac}\langle T^{-}J^{+}J^{-}\rangle_{nh}\,, \qquad&\langle \Tilde{T}^{+}\Tilde{J}^{-}\Tilde{J}^{+}\rangle&=0\notag\,, \\
    &\langle \Tilde{T}^{-}\Tilde{J}^{+}\Tilde{J}^{+}\rangle=0\,, &\langle \Tilde{T}^{+}\Tilde{J}^{-}\Tilde{J}^{-}\rangle&=0.
\end{align}
Equation \eqref{TJJantichiral} contains only the net negative helicity configurations of the correlators and is precisely the definition of the anti-chiral sector, showing that the limit \eqref{anti-chirallimit} is as we desired. In \cite{Jain:2024bza}, the authors show that this works for any three point function of the higher spin currents including those that involves the $O_2$ scalar operator.

Thus, what we see is that we have uncovered a subsector that is dual to anti-Chiral higher-spin gravity. Indeed, at the three-point level, the main signature of (anti-)Chiral higher-spin gravity is that it features cubic interactions with the total helicity being (negative) positive. 

\subsection{Beyond three-points}
Given the fact that we potentially have the results of all four and higher point functions in the CS+matter theory, we can try to take the chiral limit for these quantities.

The scalar correlator in the CS+fermionic matter theory is given by \cite{Bedhotiya:2015uga,Turiaci:2018nua},
\begin{align}
    \langle \Tilde{O}_2\Tilde{O}_2\Tilde{O}_2\Tilde{O}_2\rangle=\frac{\cos^4(\theta)(1+\Tilde{\lambda}^2)^2}{\Tilde{N}}\langle O_2O_2O_2O_2\rangle_{FF}=\frac{1}{\Tilde{N}}\langle O_2O_2O_2O_2\rangle_{FF}\,,
\end{align}
where we used $\Tilde{\lambda}=\tan(\theta)$. Therefore, in the chiral limit and the anti-chiral limits \eqref{chirallimit} and \eqref{anti-chirallimit}, this correlator goes to zero.

Lets now consider a correlator of a spin-s current with three scalars. This correlator in the two helicity configurations is given by \cite{Li:2019twz,Silva:2021ece},
\begin{align}
    &\langle \Tilde{J}_s^{-}\Tilde{O}_2\Tilde{O}_2\Tilde{O}_2\rangle=\frac{e^{i\theta}}{\Tilde{N}}\big(\cos(\theta)\langle J_s^{-}O_2O_2O_2\rangle_{FF}+\sin(\theta)\langle J_s^{-}O_2O_2O_2\rangle_{CB}\big)\notag\,,\\
    &\langle \Tilde{J}_s^{+}\Tilde{O}_2\Tilde{O}_2\Tilde{O}_2\rangle=\frac{e^{-i\theta}}{\Tilde{N}}\big(\cos(\theta)\langle J_s^{+}O_2O_2O_2\rangle_{FF}+\sin(\theta)\langle J_s^{+}O_2O_2O_2\rangle_{CB}\big)\,.
\end{align}
If we now take the anti-chiral limit \eqref{anti-chirallimit} we see that
\begin{align}\label{JsO2O2O2aclimit}
    &\langle \Tilde{J}_s^{-}\Tilde{O}_2\Tilde{O}_2\Tilde{O}_2\rangle=\frac{g^2}{2}\big(\langle J_s^{-}O_2O_2O_2\rangle_{FF}-i\langle J_s^{-}O_2O_2O_2\rangle_{CB}\big)\,,&&\langle \Tilde{J}_s^{+}O_2O_2O_2\rangle=0\,,
\end{align}
thus showing that the anti-chiral limit \eqref{anti-chirallimit} is well defined and indeed picks out just the anti-chiral sector for these types of four-point functions.

In the above examples, we notice a very nice separation between the chiral (net negative helicity) and anti-chiral (net positive helicity) sectors. However, by investigating several more general four-point functions, one finds that in addition to taking the (anti-)chiral limit, one needs to throw away some extra pieces to obtain finite results. A more detailed investigation remains a subject of a  future work. However, the bulk makes a lot of interesting predictions which we briefly discuss now. Lets consider a spin-2 four point function (four graviton scattering). The exchange diagram for such a process involves two cubic vertices. Lets take all gravitons to have negative helicity. Then, the only possible exchanges are the scalar and a graviton (the intermediate propagator connects minus to plus helicity). Any higher spin exchange would lead to one of the vertices having zero or positive helicity. Similarly, one can see that for fixed external spins, only finitely many exchanges are possible. For example, we can use this to show that some correlators are zero. Consider the scattering of a graviton and three scalars. The scalar exchange does not contribute as the scalar three point function is zero. The spin-1 exchange does not contribute since there are no such interaction vertices. The spin-2 exchange itself will also lead to a net positive or zero helicity vertex which are both zero in the anti-chiral theory. 
\begin{exercise}
    Analyze the loop expansion in AdS chiral higher spin theory focusing on two and four point functions. 
    \hint{You can find details in \cite{Jain:2024bza}.}
\end{exercise}

\section{Aspects of Lorentzian CFT}\label{sec:LorentzianCFT}
So far we have confined ourselves mostly to Euclidean signature. The following section will deal with Lorentzian signature CFTs. The reason is that the twistor formulation that is the subject of the second and third part of these notes is most naturally formulated in $\mathbb{R}^{2,1}$ rather than its Euclidean counterpart. We recap the discussion of Lorentzian signature spinors and discuss Wightman functions of conserved currents which are our observables of interest. References on the analysis of various aspects of Lorentzian CFTs include \cite{Gillioz:2018mto,Gillioz:2019lgs,Gillioz:2019iye,Bautista:2019qxj,Gillioz:2020mdd,Gillioz:2020wgw,Karateev:2020axc,Gillioz:2021sce,Gillioz:2021kee,Nishikawa:2023zcv}. Our treatment will closely follow \cite{Bala:2025gmz}.
\subsection{CPT}\label{subsec:CPT}
The beginning of our discussion in this part was on spinors in $\mathbb{R}^{2,1}$. We found that a general 3-momentum vector can be written as,
\begin{align}
    p_{ab}=\lambda_{(a}\Bar{\lambda}_{b)},
\end{align}
with real and independent $\lambda,\Bar{\lambda}$. An important feature of many of the Lorentzian field theories we study is $CPT$ invariance. Lets see how $CPT$ acts on spinor variables. Since the correlators that we are interested in that are invariant under charge conjugation, we focus on parity and time-reversal.
\subsubsection*{Parity and time reversal in spinor helicity variables}
We work in $\mathbb{R}^{2,1}$ with the Minkowski coordinates $(t,x,z)$. A parity transformation is a flip of one of the spatial directions say,
\begin{align}\label{parity1}
    P:(t,x,z)\to (t,x,-z).
\end{align}
Time-reversal flips $t\to -t$ while keeping $x,z$ unchanged.
\begin{align}\label{timereversal1}
    T:(t,x,z)\to (-t,x,z).
\end{align}
Lets translate this statements to the spinor helicity variables. Our starting point is,
\begin{align}
   p_a^b=\frac{1}{2}\big(\lambda_a\Bar{\lambda}^b+\Bar{\lambda}_a\lambda^b\big).
\end{align}
In matrix notation,
\begin{align}\label{momentummatrix1a}
    p_a^b=\begin{pmatrix}
        p_z&&p_x-p_t\\
        p_x+p_t&& -p_z
    \end{pmatrix}=\begin{pmatrix}
        \frac{\lambda_1\Bar{\lambda}_2+\Bar{\lambda}_1\lambda_2}{2}&&-\lambda_1\Bar{\lambda}_1\\
        \lambda_2\Bar{\lambda}_2&&-\frac{\lambda_2\Bar{\lambda}_1+\Bar{\lambda}_2\lambda_1}{2}
    \end{pmatrix}.
\end{align}
The subscripts $1,2$ here refer to the spinor components.
We shall now derive the action of $P$ and $T$ on the spinors $\lambda$ and $\Bar{\lambda}$ by considering the action of parity and time-reversal on the momentum matrix .
\subsubsection*{Parity}
The parity transformation \eqref{parity1} clearly will act on the momentum matrix by flipping $p_z\to -p_z$:
\begin{align}
    P:\begin{pmatrix}
        p_z&&p_x-p_t\\
        p_x+p_t&& -p_z
    \end{pmatrix}\to \begin{pmatrix}
        -p_z&&p_x-p_t\\
        p_x+p_t&& p_z
    \end{pmatrix}.
\end{align}
Using the expressions of each component of this matrix in terms of the spinors \eqref{momentummatrix1a}, we can write this transformation as follows:
\begin{align}
    P:\big\{\lambda_1\Bar{\lambda}_2+\Bar{\lambda}_1\lambda_2\to-\lambda_1\Bar{\lambda}_2-\Bar{\lambda}_1\lambda_2,\lambda_1\Bar{\lambda}_1\to\lambda_1\Bar{\lambda}_1,\lambda_2\Bar{\lambda}_2\to \lambda_2\Bar{\lambda}_2,\lambda_2\Bar{\lambda}_1+\Bar{\lambda}_2\lambda_1\to -\lambda_2\Bar{\lambda}_1-\Bar{\lambda}_2\lambda_1\big\}.
\end{align}
This can be implemented by,
\begin{align}\label{parityonspinors}
    P:\begin{pmatrix}
        \lambda_1,&\lambda_2
    \end{pmatrix}\to\begin{pmatrix}
        -\Bar{\lambda}_1,&\Bar{\lambda}_2
    \end{pmatrix},\begin{pmatrix}
        \Bar{\lambda}_1,&\Bar{\lambda}_2
    \end{pmatrix}\to\begin{pmatrix}
        -\lambda_1,&\lambda_2
    \end{pmatrix}.
\end{align}
\subsubsection*{Time-reversal}
The time-reversal transformation \eqref{timereversal1} flips $p_z\to -p_z$ and $p_x\to -p_x$. Also, it is important to remember that time-reversal is an anti-unitary operator and does not flip the time component of the momentum (energy).
\begin{align}
    T:\begin{pmatrix}
        p_z&&p_x-p_t\\
        p_x+p_t&& -p_z
    \end{pmatrix}\to \begin{pmatrix}
        -p_z&&-p_x-p_t\\
        -p_x+p_t&& p_z
    \end{pmatrix}.
\end{align}
Using the expressions of each component of this matrix in terms of the spinors \eqref{momentummatrix1a}, we can write this transformation as follows:
\begin{align}
    T:\big\{\lambda_1\Bar{\lambda}_2+\Bar{\lambda}_1\lambda_2\to-\lambda_1\Bar{\lambda}_2-\Bar{\lambda}_1\lambda_2,\lambda_1\Bar{\lambda}_1\to\lambda_2\Bar{\lambda}_2,\lambda_2\Bar{\lambda}_2\to \lambda_1\Bar{\lambda}_1,\lambda_2\Bar{\lambda}_1+\Bar{\lambda}_2\lambda_1\to -\lambda_2\Bar{\lambda}_1-\Bar{\lambda}_2\lambda_1\big\},
\end{align}
which can be implemented via,
\begin{align}
     T:\begin{pmatrix}
        \lambda_1,&\lambda_2
    \end{pmatrix}\to\begin{pmatrix}
        \lambda_2,&-\lambda_1
    \end{pmatrix},\begin{pmatrix}
        \Bar{\lambda}_1,&\Bar{\lambda}_2
    \end{pmatrix}\to\begin{pmatrix}
        \Bar{\lambda}_2,&-\Bar{\lambda}_1
        \end{pmatrix}.
\end{align}
\subsubsection*{PT}
\begin{exercise}
    Show that under parity and time-reversal that the spinor brackets transform as,
   \begin{align}\label{PandTspinorsfinal}&P\big(\langle \lambda \chi\rangle\big)=\langle \Bar{\chi}\Bar{\lambda}\rangle~,~P\big(\langle \Bar{\lambda}\Bar{\chi}\rangle\big)=\langle \chi \lambda\rangle~,~P\big(\langle \lambda \Bar{\lambda}\rangle\big)=\langle \lambda \Bar{\lambda}\rangle,\notag\\
        & T\big(\langle \lambda \chi\rangle\big)=\langle \lambda \chi\rangle~,~T\big(\langle \Bar{\lambda} \Bar{\chi}\rangle\big)=\langle \Bar{\lambda} \Bar{\chi}\rangle~,~T\big(\langle \lambda \Bar{\lambda}\rangle\big)=\langle \lambda \Bar{\lambda}\rangle.
    \end{align}
Using these results, show that,
\begin{align}\label{PTspinors}
        PT\big(\langle \lambda \chi\rangle\big)=\langle \Bar{\chi}\Bar{\lambda}\rangle~,~PT\big(\langle\Bar{\lambda}\Bar{\chi}\rangle\big)=\langle \chi \lambda\rangle~,~ PT\big(\langle \lambda \Bar{\lambda}\rangle\big)=\langle \lambda \Bar{\lambda}\rangle.
    \end{align}
\end{exercise}
This concludes our discussion on how $PT$ acts on the spinor helicity variables. The next thing to do is to consider the observables of interest to us. These are Wightman functions of conserved currents in the helicity basis. To start, lets first precisely define what a Wightman function really is. For more details, please refer to David Duffins' TASI lecture notes on Conformal Field Theory in Lorentzian Signature available on his \href{http://theory.caltech.edu/~dsd/.}{Caltech home page}.

\subsection{Definitions and generalities}\label{subsec:Definitions}

A Wightman function (in the Vacuum state) is the expectation value of a product of operators inserted at different locations:
\begin{align}\label{WightmanFunction}
    W_n(x_1,\cdots, x_n)=\langle 0|O_{1}(t_1,\Vec{x}_1)\cdots O_n(t_n,\Vec{x}_n)|0\rangle.
\end{align}
 $O_i$, $i=1,\cdots,n$ are generic operators with or without spin. Clearly, there are $n!$ distinct Wightman functions due to that many possible orderings of operators. However, if any two operators are space-like separated they commute as a consequence of microcausality. In the Heisenberg picture (and assuming that the Vacuum is Poincare invariant) we can write \eqref{WightmanFunction} as,
\small
\begin{align}
    &W(x_1,\cdots,x_n)\notag\\&=\langle 0|O_1(0)e^{-iH(t_1-t_2)-H(\epsilon_1-\epsilon_2)+i \Vec{P}\cdot(\Vec{x}_1-\Vec{x}_2)}O_2(0)\cdots e^{-iH(t_{n-1}-t_n)-H(\epsilon_{n-1}-\epsilon_n)+i\Vec{P}\cdot(\Vec{x}_{n-1}-\Vec{x}_n)}O_n(0)|0\rangle,
\end{align}
\normalsize
where we have used the $i\epsilon$ prescription with the constraint $\epsilon_1>\epsilon_2>\cdots>\epsilon_n$ which defines this Wightman function has a particular operator ordering. This prescription serves to exponentially suppress contributions from high energy states. The limit $\epsilon_i\to 0$ (keeping the ordering fixed which) is only to be taken after smearing with the Schwartz functions, see the notes by David Duffins available at his \href{http://theory.caltech.edu/~dsd/.}{Caltech home page}. One can form other correlation functions of interest if we have obtained the Wightman functions. For instance, the $n$ point time ordered correlator is a sum of the $n!$ different Wightman functions multiplied by the Heaviside theta functions that enforce the time ordering:
\small
\begin{align}\label{TimeOrderedCorr}
    \langle 0|T\{O_1(t_1,\Vec{x}_1)O_2(t_2,\Vec{x}_2)\cdots O_n(t_n,\Vec{x}_n)\}|0\rangle=\theta(t_1>t_2>\cdots>t_n)\langle 0|O_1(t_1,\Vec{x}_1)O_2(t_2,\Vec{x}_2)\cdots O_n(t_n,\Vec{x}_n)|0\rangle+\cdots.
\end{align}
\normalsize
\begin{exercise}
    Consider the Wightman function of a $U(1)$ current and two scalars $\phi$ and $\chi$ with charges $q_{\phi}$ and $q_{\chi}$. As a consequence of conservation we have,
\begin{align}\label{WightmanZeroWT}
    \partial_{1\mu}\langle 0|J^\mu(x_1)\phi(x_2)\chi(x_3)|0\rangle=0.
\end{align}
Show that the Ward-Takahashi identity for a time-ordered correlator can be derived from its definition in terms of Wightman functions \eqref{TimeOrderedCorr}:
\begin{align}\label{timeorderedcorrWT}
    \partial_{1\mu}\langle 0|T\{J^\mu(x_1)\phi(x_2)\chi(x_3)\}|0\rangle=(q_\phi \delta^d(x_1-x_2)+q_\chi \delta^d(x_1-x_3))\langle 0|T\{\phi(x_2)\chi(x_3)\}|0\rangle.
\end{align} 
Integrating the above equation with respect to $x_1$ over the entire manifold $\mathbb{R}^{1,d-1}$ and using Gauss' theorem show that,
\begin{align}
    0=(q_{\phi}+q_{\chi})\langle 0|T\{\phi(x_2)\chi(x_3)\}|0\rangle\implies q_\chi=-q_\phi,
\end{align}
thus imposing conservation of charge.
\end{exercise}

\subsection{Euclidean$\to$Wightman correlator}\label{subsec:EuclidtoWightmangeneral}
It is useful to understand how to compute the Wightman correlator given a Euclidean space correlator since the latter is naturally computed using the standard path integral for instance. Given a Euclidean correlator
\begin{align}
    \langle O_{1}(\tau_1,\Vec{x}_1)\cdots O_{n}(\tau_n,\Vec{x}_n)\rangle,
\end{align}
one can obtain the Wightman function \eqref{WightmanFunction} as follows. Analytically continue the Euclidean times $\tau_i=it_i+\epsilon_i$. We now take $\epsilon_i\to 0$ keeping fixed the ordering $\epsilon_1>\epsilon_2>\cdots>\epsilon_n$ to maintain the Euclidean time ordering that serves to damp contributions from high energy states. This results in the Wightman function \eqref{WightmanFunction}. 
\begin{align}\label{PosSpaeEuclidToWightman}
    W(x_1,\cdots,x_2)= \langle O_{1}(t_1+i\epsilon_1,\Vec{x}_1)\cdots O_{n}(t_n+i\epsilon_n,\Vec{x}_n)\rangle~,\epsilon_1>\epsilon_2>\cdots>\epsilon_n.
\end{align}
Since our interest is ultimately to work in momentum space, one can attempt to perform this Wick rotation directly there. At the level of two and three points, this process is clear. The situation at higher points is more complicated and to the best of my knowledge is not worked out explicitly anywhere. When we say $n-$point below, we actually mean $n=2,3$ and are only schematic for higher points.

Consider an $n-$point momentum space Euclidean correlator,
\begin{align}\label{EuclidAnsatz}
    \langle J_{s_1}(z_1,p_1)\cdots J_{s_n}(z_n,p_n)\rangle=\sum_{\mathcal{I}}\mathcal{T}_{\mathcal{I}}(\{z_j\cdot z_k,z_j\cdot p_k,\cdots\})\mathcal{F}_{\mathcal{I}}(\{p_j\cdot p_k\}),
\end{align}
where the sum runs over linearly independent tensor structures $\mathcal{T}_{\mathcal{I}}$ and $\mathcal{F}_{\mathcal{I}}$ are the associated form factors that are a function of a linearly independent set of the Lorentz invariants constructed out of the momenta. We also assume conformal invariance in the above and what is to follow.
Given the knowledge about the analytic structure of the Euclidean CFT correlators, one can obtain the analogous relation in momentum space.
This is essentially a two step process:
\begin{itemize}
\item Analytically continue $p^0\to i p^0.$

\item The next step is to take the discontinuity of the Wick rotated Euclidean form factors with respect to the Lorentz invariant quantities constructed out of the momenta. This yields the corresponding Wightman function form factors.
Consider the Euclidean correlator form factors $\mathcal{F}_{\mathcal{I}}(\{p_j\cdot p_k\})$ \eqref{EuclidAnsatz}. Let us focus on the Lorentz invariant $p_1^2$. When Wick rotated, the form factor has a branch cut starting at $p_1^2=0$. Thus, we must take a discontinuity with respect to this variable. Proceeding this way, one obtains discontinuities with respect to the other Lorentz invariants. Note that the tensor structures $\mathcal{T}_{\mathcal{I}}$ do not depend on these quantities and hence their analytic continuation is trivial. The interested reader can refer \cite{Bautista:2019qxj,Baumann:2024ttn,Bala:2025gmz} for more details.
\end{itemize}
\subsection{Example of a Wightman function}\label{subsec:WightmanExample}
Since the Wightman functions of currents are identically conserved (no contact terms in Ward-Takahashi identities), their structure is much simpler than their Euclidean counterparts at the level of three points at least. For example we have \cite{Bala:2025gmz},
\begin{align}\label{TJJhelicities}
     &\langle 0|T^{-}J^{-}J^{-}|0\rangle=(c_{211}^{(h)}-i c_{211}^{odd})\frac{\langle 1 2\rangle^2\langle 3 1\rangle^2 p_1}{E^4}+4c_{211}^{(nh)}\frac{\langle 12\rangle^2\langle 3 1\rangle^2 p_1}{(E-2p_3)^2(E-2p_2)^2},\notag\\&\langle 0|T^{-}J^{-}J^{+}|0\rangle=(c_{211}^{(h)}-i c_{211}^{odd})\frac{\langle 1 2\rangle^2\langle \Bar{3}1\rangle^2 p_1}{(E-2p_3)^4}+4c_{211}^{(nh)}\frac{\langle 12\rangle^2\langle \Bar{3} 1\rangle^2 p_1}{(E-2p_1)^2 E^2},\notag\\
     &\langle 0|T^{-}J^{+}J^{-}|0\rangle=(c_{211}^{(h)}-i c_{211}^{odd})\frac{\langle 1 \Bar{2}\rangle^2\langle 3 1\rangle^2 p_1}{(E-2p_2)^4}+4c_{211}^{(nh)}\frac{\langle 1\Bar{2}\rangle^2\langle 3 1\rangle^2 p_1}{(E-2p_1)^2 E^2},\notag\\&\langle 0|T^{+}J^{-}J^{-}|0\rangle=(c_{211}^{(h)}-i c_{211}^{odd})\frac{\langle \Bar{1} 2\rangle^2\langle 3 \Bar{1}\rangle^2 p_1}{(E-2p_1)^4}+4c_{211}^{(nh)}\frac{\langle \Bar{1}2\rangle^2\langle 3 \Bar{1}\rangle^2 p_1}{(E-2p_3)^2(E-2p_2)^2},\notag\\
     &\langle 0|T^{+}J^{+}J^{+}|0\rangle=(c_{211}^{(h)}+ic_{211}^{odd})\frac{\langle \Bar{1}\Bar{2}\rangle^2\langle\Bar{3}\Bar{1}\rangle^2 p_1}{E^4}+4c_{211}^{(nh)}\frac{\langle \Bar{1}\Bar{2}\rangle^2\langle \Bar{3} \Bar{1}\rangle^2 p_1}{(E-2p_3)^2(E-2p_2)^2},\notag\\&\langle 0|T^{+}J^{+}J^{-}|0\rangle=(c_{211}^{(h)}+ic_{211}^{odd})\frac{\langle \Bar{1}\Bar{2}\rangle^2\langle3\Bar{1}\rangle^2 p_1}{(E-2p_3)^4}+4c_{211}^{(nh)}\frac{\langle \Bar{1}\Bar{2}\rangle^2\langle 3 \Bar{1}\rangle^2 p_1}{(E-2p_1)^2 E^2},\notag\\
     &\langle 0|T^{+}J^{-}J^{+}|0\rangle=(c_{211}^{(h)}+ic_{211}^{odd})\frac{\langle \Bar{1}2\rangle^2\langle\Bar{3}\Bar{1}\rangle^2 p_1}{(E-2p_2)^4}+4c_{211}^{(nh)}\frac{\langle \Bar{1}2\rangle^2\langle \Bar{3} \Bar{1}\rangle^2 p_1}{(E-2p_1)^2 E^2},\notag\\&\langle 0|T^{-}J^{+}J^{+}|0\rangle=(c_{211}^{(h)}+ic_{211}^{odd})\frac{\langle 1\Bar{2}\rangle^2\langle\Bar{3}1\rangle^2 p_1}{(E-2p_1)^4}+4c_{211}^{(nh)}\frac{\langle 1\Bar{2}\rangle^2\langle \Bar{3} 1\rangle^2 p_1}{(E-2p_2)^2 (E-2p_3)^2}.
\end{align}
One can also start from the Wightman correlator and obtain the corresponding Euclidean correlators. We wont have time to go into this but the interested reader can see \cite{Baumann:2024ttn,Bala:2025gmz}.
\begin{exercise}
    Consider the free  $U(1)$ bosonic and fermionic theories. Construct the stress tensor and the $U(1)$ conserved currents in these theories. Compute the Euclidean space $\langle TJJ\rangle$ correlators. Form the homogeneous and non-homogeneous Euclidean correlators using \eqref{handnhinfreetheory}. Wick rotate these Euclidean correlators to their Wightman counterparts using the methods outlined in this section and verify the above expressions in the eight helicities. 
    \bonus{Obtain the parity odd correlator using the epsilon transform of the homogeneous correlator \eqref{hoddeptofheven} and obtain its Wightman counterpart. A useful reference is \cite{Bala:2025gmz}.}
\end{exercise}
This concludes the first part of these notes. We now move on to a discussion about twistors and their application to 3d CFT. 

\part{Twistors}\label{part:Twistors}
\section{The Geometry of Twistor Space}\label{sec:TwistorGeometry}
The subject of the second part of these notes is a Twistor formulation for 3d CFT. The flow and contents are heavily based on \cite{Bala:2025qxr} which contains more detail than is covered here. Now, without any further ado, let us begin.

The spacetime of interest to us is $\mathbb{R}^{2,1}$. The \textit{real Twistor space} associated to it is a subset $\mathbb{PT}$ of the real projective space $\mathbb{RP}^3$. It is spanned by projective coordinates $Z^A$ which are in the fundamental representation of Sp$(4)$, the double cover of the $2+1$ dimensional conformal group $SO(3,2)$. It can be written as a direct sum of fundamental representations of $SL(2,\mathbb{R})$  which is the double cover of the $2+1$ dimensional Lorentz group $SO(2,1)$:
\begin{align}\label{defoftwistorZA}
    Z^A=(\lambda^a,\Bar{\mu}_{a'}).
\end{align}
The projectiveness of these coordinates implies that we can construct the following charts to span the space:
\begin{align}
    U_i=\{(Z^i)^{-1}Z^A,Z^i\ne 0\},i\in\{1,2,3,4\}.
\end{align}
The connection to $\mathbb{R}^{2,1}$ is provided by the \textit{incidence relations},
\begin{align}\label{IncidenceRelation}
    \Bar{\mu}_a=-x_{ab} \lambda^b.
\end{align}
Here, $x_{ab}=(\sigma_\mu)_{ab}x^\mu$ is the usual contraction of the position vector $x^\mu$ with the Pauli matrices $\sigma_\mu$ to result in a symmetric $2\times 2$ matrix.\\
Lets analyze the incidence relations \eqref{IncidenceRelation}. Since $\lambda$ and $\Bar{\mu}$ are both valued in $\mathbb{R}^2$, \eqref{IncidenceRelation} defines a $\mathbb{R}^2\in \mathbb{R}^4$. However, we must take into account the fact that the coordinates are projective. Modding out by projective rescalings defines a $\frac{\mathbb{R}^2}{\mathbb{R}-\{0\}}\cong \mathbb{RP}^{1}\subset \mathbb{RP}^3$. Thus, given any point $x_{ab}\in \mathbb{R}^{2,1}$, we can associate a $\mathbb{RP}^{1}$ in twistor space which is topologically equivalent to a circle.\\
The non-locality of this correspondence works the other way around too. A point in twistor space can be defined as the intersection of two lines \eqref{IncidenceRelation} associated to points $x,y\in\mathbb{R}^{2,1}$. We have,
\begin{align}\label{xablambdabeq0}
x_{ab}\lambda^b=y_{ab}\lambda^b\implies (x-y)_{ab}\lambda^b=0.
\end{align}
Given the fact that $x_{ab}$ and $y_{ab}$ are symmetric matrices, it is easy to show that,
\begin{align}\label{nullpoints1}
     (x-y)_{ab}=\omega~\lambda^a \lambda^b,
\end{align}
for some $\omega\in \mathbb{R}$. Squaring this quantity results in zero showing that $x$ and $y$ are null separated. To summarize,
\begin{align}
    &\textbf{Point in}~\mathbb{R}^{2,1}\to \mathbb{RP}^1\in \mathbb{PT},\notag\\
    &\textbf{Point in}~\mathbb{PT}\to \textbf{Null ray in}~\mathbb{R}^{2,1}.
\end{align}
 Since null lines/light rays are invariant under conformal transformations, twistor space as it stands is not sensitive to a conformal factor multiplying the metric. This is where the \textit{infinity} twistor enters the game.  It breaks the conformal invariance and encodes the structure of spacetime at infinity. First, let us see how we can construct the $\mathbb{R}^{2,1}$ metric up to an overall scale from two twistors $Z_1^A,Z_2^A$ associated to a single point $x$ using the incidence relations \eqref{IncidenceRelation}. We construct the skew combination,
\begin{align}\label{XAB}
        &X^{AB}=Z_1^{[A}Z_2^{B]}=\frac{1}{2}\begin{pmatrix}
            \lambda_1^a\lambda_2^b-\lambda_1^b\lambda_2^a &&\lambda_1^a\Bar{\mu}_{2b'}-\Bar{\mu}_{1b'}\lambda_2^a\\
            \Bar{\mu}_{1a'}\lambda_2^b-\lambda_1^b\Bar{\mu}_{2a'}&&\Bar{\mu}_{1a'}\Bar{\mu}_{2b'}-\Bar{\mu}_{1b'}\Bar{\mu}_{2a'}.
        \end{pmatrix}=\frac{\langle 1 2\rangle}{2}\begin{pmatrix}
            \epsilon^{ab}&&-x^{a}_{b'}\\
            x^{b}_{a'}&& -\epsilon_{a'b'}x^2
        \end{pmatrix},
\end{align}
\begin{exercise}\label{ex:5sec3}
    Using the definition of the twistor \eqref{defoftwistorZA} and the incidence relation \eqref{IncidenceRelation} derive \eqref{XAB}. You also need to show that for any pair of spinors $\lambda^a$, $\chi^b$ that the following equality holds:
    \begin{align}
        \lambda^a \chi^b-\lambda^b \chi^a=\langle \lambda \chi\rangle \epsilon^{ab},
    \end{align}
    where $\langle \lambda \chi\rangle=\lambda_a \chi^a$.
    \bonus{Show that $X^2=\frac{1}{2}\epsilon_{ABCD}X^{AB}X^{CD}=0$ where $\epsilon_{ABCD}$ is given in \eqref{OmegaandEpsilon}. Therefore, we have $X^2=0$ and the fact that $X^{AB}\sim r X^{AB},r\in\mathbb{R}$. To the reader familiar with CFT, this should immediately bring to mind the embedding space approach with $X^{AB}$ interpreted as its coordinates. These conditions define a section of the null cone in $3+2$ dimensional space-time. Twistors from this perspective are sort of the square roots of the embedding space positions. For an excellent and foundational perspective on this approach to Twistors for $\mathbb{R}^{2,1}$, please refer to \cite{Baumann:2024ttn}.}
\end{exercise}
Using this quantity, one can construct the following natural line element
\begin{align}\label{conformalmetric}
    d\tilde{s}^2=\frac{1}{2}\epsilon_{ABCD}dX^{AB}dX^{CD}=\frac{1}{2}\text{Pf}\big(dX\big).
\end{align}
The four index Levi-Civita symbol in $\mathbb{RP}^{3}$ is constructed out of the Sp$(4)$ invariant anti-symmetric tensor $\Omega$ via,
\begin{align}\label{OmegaandEpsilon}
    \epsilon_{ABCD}=-\bigg(\Omega_{AB}\Omega_{CD}-\Omega_{AC}\Omega_{BD}+\Omega_{AD}\Omega_{BC}\bigg),~~ \Omega_{AB}=\begin{pmatrix}
        0&&\delta_{a}^{b'}\\
        -\delta_{a'}^b &&0
    \end{pmatrix}.
\end{align}
$\text{Pf}(dX)$ is the Pfaffian of the matrix $dX^{AB}$ formed using \eqref{XAB}.
Opening up \eqref{conformalmetric} using the variation of \eqref{XAB} and \eqref{OmegaandEpsilon} we obtain,
\begin{align}\label{conformalmetric1}
    d\tilde{s}^2=\langle 1 2\rangle^2(-dt^2+dx^2+dz^2)=\langle 1 2\rangle^2\eta_{\mu\nu}dx^\mu dx^\nu.
\end{align}
Note that \eqref{conformalmetric1} depends on $\langle 12\rangle^2$ and thus reproduces the Minkowski metric up to an overall conformal factor. To obtain the flat metric, we must introduce another bi-twistor $I_{AB}$ to cancel out this factor. Define instead of \eqref{conformalmetric},
\begin{align}\label{conformalmetric2}
    ds^2=\frac{\epsilon_{ABCD}dX^{AB}dX^{CD}}{2(I_{AB}X^{AB})^2}=\frac{\text{Pf}(dX)}{2(I\cdot X)^2}.
\end{align}
This also ensures that this metric is invariant under projective rescalings $Z\to r Z, r\in\mathbb{R}$ as is appropriate in $\mathbb{RP}^3$.
If we choose,
\begin{align}\label{infinitytwistor}
    I_{AB}=\begin{pmatrix}
        \epsilon_{ab}&&0\\
        0&&0,
    \end{pmatrix}
\end{align}
we obtain,
\begin{align}
    (I_{AB}X^{AB})^2=\langle 1 2\rangle^2.
\end{align}
Substituting this in \eqref{conformalmetric2} and using \eqref{conformalmetric1} results in the flat Minkowski metric of $\mathbb{R}^{2,1}$ viz,
\begin{align}
    ds^2=-dt^2+dx^2+dz^2.
\end{align}
Thus, we see that \eqref{infinitytwistor} is the infinity twistor that breaks the Sp$(4)$ conformal invariance and picks out the  $\mathbb{R}^{2,1}$ flat Minkowski metric that has only the usual Poincare invariance. This concludes our discussion on the basics of the geometry of real twistor space. We'll now introduce one of the main ingridients of twistor theory: The \textit{Penrose transform}.

\section{The Penrose Transform}\label{sec:PenroseTransform}
The aim of this subsection is to introduce the Penrose transform for symmetric traceless conserved currents \cite{Baumann:2024ttn,Bala:2025qxr}. We will generalize to generic operators subsequently. Given a symmetric traceless conserved current $J_s^{a_1\cdots a_{2s}}(x)$ we can represent it as follows to make manifest its conservation:
\begin{align}\label{PenroseTransform}
    J_s^{a_1\cdots a_{2s}}(x) = \int \langle \lambda d\lambda \rangle \lambda^{a_1} \cdots \lambda^{a_{2s}} \hat{J}_s^{+}(\lambda, \bar{\mu}) \big|_X,
\end{align}
where $X$ denotes imposing the incidence relation \eqref{IncidenceRelation} and the measure is the natural one on $\mathbb{RP}^1$. An important point to note about \eqref{PenroseTransform} is that the integrand should be invariant under projective rescalings since $\mathbb{PT}$ is a projective space. This in turn demands,
\begin{align}\label{Jsprojective}
    \hat{J}_s^{+}(r \lambda,r \Bar{\mu})=\frac{1}{r^{2s+2}}\hat{J}_s^{+}(\lambda,\Bar{\mu}).
\end{align}
Let us check that \eqref{PenroseTransform} represents a symmetric traceless conserved current. Tracelessness is obvious. Lets check conservation by taking the divergence of the current. This yields,
\begin{align}\label{PenroseTransformConservation}
    \frac{\partial}{\partial x^{a_1 a_2}}J_s^{a_1\cdots a_{2s}}(x)=\int \langle \lambda d\lambda\rangle \lambda^{a_1}\cdots\lambda^{a_{2s}}\frac{\partial \Bar{\mu}^a}{\partial x^{a_1a_2}}\frac{\partial}{\partial \Bar{\mu}^a}\hat{J}_s^{+}(\lambda,\Bar{\mu})|_{X}
\end{align}
Using the incidence relation \eqref{IncidenceRelation} we find that,
\begin{align}\label{mubarxderivative}
   \frac{\partial \Bar{\mu}^a}{\partial x^{a_1a_2}}=\lambda_b\big(-2\delta^b_{a_2}\delta^a_{a_1}+\epsilon_{a_1a_2}\epsilon^{ab}\big)=-2\lambda_{a_2}\delta^{a}_{a_1}+\epsilon_{a_1a_2}\lambda^a.
\end{align}
Substituting \eqref{mubarxderivative} in \eqref{PenroseTransformConservation} yields,
\begin{align}
     \frac{\partial}{\partial x^{a_1 a_2}}J_s^{a_1\cdots a_{2s}}(x)=\int \langle \lambda d\lambda\rangle \lambda^{a_1}\cdots\lambda^{a_{2s}}\big(-2\lambda_{a_2}\delta^a_{a_1}+\epsilon_{a_1a_2}\lambda^a)\frac{\partial}{\partial \Bar{\mu}^a}\hat{J}_s^{+}(\lambda,\Bar{\mu})|_{X}=0,
\end{align}
since $\lambda^{a_2}\lambda_{a_2}=0$ and $\epsilon_{a_1a_2}\lambda^{a_1}\lambda^{a_2}=0$. Note that this is true for any $J_s^{+}(\lambda,\Bar{\mu})$, thus showing that the Penrose transform \eqref{PenroseTransform} allows us to represent conserved currents in an unconstrained way similar to null vectors as we saw earlier \eqref{nullpoints1}. Note that the integrand of the Penrose transform \eqref{PenroseTransform} involves the object $\hat{J}_s^{+}(\lambda,\Bar{\mu})$. As well shall make much more precise subsequently, this object is none other than the twistor space counterpart of the positive helicity component of the current. There also exists a version of the Penrose transform for the negative helicity component which is,
\begin{align}\label{negativehelicityPenroseTransform}
    J_s^{a_1\cdots a_{2s}}(x)=\int \langle \lambda d\lambda\rangle \frac{\partial}{\partial \Bar{\mu}_{a_1}}\cdots \frac{\partial}{\partial \Bar{\mu}_{a_{2s}}}\hat{J}_s^{-}(\lambda,\Bar{\mu})|_X,~\text{with}~\hat{J}_s^{-}(r\lambda,r\Bar{\mu})=\frac{1}{r^{-2s+2}}\hat{J}_s^{-}(\lambda,\Bar{\mu}).
\end{align}
Finally, there exist Penrose transforms involving currents in the dual of Twistor space that form the subject of the below exercise:
\begin{exercise}
    Show that the dual Twistor space Penrose transforms,
    \begin{align}\label{dualPenroseTransform}
        &J_s^{a_1\cdots a_{2s}}(x)=\int \langle \Bar{\lambda}d\Bar{\lambda}\rangle\frac{\partial}{\partial \Bar{\mu}_{a_1}}\cdots\frac{\partial}{\partial\Bar{\mu}_{a_{2s}}}\hat{J}_s^{+}(\mu,\Bar{\lambda})|_X,\notag\\
        &J_s^{a_1\cdots a_{2s}}(x)=\int \langle \Bar{\lambda}d\Bar{\lambda}\rangle\Bar{\lambda}^{a_1}\cdots \Bar{\lambda}^{a_{2s}}\hat{J}_s^{-}(\mu,\Bar{\lambda})|_X,
    \end{align}
    with the dual incidence relations $\mu_a=x_{ab}\Bar{\lambda}^b$ also define a symmetric traceless conserved current. How should $\hat{J}_s^{\pm}(\mu,\Bar{\lambda})$ behave under projective rescalings? Contrast this with \eqref{Jsprojective} and \eqref{negativehelicityPenroseTransform}. The coordinates $W_A=(\mu,\Bar{\lambda})$ are called \text{dual-twistor} coordinates and are Fourier conjugate to $Z^A=(\lambda,\Bar{\mu})$.
    \bonus{By definition we have,
    \begin{align}\label{ZtoWfullFourier}
    \hat{J}_s^{\pm}(Z)=\int \frac{d^4 W}{(2\pi)^2}e^{i W\cdot Z}\hat{J}_s^{\pm}(W).
\end{align}
Using this formula, derive \eqref{dualPenroseTransform} starting from \eqref{PenroseTransform} for the positive helicity and \eqref{negativehelicityPenroseTransform} for its negative helicity counterpart.}
\end{exercise}
Before we proceed, let us note the interesting fact that the Penrose transforms \eqref{PenroseTransform} and \eqref{negativehelicityPenroseTransform} only require a single helicity to reconstruct the total position space operator! Thus, we can and will choose to work with just \eqref{PenroseTransform} due to its simpler form.

\section{Witten's half Fourier transform}\label{sec:WittenTransform}
How does all of this connect with the first part of the lectures on spinor helicity variables? Before we answer that question, we must first take a different route to twistor space: Through the vessel of Witten's half Fourier transform.\\
The operation of a half-Fourier transform to twistor space due to Witten \cite{Witten:2003nn} developed in the context of four dimensional scattering amplitudes can be adapted to conserved currents in three dimensions as follows:
\begin{align}\label{WittenTransform}
   \hat{J}_s^{\pm}(Z)=\hat{J}_s^{+}(\lambda,\Bar{\mu})=\int \frac{d^2 \Bar{\lambda}}{(2\pi)^2}e^{i\Bar{\lambda}\cdot \Bar{\mu}}\frac{J_s^{\pm}(\lambda,\Bar{\lambda})}{p^{s-1}}=\int\frac{d^2\Bar{\lambda}}{(2\pi)^2}e^{i\Bar{\lambda}\cdot\Bar{\mu}}\hat{J}_s^{\pm}(\lambda,\Bar{\lambda}),
\end{align}
where $p=-\frac{1}{2}\langle \lambda \Bar{\lambda}\rangle$ is the momentum magnitude in spinor helicity variables. As usual, we have defined the rescaled current $\hat{J}_s^{\pm}=\frac{J_s^{\pm}}{p^{s-1}}$ so that special conformal transformations act in a simple manner. One immediate consequence of the fact that $\hat{J}_s^{\pm}$ has helicity $\pm s$ translates through \eqref{WittenTransform} to $J_s^{\pm}$ satisfying,
\begin{align}
    \hat{J}_s^{\pm}(r Z)=\frac{1}{r^{\pm 2 s+2}}\hat{J}_s^{\pm}(Z),
\end{align}
which is nothing but the projective rescaling property of the currents in the Penrose transforms \eqref{PenroseTransform} and \eqref{negativehelicityPenroseTransform} indicating the equivalence of the two approaches. One important point to note is that for this transformation to make sense, we require that $\lambda$ and $\Bar{\lambda}$ are real and independent. This in turn requires the space-time to be the Lorentzian $\mathbb{R}^{2,1}$ as we discussed earlier \eqref{LorentzianReality}. 

Finally, similar to the Penrose transforms \eqref{PenroseTransform}, \eqref{negativehelicityPenroseTransform} having alternate dual versions \eqref{dualPenroseTransform}, the Witten transform \eqref{WittenTransform} also has a dual counterpart.
\begin{align}\label{WittendualTransform}
    \hat{J}_s^{\pm}(W)=\hat{J}_s^{+}(\mu,\Bar{\lambda})=\int \frac{d^2 \lambda}{(2\pi)^2}e^{-i\lambda\cdot\mu}\frac{J_s^{\pm}(\lambda,\Bar{\lambda})}{p^{s-1}}=\int \frac{d^2 \lambda}{(2\pi)^2}e^{-i\lambda\cdot\mu}\hat{J}_s^{\pm}(\lambda,\Bar{\lambda}).
\end{align}
\begin{exercise}
    Derive \eqref{ZtoWfullFourier} using \eqref{WittenTransform} and \eqref{WittendualTransform}.
\end{exercise}

\section{A Tale of Three Transforms}\label{sec:PenroseFromWittenandFourier}

The transform that all of us learn at kindergarten is none other than the Fourier transform. What we are going to do now is combine it with Witten's half Fourier transform \eqref{WittenTransform} and see what comes out. 

The Fourier transform for a spin-s symmetric traceless conserved current is given by,
\begin{align}\label{FourierTrans1}
    J_s^{a_1\cdots a_{2s}}(x^\mu)=\int \frac{d^3 p}{(2\pi)^3}e^{-2ip\cdot x} J_s^{a_1\cdots a_{2s}}(p^\mu).
\end{align}
Lets now express the momentum in spinor helicity variables. This also requires knowledge of how the measure transforms.
\begin{exercise}
    Using the fact that $p^\mu=\frac{1}{2}(\sigma^\mu)^a_b \lambda_a \Bar{\lambda}^b$ show,
    \begin{align}\label{projvsnonproj1}
         \int d^3 p~f(p^\mu)=\frac{1}{4\text{Vol}(GL(1,\mathbb{R}))}\int d^2\lambda d^2\Bar{\lambda}|\lambda\cdot\Bar{\lambda}|f(\lambda,\Bar{\lambda}).
    \end{align}
    where the volume of $GL(1,\mathbb{R})$ is defined as,
    \begin{align}
        \text{Vol}(GL(1,\mathbb{R}))=\int_{-\infty}^{\infty} \frac{dc}{|c|}.
    \end{align}
    This quantity is roughly speaking, the length of the real line. It is of course, infinite, but the integral on the RHS of \eqref{projvsnonproj1} is divergent as well which cancels out this volume factor leaving behind a finite answer provided $f$ is a well behaved function which is anyway assumed when writing the LHS of \eqref{projvsnonproj1}.
    \hint{Expand both sides of \eqref{projvsnonproj1} into their components.}
\end{exercise}
Using the result of this exercise, \eqref{FourierTrans1} becomes,
\begin{align}\label{FouriertoPenrosestep1}
    J_s^{a_1\cdots a_{2s}}(x^\mu)&=\frac{1}{4\text{Vol}(GL(1,\mathbb{R}))}\int \frac{d^2\lambda d^2\Bar{\lambda}}{(2\pi)^3}|\lambda\cdot \Bar{\lambda}|e^{i\Bar{\lambda}_a\lambda_b x^{ab}}J_s^{a_1\cdots a_{2s}}(\lambda,\Bar{\lambda})\notag\\
    &=\frac{1}{4\text{Vol}(GL(1,\mathbb{R}))}\int \frac{d^2\lambda d^2\Bar{\lambda}}{(2\pi)^3}|\lambda\cdot \Bar{\lambda}|e^{i\Bar{\lambda}_a\lambda_b x^{ab}}\epsilon^{a_1b_1}\cdots \epsilon^{a_{2s}b_{2s}}J_{sb_1\cdots b_{2s}}(\lambda,\Bar{\lambda}),
\end{align}
where in the second line we used the Levi-Civita symbol to lower the indices of the current. Let us now use the identity we derived in exercise \ref{ex:5sec3},
\begin{align}\label{epsilonabid}
    \epsilon^{ab}=\frac{\lambda^a\Bar{\lambda}^b-\Bar{\lambda}^a\lambda^b}{\lambda\cdot \Bar{\lambda}}.
\end{align}
Substituting this in \eqref{FouriertoPenrosestep1} yields,
\small
\begin{align}\label{FouriertoPenroseStep1next}
    J_s^{a_1\cdots a_{2s}}(x)=\frac{1}{4\text{Vol}(GL(1,\mathbb{R}))}\int \frac{d^2\lambda d^2\Bar{\lambda}}{(2\pi)^3}|\lambda\cdot \Bar{\lambda}|e^{i\Bar{\lambda}_a\lambda_b x^{ab}}\frac{(\lambda^{a_1}\Bar{\lambda}^{b_1}-\Bar{\lambda}^{a_1}\lambda^{b_1})\cdots (\lambda^{a_{2s}}\Bar{\lambda}^{b_{2s}}-\Bar{\lambda}^{a_{2s}}\lambda^{b_{2s}})}{((\lambda\cdot\Bar{\lambda})^2)^s}J_{sb_1\cdots b_{2s}}(\lambda,\Bar{\lambda}).
\end{align}
\normalsize
Using the definitions of the positive and negative helicity components of a current in \eqref{Jspm} and the fact that the mixed helicity components are zero, we can re-write the above as,
\begin{align}\label{FouriertoPenroseinterstep}
    J_s^{a_1\cdots a_{2s}}(x)=\frac{1}{4\text{Vol}(GL(1,\mathbb{R}))}\int \frac{d^2\lambda d^2\Bar{\lambda}}{(2\pi)^3}\frac{|\lambda\cdot \Bar{\lambda}|}{((\lambda\cdot\Bar{\lambda})^2)^s}&e^{i\Bar{\lambda}_a\lambda_b x^{ab}}\bigg(\lambda^{a_1}\cdots \lambda^{a_{2s}}\Bar{\lambda}^{b_1}\cdots \Bar{\lambda}^{b_{2s}}J_{sb_1\cdots b_{2s}}(\lambda,\Bar{\lambda})\notag\\
    &+(-1)^{2s}\Bar{\lambda}^{a_1}\cdots \Bar{\lambda}^{a_{2s}}\lambda^{b_1}\cdots \lambda^{b_{2s}}J_{sb_1\cdots b_{2s}}(\lambda,\Bar{\lambda})\bigg).
\end{align}
We can now re-label $\lambda\leftrightarrow \Bar{\lambda}$ in the second term which yields,
\begin{align}\label{FouriertoPenrosestep2}
    J_s^{a_1\cdots a_{2s}}(x)&=\frac{1}{2\text{Vol}(GL(1,\mathbb{R}))}\int \frac{d^2\lambda d^2\Bar{\lambda}}{(2\pi)^3}\lambda^{a_1}\cdots \lambda^{a_{2s}}e^{i\Bar{\lambda}_a\lambda_b x^{ab}}\frac{\Bar{\lambda}^{b_1}\cdots \Bar{\lambda}^{b_{2s}}}{|\lambda\cdot \Bar{\lambda}|^{s}}\frac{J_{sb_1\cdots b_{2s}}(\lambda,\Bar{\lambda})}{|\lambda\cdot \Bar{\lambda}|^{s-1}}\notag\\
    &=\frac{1}{2^{2s}\text{Vol}(GL(1,\mathbb{R}))}\int \frac{d^2\lambda d^2\Bar{\lambda}}{(2\pi)^3}\lambda^{a_1}\cdots \lambda^{a_{2s}}e^{i\Bar{\lambda}_a\lambda_b x^{ab}}\hat{J}_s^{+}(\lambda,\Bar{\lambda}),
\end{align}
where we have used the definition of the polarizations \eqref{3dpolarizationspinors} and identified the rescaled positive helicity current,
\begin{align}
    \hat{J}_s^{+}(\lambda,\Bar{\lambda})=\zeta_{+}^{b_1}\cdots \zeta_{+}^{b_{2s}}\frac{J_{sb_1\cdots b_{2s}}}{|\frac{\lambda\cdot \Bar{\lambda}}{2}|^{s-1}},
\end{align}
This rescaled current is precisely what features in the integrand of Witten's transform \eqref{WittenTransform}. Note that this quantity satisfies,
\begin{align}\label{rescaledJsprojective}
    \hat{J}_s^{+}(r \lambda,\frac{\Bar{\lambda}}{r})=\frac{1}{r^{2s}}\hat{J}_s^{+}( \lambda,\Bar{\lambda})
\end{align}
Therefore we can use the inverse of the Witten transform \eqref{WittenTransform},
\begin{align}\label{JsinverseWitten1}
    \hat{J}_s^{+}(\lambda,\Bar{\lambda})=\int d^2\Bar{\mu}~e^{-i\Bar{\lambda}\cdot \Bar{\mu}}\hat{J}_s^{+}(\lambda,\Bar{\mu}). 
\end{align}
Using \eqref{rescaledJsprojective}, this leads to the projective property,
\begin{align}\label{Jshatprojective1}
    \hat{J}_s^{+}(r\lambda,r\Bar{\mu})=\frac{1}{r^{2s+2}}\hat{J}_s^{+}(\lambda,\Bar{\mu}),
\end{align}
which is also what the current in the Penrose transform satisfies \eqref{Jsprojective}. Using \eqref{JsinverseWitten1} in \eqref{FouriertoPenrosestep2} results in,
\begin{align}\label{FouriertoPenrosestep3}
    J_s^{a_1\cdots a_{2s}}(x)&=\frac{1}{2^{2s}\text{Vol}(GL(1,\mathbb{R})}\int \frac{d^2\lambda d^2\Bar{\mu} d^2\Bar{\lambda}}{(2\pi)^3}\lambda^{a_1}\cdots \lambda^{a_{2s}}e^{-i\Bar{\lambda}_a(\Bar{\mu}^a-\lambda_b x^{ab})}\hat{J}_s^{+}(\lambda,\Bar{\mu})\notag\\
    &=\frac{1}{2^{2s}2\pi\text{Vol}(GL(1,\mathbb{R})}\int d^2\lambda d^2\Bar{\mu}\lambda^{a_1}\cdots \lambda^{a_{2s}}\delta^2(\Bar{\mu}^a-x^{ab}\lambda_b)\hat{J}_s^{+}(\lambda,\Bar{\mu})\notag\\
    &=\frac{1}{2^{2s}2\pi\text{Vol}(GL(1,\mathbb{R})}\int d^2\lambda \lambda^{a_1}\cdots \lambda^{a_{2s}} \hat{J}_s^{+}(\lambda,\Bar{\mu})|_X,
\end{align}
Thus we see that the incidence relation $X$ we discussed earlier \eqref{IncidenceRelation} appears naturally in \eqref{FouriertoPenrosestep3}. To bring this to the form of the Penrose transform \eqref{PenroseTransform}, we require one final projective integral identity. 
\begin{exercise}
Show that,
\begin{align}\label{projID1}
    f(r \lambda,r\Bar{\mu})=\frac{1}{r^2}f(\lambda,\Bar{\mu})\implies \frac{1}{\text{Vol}(GL(1,\mathbb{R}))}\int d^2\lambda f(\lambda,\Bar{\mu})=\int \langle \lambda d\lambda\rangle f( \lambda,\Bar{\mu}).
\end{align}
\end{exercise}
 We see that the integrand in \eqref{FouriertoPenrosestep3} satisifes this property by virtue of \eqref{Jshatprojective1} resulting in,
\begin{align}
    J_s^{a_1\cdots a_{2s}}(x)=\frac{1}{2^{2s}2\pi}\int \langle \lambda d\lambda\rangle \lambda^{a_1}\cdots \lambda^{a_{2s}}\hat{J}_s^{+}(\lambda,\Bar{\mu})|_X,
\end{align}
which is exactly the Penrose transform \eqref{PenroseTransform}! Therefore, the twistor space described by Witten's transform \eqref{WittenTransform} is exactly the same as the twistor space of Penrose that features in the Penrose transform \eqref{PenroseTransform}. To summarize \cite{Bala:2025gmz},
\begin{align}
    \textbf{Fourier+Witten Transform}\implies \textbf{Penrose Transform}.
\end{align}

\section{The Conformal generators and Ward identities}\label{sec:conformalWardIdTwistor}
\subsection{The Generators of Sp$(4)$}
Lets now figure out how the conformal generators act on $\hat{J}_s^{\pm}(Z)$ and $\hat{J}_s^{\pm}(W)$. This is easiest done through Witten's half-Fourier transforms \eqref{WittenTransform}, \eqref{WittendualTransform}. On the Twistor space currents $\hat{J}_s^{\pm}$, they act as \cite{Bala:2025gmz},
\begin{align}\label{action1twistora}
    P_{ab}=i\lambda_{(a}\frac{\partial}{\partial\Bar{\mu}^{b)}},&\qquad K_{ab}=i\Bar{\mu}_{(a}\frac{\partial}{\partial \lambda^{b)}},\notag\\
\Tilde{M}_{ab}=i \bigg(\lambda_{(a}\frac{\partial}{\partial\lambda^{b)}}+\Bar{\mu}_{(a}\frac{\partial}{\partial\Bar{\mu}^{b)}}\bigg),&\qquad  D=\frac{i}{2}\bigg(\lambda^a\frac{\partial}{\partial\lambda^a}-\Bar{\mu}^a\frac{\partial}{\partial\Bar{\mu}^{a}}\bigg).
\end{align}
Similarly, their dual-twistor counterparts are as follows:
\begin{align}\label{action2twistora}
    P_{ab}=-i\Bar{\lambda}_{(a}\frac{\partial}{\partial\mu^{b)}},\qquad& K_{ab}=-i\mu_{(a}\frac{\partial}{\partial \Bar{\lambda}^{b)}}, \notag\\
\Tilde{M}_{ab}=i \bigg(\Bar{\lambda}_{(a}\frac{\partial}{\partial\Bar{\lambda}^{b)}}+\mu_{(a}\frac{\partial}{\partial\mu^{b)}}\bigg),\qquad& D=\frac{i}{2}\bigg(\Bar{\lambda}^a\frac{\partial}{\partial\Bar{\lambda}^a}-\mu^a\frac{\partial}{\partial\mu^{a}}\bigg).
\end{align}
\begin{exercise}
    Show that all these generators can be packaged into $T_{AB} = Z_{(A} \frac{\partial}{\partial Z^{B)}}$ by writing it in block matrix form \cite{Bala:2025gmz}:
\begin{align}
    T_{AB} = \begin{pmatrix}
    -\bar{\mu}_{(a'} \frac{\partial}{\partial \lambda^{b)}} &  -\bar{\mu}_{a'} \frac{\partial}{\partial \bar{\mu}_{b}} + \lambda^{b'} \frac{\partial}{\partial \lambda^{a}}  \\[1em]
    -\bar{\mu}_{a'} \frac{\partial}{\partial \bar{\mu}_{b}}+ \lambda^{b} \frac{\partial}{\partial \lambda^{a'}} & +\lambda^{(a} \frac{\partial}{\partial\bar{\mu}_{b')}} 
    \end{pmatrix}
    =
    \begin{pmatrix}
    \textit{i} K_{a'b} &  -\textit{i} M^b_{a'} + \frac{2}{\textit{i}}\delta_{a'}^{b} D \\
    -\textit{i} M^b_{a'}+ \frac{2}{\textit{i}}\delta^{b}_{a'} D & -\textit{i} P^{ab'}
    \end{pmatrix}.
\end{align}
where $P,\; K,\; M,\; D$ are the conformal generators as given in \eqref{action1twistora}.  Note that we used the fact that $Z_A=\Omega_{BA}Z^B$, where $Z^A$ is given in \eqref{defoftwistorZA} and the invariant symplectic form $\Omega_{AB}$ is given in \eqref{OmegaandEpsilon}. 
Similarly, one can obtain the components of the dual twistor generator $W_{(A}\frac{\partial}{\partial W^{B)}}$ which takes a similar form but in terms of the dual-twistor component generators \eqref{action2twistora}.
\end{exercise}
Therefore, the action of $T_{AB}$ on the rescaled currents is as follows:
\begin{align}\label{TABtwistoranddual}
    [T_{AB},\hat{J}_s^{\pm}(Z)]=Z_{(A}\frac{\partial}{\partial Z^{B)}}\hat{J}_s^{\pm}(Z),\qquad[T_{AB},\hat{J}_s^{\pm}(W)]=W_{(A}\frac{\partial}{\partial W^{B)}}\hat{J}_s^{\pm}(W).
\end{align}
 \begin{exercise}
  Show that the $T_{AB}$ obey the $Sp(4)$ algebra:
\begin{align}
    [T_{AB},T_{CD}]=\Omega_{AC}T_{BD}+\Omega_{AD}T_{BC}+\Omega_{BC}T_{AD}+\Omega_{BD}T_{AC}.
\end{align}
This actually proves that the conformal algebra $\mathfrak{so(3,2)}$ is isomorphic to $\mathfrak{sp(4;\mathbb{R})}$.
 \end{exercise}
 \subsection{The Conformal Ward identities}
 The Twistor space conformal Ward identities take the form,
 \begin{align}\label{manifesttwistorconformalWard}
    \langle 0|\cdots [T_{AB},J_s^{\pm}]\cdots|0\rangle=0.
\end{align}
Recall however, that the action of the spinor helicity special conformal generator on a correlator is not zero but is proportional to the Ward-Takahashi identity. To circumvent dealing with this, we focus on Wightman functions which have a zero Ward-Takahashi identity as we discussed earlier (hence the notation of a ordered vacuum expectation value).
The helicity counting identities translate to,
\begin{align}\label{manifesttwistorhelicitycount}
    &h_j\langle 0|\cdots \hat{J}_{s_j}^{\pm}(Z_j)\cdots |0\rangle=-\frac{1}{2}\big(Z_j^A\frac{\partial}{\partial Z_j^A}+2\big)\langle 0|\cdots \hat{J}_{s_j}^{\pm}(Z_j)\cdots|0\rangle=\pm s_j\langle 0|\cdots \hat{J}_{s_j}^{\pm}(Z_j)\cdots|0\rangle,\notag\\
    &h_j\langle 0|\cdots \hat{J}_{s_j}^{\pm}(W_j)\cdots|0\rangle=\frac{1}{2}\big(W_j^A\frac{\partial}{\partial W_j^A}+2\big)\langle 0|\cdots \hat{J}_{s_j}^{\pm}(W_j)\cdots|0\rangle=\pm s_j\langle 0|\cdots \hat{J}_{s_j}^{\pm}(W_j)\cdots|0\rangle,
\end{align}
whose finite version is,
\begin{align}\label{helicityIdZandW}
    &\langle 0|\cdots \hat{J}_{s_j}^{\pm}(r~Z_j)\cdots\rangle=\frac{1}{r^{\pm 2 s_j+2}}|0\langle \cdots \hat{J}_{s_j}^{\pm}(Z_j)\cdots|0\rangle\notag\\
    &\langle 0|\cdots \hat{J}_{s_j}^{\pm}(r~W_j)\cdots|0\rangle=\frac{1}{r^{\mp 2 s_j+2}}\langle \cdots \hat{J}_{s_j}^{\pm}(W_j)\cdots|0\rangle,
\end{align}
which is exactly the projective rescaling property of the Penrose transform integrands \eqref{PenroseTransform}, \eqref{negativehelicityPenroseTransform}, \eqref{dualPenroseTransform}.

\subsection{Solutions to the Ward identities}
The solutions to the Twistor space conformal Ward identities \eqref{manifesttwistorconformalWard} come in two flavors: In \cite{Baumann:2024ttn,Bala:2025gmz}, the authors construct conformally invariant solutions by taking them to depend only on twistor symplectic dot products which belong to the set,
\begin{align}\label{twistordotprods}
    \big\{Z_i\cdot Z_j=-Z_{i}^{A}\Omega_{AB}Z_j^B~,~W_i\cdot W_j=W_{iA}\Omega^{AB}W_{jB}~,~W_i\cdot Z_j=W_{iA}Z_{j}^{A}\big\}.
\end{align}
Since $\Omega$ is an invariant of the conformal group Sp$(4)$, these dot products are clearly conformally invariant.

 There also exist another class of natural invariants that were considered in \cite{Bala:2025qxr}: Projective delta functions.
The object,
\begin{align}\label{firstdefdelta4}
    \delta^4(c_1 Z_1+\cdots+c_{n-1}Z_{n-1}+ Z_n),
\end{align}
is an invariant of Sp$(4)$ (It is actually an invariant of the larger $SL(4,\mathbb{R})$ group). To show this, consider an Sp$(4)$ transformation $Z_i\to M Z_i, M\in \text{Sp}(4)$. We have,
\begin{align}
    &\delta^4(c_1 Z_1+\cdots+c_{n-1}Z_{n-1}+Z_n)\to \delta^4(c_1 M Z_1+\cdots+c_{n-1}M Z_{n-1}+M Z_n)\notag\\&=\frac{1}{|\text{Det}(M)|} \delta^4(c_1 Z_1+\cdots+c_{n-1}Z_{n-1}+Z_n)=\delta^4(c_1 Z_1+\cdots+c_{n-1}Z_{n-1}+Z_n),
\end{align}
since symplectic transformations are subgroups of special linear transformations that preserve the volume element and hence $\text{Det}(M)=1$.
However, we still need to erase the arbitrariness in the parameters $c_i$ which we integrate over on the support of a function $f(c_1,c_2,\cdots c_{n-1})$,
\begin{align}\label{nptansatz}
    \mathcal{F}(Z_1,\cdots Z_n)=\int dc_1\cdots dc_{n-1}~ f(c_1,\cdots,c_{n-1})\delta^4(c_1 Z_1+\cdots  Z_n),
\end{align}
where the function $f$ can also depend on symplectic dot products of twistors \eqref{twistordotprods}. Since we are working in the real projective space $\mathbb{RP}^3$, we need to ensure that this quantity has good projective properties. More precisely we demand,
\begin{align}\label{nptprojective}
    \mathcal{F}(Z_1,\cdots,r Z_k,\cdots Z_n)=\frac{1}{r^{2\alpha_k+2}}\mathcal{F}(Z_1,\cdots,Z_k,\cdots Z_n),
\end{align}
for some $\alpha_k\in \mathbb{R}$. When constructing Wightman functions of conserved currents this amounts to imposing the helicity counting identity \eqref{helicityIdZandW} where $\alpha_k$ is identified with the helicity of the current with argument $Z_k$. One can also form these delta function invariants with just dual twistors or a mix of twistors and dual twistors since the symplectic form \eqref{OmegaandEpsilon} can be used to raise and lower indices as we please. Therefore, the general solution to the twistor space Ward identities is a sum of a function of the dot products \eqref{twistordotprods} and projective delta functions \eqref{nptprojective} with the helicity counting identity \eqref{helicityIdZandW} imposed to further restrict the functional form. We'll see this explicitly for two and three point functions now.
\subsection{Two and Three point functions of currents}
Lets start with two point functions. We'll work out the positive helicity case based on \cite{Baumann:2024ttn} in detail and leave the rest as exercises. Conformal invariance is satisfied if the correlator depends on the dot product $Z_1\cdot Z_2$ (see \eqref{twistordotprods}). However, since this treats $Z_1$ and $Z_2$ on equal footing, the two point function also requires that the two currents are identical which is a familiar fact from traditional CFT. Therefore we have,
\begin{align}
    \langle 0|\hat{J}_{s}^{+}(Z_1)\hat{J}_{s}^{+}(Z_2)|0\rangle=F(Z_1\cdot Z_2)
\end{align}
The helicity counting identity \eqref{helicityIdZandW} allows for two classes of solutions:
\begin{align}
    F(Z_1\cdot Z_2)\in\bigg\{\frac{1}{(Z_1\cdot Z_2)^{2s+2}},\delta^{[2s+1]}(Z_1\cdot Z_2)\bigg\}.
\end{align}
The first of these, it turns out, is the correct solution for the parity even two point function. The latter solution does not even scale properly under rescalings by negative numbers. It shall however return, accompanied by extra ingredients in a later section. Therefore we have,
\begin{align}\label{JsJsevenTwopointTwistor}
     \langle 0|\hat{J}_{s}^{+}(Z_1)\hat{J}_{s}^{+}(Z_2)|0\rangle=\frac{i^{2s+2}c_s}{(Z_1\cdot Z_2)^{2s+2}}.
\end{align}
\begin{exercise}
    In this exercise, we validate the solution \eqref{JsJsevenTwopointTwistor} through both the Witten \eqref{WittenTransform} and Penrose transforms \eqref{PenroseTransform}. Lets do the Witten transform first as its the easier calculation out of the two. Show that,
    \begin{align}
        \int d^2 \Bar{\mu}_1 \int d^2\Bar{\mu}_2 e^{-i\Bar{\lambda}_1\cdot \Bar{\mu}_1-i\Bar{\lambda}_2\cdot \Bar{\mu}_2}~\frac{i^{2s+2}~c_s}{(Z_1\cdot Z_2)^{2s+2}}\propto c_s\frac{\langle \Bar{1}\Bar{2}\rangle^{2s}}{p_1^{2s-1}}\delta^3(p_1+p_2),
    \end{align}
    which is indeed the correct two point function. Next, show that,
    \begin{align}
        \int \langle\lambda_1 d\lambda_1\rangle\langle \lambda_2 d\lambda_2\rangle (\zeta_1\cdot \lambda_1)^{2s}(\zeta_2\cdot \lambda_2)^{2s}\frac{c_s}{(Z_1\cdot Z_2)^{2s+2}}|_X=c_s\frac{(\zeta_{1a}(x_1-x_2)^{ab}\zeta_{2b})^{2s}}{|x_1-x_2|^{4s+2}}.
    \end{align}
    To calculate these integrals, the embedding space approach of \cite{Baumann:2024ttn} is quite useful and we refer the reader to their paper. An alternate, perhaps less covariant way is to locally parametrize $\lambda_1=(1,\xi_1)$ and $\lambda_2=(1,\xi_2)$ and simply carry out the integrals which are now simply  ordinary 1d integrals from $-\infty$ to $\infty$. A word of caution: Use a cutoff for the integrals and then take the cutoff to infinity at the end of the calculation.
\end{exercise}
\begin{exercise}
    (a)~Show that the negative helicity two point function is given by,
    \begin{align}\label{JsJsevenTwopointTwistormm}
        \langle 0|\hat{J}_s^{-}(Z_1)\hat{J}_s^{-}(Z_2)|0\rangle=\frac{i^{-2s+2}c_s}{(Z_1\cdot Z_2)^{-2s+2}}.
    \end{align}
    Validate this by converting this to position space and spinor helicity variables.\\
    (b)~Obtain the dual-Twistor space two point functions either directly or by using the twistor Fourier transform \eqref{ZtoWfullFourier} and the twistor space results \eqref{JsJsevenTwopointTwistor}, \eqref{JsJsevenTwopointTwistormm}. 
    \bonus{What is the relationship between twistor and dual-twistor space correlators? There is of course the twistor Fourier transform \eqref{ZtoWfullFourier} that converts a function of $Z$ to a function of $W$ and vice versa but does not change the helicity. However, what is interesting they are also related by the discrete CPT transformation. under a CPT transormation, helicities flip which is enacted (schematically) by $\lambda \to \Bar{\lambda}$ and in Twistor space also $\Bar{\mu}\to \mu$. Therefore, not only are $Z$ and $W$ interchanged under a CPT transformation but so are positive and negative helicity currents. After deriving the precise rules for a CPT transformation starting from spinor helicity variables and the half-Fourier transforms \eqref{WittenTransform}, \eqref{WittendualTransform}, derive the CPT operation in Twistor space.}
\end{exercise}
Lets now add a third current and consider,
\begin{align}
    \langle 0|\hat{J}_{s_1}^{+}(Z_1)\hat{J}_{s_2}^{+}(Z_2)\hat{J}_{s_3}^{+}(Z_3)|0\rangle.
\end{align}
Lets look at the classes of solutions that depend on the twistor dot products \eqref{twistordotprods}. Satisfying the helicity counting identity \eqref{helicityIdZandW} leads to the following result:
\begin{align}
   \langle 0|\hat{J}_{s_1}^{+}(Z_1)\hat{J}_{s_2}^{+}(Z_2)\hat{J}_{s_3}^{+}(Z_3)|0\rangle\supset \bigg(\frac{a_{12}}{(Z_1\cdot Z_2)^{s_1+s_2-s_3+1}}+b_{12} \delta^{[s_1+s_2-s_3]}(Z_1\cdot Z_2)\bigg)\times \text{Perms}.
\end{align}
Similar to the two point case, it turns out that the correct solution does not involve both the polynomial and delta function terms \cite{Baumann:2024ttn}. However, what is different from the two point case is that it is the delta function solution that figures here and not the polynomial. Therefore we have one solution,
\begin{align}
    \langle 0|\hat{J}_{s_1}^{+}(Z_1)\hat{J}_{s_2}^{+}(Z_2)\hat{J}_{s_3}^{+}(Z_3)|0\rangle_h=i^{s_1+s_2+s_3}\delta^{[s_1+s_2-s_3]}(Z_1\cdot Z_2)\delta^{[s_2+s_3-s_1]}(Z_2\cdot Z_3)\delta^{[s_3+s_1-s_2]}(Z_3\cdot Z_1).
\end{align}
What other solutions are possible? Well, lets test out the projective delta function \eqref{nptprojective}. At the level of three points, it turns out that there is basically a unique invariant we can form, that being,
\small
\begin{align}\label{delta4def}
&\delta^3(Z_1,Z_2,Z_3;\alpha_{12},\alpha_{23},\alpha_{31})=(-i)^{-\alpha_{12}-\alpha_{23}-\alpha_{31}}\delta^{[-\alpha_{12}-\alpha_{23}-\alpha_{31}]}(Z_1\cdot Z_2)\int dc_{23}dc_{31}c_{23}^{\alpha_{23}}c_{31}^{\alpha_{31}}\delta^4(c_{23}Z_1+c_{31}Z_2+Z_3)\notag\\&=\int dc_{12}dc_{23}dc_{31}c_{12}^{\alpha_{12}}c_{23}^{\alpha_{23}}c_{31}^{\alpha_{31}}\delta^4(c_{23} Z_1+c_{31} Z_2+c_{12} Z_3)e^{\frac{i}{3}\big(\frac{Z_1\cdot Z_2}{c_{12}}+\frac{Z_2\cdot Z_3}{c_{23}}+\frac{Z_3\cdot Z_1}{c_{31}}\big)}.
\end{align}
\normalsize
If we take $\alpha_{12}=-s_1-s_2+s_3,\alpha_{23}=-s_2-s_3+s_1,\alpha_{31}=-s_3-s_1+s_2$, the helicity counting identity \eqref{helicityIdZandW} is satisfied. Thus, we have another solution,
\begin{align}
     \langle 0|\hat{J}_{s_1}^{+}(Z_1)\hat{J}_{s_2}^{+}(Z_2)\hat{J}_{s_3}^{+}(Z_3)|0\rangle_{nh}=\delta^3(Z_1,Z_2,Z_3;-s_1-s_2+s_3,-s_2-s_3+s_1,-s_3-s_1+s_2).
\end{align}
Having exhausted both classes of invariants we can write the most general (parity-even) three point function of currents:
\begin{align}\label{3pointparityevenTwistorppp}
     \langle 0|\hat{J}_{s_1}^{+}(Z_1)\hat{J}_{s_2}^{+}(Z_2)\hat{J}_{s_3}^{+}(Z_3)|0\rangle=c_{s_1s_2s_3}^{(h)} \langle 0|\hat{J}_{s_1}^{+}(Z_1)\hat{J}_{s_2}^{+}(Z_2)\hat{J}_{s_3}^{+}(Z_3)|0\rangle_h+c_{s_1s_2s_3}^{(nh)} \langle 0|\hat{J}_{s_1}^{+}(Z_1)\hat{J}_{s_2}^{+}(Z_2)\hat{J}_{s_3}^{+}(Z_3)|0\rangle_{nh}.
\end{align}
Lets now explain the meaning of the subscripts $h$ and $nh$. They stand for homogeneous and non-homogeneous respectively. As we discussed in part \ref{part:SH} of these notes, these quantities from the AdS bulk perspective correspond to minimally coupled and higher derivative interaction contributions to three point functions when $s_i+s_j\ge s_k,i\ne j\ne k, i,j,k\in\{1,2,3\}$. The $h$ Wightman functions when Wick rotated to Euclidean space are still identically conserved whereas the $nh$ Wightman functions saturate the current-conservation Ward-Takahashi identity of the correlator. When the spin triangle inequality is not obeyed, we instead have two non-homogeneous solutions that when Wick rotated to Euclidean space, both have non trivial Ward-Takahashi identities.

To sum it all up, we have found two distinct solutions at the level of three point parity even correlators of conserved currents which also tallies up with our earlier spinor helicity analysis.

\begin{exercise}
    (a)For the patient reader: Convert the result \eqref{3pointparityevenTwistorppp} to spinor helicity variables. After obtaining the results for at least three other helicities by arguments similar to those above and converting to spinor helicity variables, find out the momentum space correlator that gives rise to these structures when converted to SH variables (at least for low spins). Wick rotate to Euclidean space by performing dispersive integrals following \cite{Bala:2025gmz} and show that it gives us the known answers and verify the conformal Ward identity.\\
    (b) For the patient and tenacious reader: Perform a Penrose transform \eqref{PenroseTransform} of the three point function \eqref{3pointparityevenTwistorppp} at least for some simple examples like $s_1=2,s_2=s_3=1$. Show that this gives rise to the known position space results.
    \hint{For the latter exercise, it is a good idea to use the elegant methods of \cite{Baumann:2024ttn} rather than to attempt a brute force approach.}
\end{exercise}

\section{The parity odd sector: The Return of the Infinity Twistor}\label{sec:TwistorParityodd}
So far, we have been dealing with the parity even sector of 3d CFT. However, as we know, the parity odd sector is also quite important. As we saw earlier, a distinct feature of 3d CFT is that it can appear at the level of two points and indeed played an important role in our construction of the candidate for a holographic dual to chiral higher spin theory. As we shall see, our old friend, the infinity twistor \eqref{infinitytwistor}, having played no role for parity even correlators of currents, shall make a return and integrate itself into the parity odd structures \cite{Bala:2025qxr}.
\subsection{The parity odd version of the Penrose transform}
 Recall that given a current $J^{\mu_1\cdots \mu_s}(x)$, its epsilon transform is defined as,
\begin{align}\label{EPT}
    (\epsilon\cdot J)^{\mu_1\cdots\mu_s}(x)=\epsilon^{\nu\sigma (\mu_1 }\int \frac{d^3 y}{|y-x|^2}\frac{\partial}{\partial y^{\nu}}J^{\mu_2\cdots \mu_s)}_{\nu}(y).
\end{align}
This is a conformally covariant transformation that preserves the scaling dimension and spin but flips the parity label when used on a correlator. In spinor notation (which also allows the simple generalization of  the transform to half-integer spin currents) the epsilon transform is given by,
\begin{align}\label{EPTspinornotation}
    (\epsilon\cdot J)^{a_1 a_2\cdots a_{2s}}(x)=-i\int \frac{d^3 y}{|y-x|^2}\frac{\partial}{\partial y^c_{(b_1}}J^{a_1 a_2\cdots a_{2s}) c}(y).
\end{align}
Given the usual Penrose transform \eqref{PenroseTransform}, we can perform an epsilon transform \eqref{EPTspinornotation} of the whole equation. This results in the following idenification of the twistor space epsilon transform:
\begin{align}\label{twistorEPTspins}
    (\epsilon \cdot \hat{J}_s)^+(Z) = 
    -i \int \frac{d^3 y}{y^2} \,
    Z^A I_{AB} \frac{\partial}{\partial Z_B}
    \hat{J}_s^+(\lambda, \bar{\mu}) \bigg|_{\bar{\mu}^a \to \bar{\mu}^a + y^{ab} \lambda_b}.
\end{align}
The ``parity odd" counterpart of the Penrose transform is given by,
\begin{align}\label{parityoddPenrosetransform1}
    (\epsilon\cdot J_s)^{a_1\cdots a_{2s}}(x)=\int \langle\lambda d\lambda\rangle\lambda^{a_1}\cdots \lambda^{a_{2s}} (\epsilon \cdot \hat{J}_s)^+(Z)|_{X}.
\end{align}
The result \eqref{twistorEPTspins} is true in any 3d QFT with a conserved current and in particular, in CFTs! This might be quite alarming since the infinity Twistor \eqref{infinitytwistor} explicitly features in the Penrose transform \eqref{parityoddPenrosetransform1} thanks to \eqref{twistorEPTspins}. However, the very point of the infinity twistor \eqref{infinitytwistor} was to break conformal invariance down to its Poincare subgroup. How can it then feature in conformal correlators? Well, lets find out.
\subsection{Two point functions}
The parity even two point function in the $(++)$ helicity configuration is given by \eqref{JsJsevenTwopointTwistor}.
To this expression, lets perform the epsilon transform using \eqref{twistorEPTspins} with respect to the first operator. This results in
\begin{align}\label{twopointodd}
   \langle 0|\hat{J}_s^{+}(Z_1)\hat{J}_s^{+}(Z_2)|0\rangle_{\text{odd}}=-i ~c_{\text{s,odd}}i^{2s+1}\text{Sgn}(\langle Z_1IZ_2\rangle)\delta^{[2s+1]}(Z_1\cdot Z_2),
\end{align}
where we have defined the infinity dot product,
\begin{align}
    \langle Z_i I Z_j\rangle=\langle i j\rangle.
\end{align}
Lets ponder over \eqref{twopointodd} for a bit. One good thing is that it identically obeys the helicity identity \eqref{helicityIdZandW} thanks to the sign factor. However, is it conformally invariant!? Well for $\langle i j\rangle>0$ and $\langle i j\rangle<0$ it identically solves the conformal Ward identity since the remaining part of the expression is made out of symplectic dot products \eqref{twistordotprods}. However, there is potential for trouble precisely at $\langle i j\rangle=0$ since the action of the generators \eqref{TABtwistoranddual} will lead to a delta function when acting on the sign function.
\begin{exercise}
    Lets resolve this by performing the Penrose transform to journey back to position space. Our starting point is,
     \begin{align}
        \langle J_{s}(x_1,\zeta_1)J_{s}(x_2,\zeta_2)\rangle_{\text{odd}}& =c_{s,\text{odd}}\int \langle\lambda_1 d\lambda_1\rangle \langle\lambda_2 d\lambda_2\rangle (\zeta_1\cdot\lambda_1)^{2s}(\zeta_2\cdot\lambda_2)^{2s} (-i)\text{Sgn}(\langle12\rangle) (i)^{2s+1} \delta^{[2s+1]}(Z_1\cdot Z_2)\big|_X.
        \end{align}
    Perform the integrals by defining a vector $v_{ab}=\lambda_{1(a}\lambda_{2b)}$ and using the projective identities \eqref{projvsnonproj1}, \eqref{projID1}, convert the above expression to an integral over $d^3 v$. For $s=1$ show that the result is,
    \begin{align}
        \langle J_{s}(x_1)J_{s}(x_2)\rangle_{\text{odd}}& =c_{s,\text{odd}}(\zeta_1\cdot \zeta_2)\zeta_{1a}\zeta_{2b}\frac{\partial}{\partial x_1^{ab}}\delta^3(x_1-x_2).
    \end{align}
    This is the well known conformally invariant contact term.
\bonus{Prove that the presence of the infinity twistor in \eqref{twopointodd} ensures that the result of a Penrose transform \eqref{PenroseTransform} is a parity odd expression. Similarly, show that the factor of $-i$ makes it odd under time-reversal. Thus, under a $PT$ transformation, the correlator is invariant as is appropriate for a unitary theory. }
\end{exercise}
\subsection{Three point functions}
Consider the homogeneous parity even three point function in \eqref{3pointparityevenTwistorppp}. Performing an epsilon transform at location $1$ leads to the following parity odd three point function:
\begin{align}\label{Js1s2s3Twistorodd}
   &\langle 0|\hat{J}_{s_1}^{+}(Z_1)\hat{J}_{s_2}^{+}(Z_2)\hat{J}_{s_3}^{+}(Z_3)|0\rangle_{\text{odd}}\notag\\&=\frac{i}{2}\int dc_{12}dc_{23}dc_{31}c_{12}^{s_1+s_2-s_3}c_{23}^{s_2+s_3-s_1}c_{31}^{s_3+s_1-s_2}\text{Sgn}\big(c_{12}\langle Z_1 IZ_2\rangle+c_{31}\langle Z_3 IZ_1\rangle\big)\big(e^{-ic_{12}Z_1\cdot Z_2-ic_{23}Z_2\cdot Z_3-ic_{31}Z_3\cdot Z_1}\big).
\end{align}
Similar to the two point function, the dependence on the infinity twistor is only through sign factors. The reason why \eqref{Js1s2s3Twistorodd} is parity odd is the sign factor which actually picks up a sign under a parity transform before turning into its CPT conjugate factor. To obtain the explicit answer, it is not yet clear how to do so by computing the Penrose integral except by a brute force method. It would be great if the reader is able to find the optimum way to evaluate these integrals, perhaps borrowing techniques from \cite{Baumann:2024ttn}. Finally, similar results in the other helicities for both two and three point functions can easily be derived given their parity even counterparts.

\section{Extension to Scalars and Generic Operators}\label{sec:ScalarsGenericTwistor}
For the Twistor space approach to be completely general and one day be a stage for the spinning conformal bootstrap, we need to describe general operators and not just conserved currents. First steps in this direction were taken in \cite{Bala:2025qxr} which we recount below.
\subsection{The Generalized Penrose transform for Scalars}
Lets start with scalars. Recall that we derived the Penrose transform from the Witten transform in section \ref{sec:PenroseFromWittenandFourier} by stacking Witten's half Fourier transform and the usual Fourier transform. Well, instead of starting with a conserved current, we could have started with a scalar operator. In particular consider the Fourier transform,
\begin{align}
    O_{\Delta}(x)=\frac{1}{\text{Vol}(GL(1,\mathbb{R})}\int \frac{d^2\lambda d^2\Bar{\lambda}}{(2\pi)^3}e^{i\lambda_a\Bar{\lambda}_b x^{ab}}|\frac{\lambda\cdot \Bar{\lambda}}{2}|O_{\Delta}(\lambda,\Bar{\lambda})=\frac{1}{\text{Vol}(GL(1,\mathbb{R})}\int \frac{d^2\lambda d^2\Bar{\lambda}}{(2\pi)^3}e^{i\lambda_a\Bar{\lambda}_b x^{ab}}\hat{O}_{\Delta}(\lambda,\Bar{\lambda}).
\end{align}
Lets define the Witten transform,
\begin{align}\label{ODeltaWitten}
    \hat{O}(Z)=\int \frac{d^2\Bar{\lambda}}{(2\pi)^2}e^{i\Bar{\lambda}\cdot \Bar{\mu}}\hat{O}_{\Delta}(\lambda,\Bar{\lambda}).
\end{align}
This leads to the Penrose transform,
\begin{align}\label{ODeltaPenrose}
    O_{\Delta}(X)=\int \langle \lambda d\lambda\rangle \hat{O}_{\Delta}(\lambda,\Bar{\mu})|_X,
\end{align}
where $X$ denotes the usual incidence relations. By dimensional analysis, we see that $\hat{O}_\Delta(Z)$ must have scaling dimension $\Delta-1$. The projectiveness of the integrand \eqref{ODeltaPenrose} also implies that $\hat{O}_{\Delta}(Z)$ has homogeneity $-2$ which is also what is obtained by setting $s=0$ in the helicity counter \eqref{helicityIdZandW}.
Lets try to write the twistor space two point function of this operator. Clearly, dimensional analysis will not be satisfied just using the symplectic dot product $Z_1\cdot Z_2$ since it is dimensionless\footnote{Recall that $Z_1\cdot Z_2=\lambda_1\cdot \Bar{\mu}_2-\lambda_2\cdot \Bar{\mu}_1$. $\lambda$ and $\Bar{\mu}$ have opposite scaling dimensions and thus this dot product is dimensionless.}. The only dimensionfull Lorentz invariant quantity we can form with two twistors is using the infinity twistor \eqref{infinitytwistor} viz $\langle Z_1 I Z_2\rangle=\langle 1 2\rangle$. Therefore we are forced to try the expression,
\begin{align}\label{ODelta2point}
    \langle 0|\hat{O}_{\Delta}(Z_1)\hat{O}_{\Delta}(Z_2)|0\rangle=\frac{i^{2\Delta}|\langle 12\rangle|^{2(\Delta-1)}}{(Z_1\cdot Z_2)^{2\Delta}}.
\end{align}
Lets immediately check this by computing its Penrose transform \eqref{ODeltaPenrose}. We have,
\begin{align}
    \langle 0|O_{\Delta}(x_1)O_{\Delta}(x_2)|0\rangle=\int \langle \lambda_1 d\lambda_1\rangle\langle \lambda_2 d\lambda_2\rangle \frac{i^{2\Delta}|\langle 12\rangle|^{2(\Delta-1)}}{(Z_1\cdot Z_2)^{2\Delta}}|_{X}.
\end{align}
To compute this integral, lets first convert the projective integrals to non-projective ones by using \eqref{projvsnonproj1}. Lets then define a vector $v_{ab}=\lambda_{1(a}\lambda_{2b)}$. This results in,
\begin{align}
    \langle 0|O_{\Delta}(x_1)O_{\Delta}(x_2)|0\rangle\propto \frac{1}{\text{Vol}(GL(1,\mathbb{R})}\int d^3 v |v|^{2\Delta-3}\int dc_{12}~c_{12}^{2\Delta}e^{2ic_{12}v\cdot x_{12}},
\end{align}
where we Schwinger parameterized the denominator, used the incidence relations and the fact that $\lambda_{1a}\lambda_{2b}x_{12}^{ab}=-2 v\cdot x_{12}$. We can then exchange the order of integrals resulting in,
\begin{align}
     \langle 0|O_{\Delta}(x_1)O_{\Delta}(x_2)|0\rangle\propto &\frac{1}{\text{Vol}(GL(1,\mathbb{R})}\int dc_{12}c_{12}^{2\Delta}\int d^3 v |v|^{2\Delta-3}e^{2i c_{12}v\cdot x_{12}}\notag\\
     &=\frac{1}{\text{Vol}(GL(1,\mathbb{R})}\int dc_{12}c_{12}^{2\Delta}\frac{1}{(c_{12}|x_{12}|)^{2\Delta}}\propto \frac{1}{|x_{12}|^{2\Delta}},
\end{align}
which is indeed the correct conformally invariant position space two point function. How do we reconcile this with the fact that the infinity twistor \eqref{infinitytwistor} features in the twistor space two point function \eqref{ODelta2point}? Lets find out now.

\subsection{Generic Operators transform in non-local representations of Sp$(4)$}
The spinor helicity variables action of the Dilatation and Special conformal generator on generic operators (suppressing possible spin indices below) with scaling dimension $\Delta$ are respectively given by,
\begin{align}\label{genDandK}
    &D=\frac{i}{2}\bigg(\Bar{\lambda}^a\frac{\partial}{\partial\Bar{\lambda}^a}+\lambda^a\frac{\partial}{\partial\lambda^a}\bigg)+i(3-\Delta)\notag\\&K^\mu=2(\sigma^\mu)^{ab}\frac{\partial^2}{\partial\lambda^a\Bar{\lambda}^b}+\frac{(\Delta-2)}{p}(\sigma^\mu)^b_a\bigg(\Bar{\lambda}^a\frac{\partial}{\partial\Bar{\lambda}^b}-\lambda^a\frac{\partial}{\partial\lambda^b}\bigg)+\frac{i(\sigma^\nu)^a_b}{p}\bigg(\Bar{\lambda}^a\frac{\partial}{\partial\Bar{\lambda}^b}-\lambda^a\frac{\partial}{\partial\lambda^b}\bigg)\mathcal{M}_{\mu\nu}.
\end{align}
The $\mathcal{M}_{\mu\nu}$ operator is zero when acting on scalars but will contribute when acting on operators with non-zero spin. The above generators can be obtained by a brute force calculation starting from position space, performing a Fourier transform to momentum space and then recasting the result in spinor helicity variables (see appendix \ref{app:CFTreview}). Our rescaled scalars are defined by,
\begin{align}
    \hat{O}_{\Delta}(\lambda,\Bar{\lambda})=p~ O_{\Delta}(\lambda,\Bar{\lambda}).
\end{align}
Using the way the above generators act on the unrescaled scalars, we can induce the action on its rescaled counterpart. Further, translating the results to twistor space via the inverse of the half-Fourier \eqref{ODeltaWitten}, we obtain the following results for the generators:
\small
\begin{align}\label{DandKODelta}
    &[D,\hat{O}_{\Delta}(\lambda,\Bar{\mu})]=\frac{i}{2}\bigg(\lambda^a\frac{\partial}{\partial\lambda^a}-\Bar{\mu}^a\frac{\partial}{\partial\Bar{\mu}^a}+2(1-\Delta)\bigg)\hat{O}_{\Delta}(\lambda,\Bar{\mu})\notag\\
    &[K_{ab},\hat{O}_{\Delta}(\lambda,\Bar{\mu})]=2i\bigg[\Bar{\mu}_{(a}\frac{\partial}{\partial\lambda^{b)}}+(\Delta-1)\bigg((\lambda\cdot\frac{\partial}{\partial\Bar{\mu}})^{-1}\big(\lambda_{(a}\frac{\partial}{\partial \lambda^{b)}}-\Bar{\mu}_{(a}\frac{\partial}{\partial\Bar{\mu}^{b)}}\big)-2(\lambda\cdot\frac{\partial}{\partial\Bar{\mu}})^{-2}\lambda_{(a}\frac{\partial}{\partial\Bar{\mu}^{b)}}\bigg)\bigg]\hat{O}_{\Delta}(\lambda,\Bar{\mu}).
\end{align}
\normalsize
\begin{exercise}
    Derive these formulae. Freely perform integration by parts and define inverse derivatives.
\end{exercise}
Note that for $\Delta=1$, the extra terms drop out and we are back to generators that act on conserved currents \eqref{TABtwistoranddual}! This is expected since taking $s \to 0$ for conserved currents which have $\Delta=s+1$ results in $\Delta=1$ scalars. Therefore, setting $s=0$ in the correlators of conserved currents results in correlators of scalars with $\Delta=1$. Lets now analyze the operators \eqref{DandKODelta}. The dilatation operator is as expected since the dimensionality of the twistor space operator is $\Delta-1$ as we previously read of from the Penrose trasnform \eqref{ODeltaPenrose}. The SCT generator on the other hand involves two distinct non-local terms: Of order $1$ and order $2$ in the non-locality. To be precise, lets define the inverse derivatives appearing in \eqref{DandKODelta} viz,
 \begin{align}\label{inversederivative}
     (\lambda\cdot\frac{\partial}{\partial\Bar{\mu}})^{-1}f(\lambda,\Bar{\mu})=-\int_0^\infty ds f(\lambda,\Bar{\mu}+s\lambda).
 \end{align}
For the inverse derivatives to make sense we require that $f(\lambda,\Bar{\mu})\not\in \text{Ker}(\lambda\cdot\frac{\partial}{\partial\Bar{\mu}})$. For the twistor space operators $\hat{O}_{\Delta}(\lambda,\Bar{\mu})$ which satisfy the unitary bound $(\Delta\ge \frac{d-2}{2})$, this requirement is indeed satisfied and therefore this operator is invertible on the subspace spanned by the twistor space operators. One can easily check using the definition \eqref{inversederivative} that it satisfies,
\begin{align}
    (\lambda\cdot\frac{\partial}{\partial\Bar{\mu}})(\lambda\cdot\frac{\partial}{\partial\Bar{\mu}})^{-1}f(\lambda,\Bar{\mu})=f(\lambda,\Bar{\mu}),
\end{align}
provided $f(\lambda,\Bar{\mu})$ vanishes when $\Bar{\mu}\to \infty$ which hopefully is true in general for any examples that we are interested in.
\begin{exercise}
    Show that the two point function \eqref{ODelta2point} is invariant under special conformal transformations with the action defined in \eqref{DandKODelta}. This exercise will reveal to the reader how to work with the non-local terms. 
\end{exercise}
To go beyond two points, one needs to be able to solve the Ward identities with the non-local SCT generator \eqref{DandKODelta}. This is still an open problem and has not yet been solved.
\subsection{The Generalized Penrose transform for any operator}
Finally, lets consider a generic spinning operator $O_{\Delta,s}$. Unlike conserved currents which have only two independent components viz positive and negative helicity, these operators have $2s+1$ independent components. For each component we define the Witten transform, \begin{align}\label{nonconsWittenTransformspins}
    \hat{\mathcal{O}}_{\Delta}^{+k}(\lambda,\Bar{\mu})=\int \frac{d^2\Bar{\lambda}}{(2\pi)^2}e^{i\Bar{\lambda}\cdot\Bar{\mu}}\frac{\mathcal{O}_{\Delta}^{+k}(\lambda,\Bar{\lambda})}{|\frac{\lambda\cdot\Bar{\lambda}}{2}|^{s-1}},
\end{align}
where $\mathcal{O}^{+k}(\lambda,\Bar{\lambda})$ contains $s+k$ positive helicity polarization spinors and $s-k$ negative  helicity polarization spinors contracted with the momentum space operator $\mathcal{O}_{\Delta a_1\cdots a_{2s}}$.
\begin{exercise}
As an application of what we have learnt so far, derive the Penrose transform for the non-conserved operator:
\begin{align}\label{PenroseTransformNonConservedSPins}
    \mathcal{O}_{\Delta,s}(x,\zeta)=\int \langle \lambda d\lambda\rangle \sum_{k=0}^{s}c_k(\zeta\cdot \lambda)^{2s-k}(\zeta\cdot \frac{\partial}{\partial\Bar{\mu}})^{k}\hat{\mathcal{O}}^{s-k}_{\Delta}(\lambda,\Bar{\mu})|_X,
\end{align}
where we have contracted the free indices with auxiliary polarization spinors $\zeta$.  $X$ denotes the incidence relation \eqref{IncidenceRelation} and $c_k$ are coefficients that can easily be determined by an explicit calculation and performing the Witten transform \eqref{nonconsWittenTransformspins} for each component.
\end{exercise}
The action of the special conformal generator will involve the extra terms in \eqref{DandKODelta} as well as a contribution from the half-Fourier transform of the spin operator in \eqref{genDandK}.

 We can obtain the two point functions for each component of the spin-s operator using dimensional analysis and invariance under projective rescalings of the Penrose transform \eqref{PenroseTransformNonConservedSPins}. For example, the component $\hat{\mathcal{O}}^k$ has scaling dimension $\Delta-(s+1)$ and scales with a factor of $\frac{1}{r^{2(s-k)+2}}$. Its two point function tthus reads,
 \begin{align}
     \langle 0|\hat{\mathcal{O}}_{\Delta}^{+k}(Z_1)\hat{\mathcal{O}}_{\Delta}^{+l}(Z_2)|0\rangle=c_{kl}\delta^{k l}\frac{\langle Z_1 I Z_2\rangle^{2(\Delta-(s+1))}}{(Z_1\cdot Z_2)^{2(\Delta-k)}}.
 \end{align}
 $c_{kl}$ are coefficients that are all related and are multiples of the two point function coefficient $c_{\mathcal{O}_{\Delta,s}}$.
 \begin{exercise}
     After fixing the coefficients carefully, perform the Penrose transform \eqref{PenroseTransformNonConservedSPins} for a spin-1 non-conserved operator. Match with the known position space result.
     \hint{It might be a good idea to first take a detour to momentum space.}
 \end{exercise}

\part{Super-Twistors}\label{part:SuperTwistors}
So far, we have mostly been dealing with conformal field theories without any extra symmetry (with the exception of the application to higher spin theories). In this part, we shall focus on CFTs that also possess supersymmetry. Some familiarity with supersymmetry is assumed so we shall not dwell over the basics and shall quickly proceed to spinor and Twistor methods. The main reference for this section is \cite{Bala:2025qxr}.

\section{The Algebra and operators of interest}\label{sec:susyalgebra}
The spinor helicity variables of the first part of these notes can be thought of as the result of taking the square root of the momentum vector. There, we took a ``Bosonic" square root and traded $p$ for the bosonic spinors $\lambda,\Bar{\lambda}$. In supersymmetry, we shall take a kind of ``Fermionic" square root. Consider a fermionic generator $Q_a$ that obeys the following anti-commutation relation:
\begin{align}\label{QQisP}
    \{Q_a,Q_b\}=i P_{ab}.
\end{align}
$Q_a$ are called the supersymmetry generators for reasons that we shall see quite shortly. They are anti-commuting spinors that are in the fundamental representation of $SL(2,\mathbb{R})$ or $SU(2)$ depending on the signature of space-time. Clearly, they have scaling dimension $\frac{1}{2}$ since their anti-commutator produces the momentum vector. Also, it turns out that the commutator of $Q_a$ with the SCT generator $K_{\mu}$ is non-trivial and leads to a new generator called $S_a$ which is known as the generator of super-special conformal transformations.
\begin{exercise}
    Using the above facts, construct the $\mathcal{N}=1$ super-conformal algebra,
\begin{align}\label{SCFTalgebra1}
[&M_{\mu\nu},M_{\rho\lambda}]=i\left(\delta_{\mu\rho}M_{\nu\lambda}-\delta_{\nu\rho}M_{\mu\lambda}-\delta_{\mu\lambda}M_{\nu\rho}+\delta_{\nu\lambda}M_{\mu\rho}\right),\notag\\
    [M_{\mu\nu},P_\alpha]&=i\left(\delta_{\mu\alpha}P_\nu-\delta_{\nu\alpha}P_\mu\right),\qquad\quad[M_{\mu\nu},K_\alpha]=i\left(\delta_{\mu\alpha}K_\nu-\delta_{\nu\alpha}K_\mu\right),\notag\\
    [D,P_\mu]&=i P_\mu,\qquad \qquad\qquad\qquad~\quad[P_\mu,K_\nu]=2i\left(\delta_{\mu\nu}D-M_{\mu\nu}\right),\notag\\
    \{Q_a,Q_b\}&=i(\sigma^\mu)_{ab}P^\mu,\qquad\qquad\qquad~ \{S_a,S_b\}=i(\sigma^\mu)_{ab}K^\mu,\notag\\
    [D,Q_a]&=\frac{i}{2}Q_a,\qquad\qquad\qquad\qquad~~[D,K_\mu]=-i K_\mu,\notag\\
    [K_\mu,Q_a]&=i(\sigma_\mu)_a^b S_b,\qquad\qquad\qquad~~~[D,S_a]=-\frac{i}{2}S_a,\notag\\
    [M_{\mu\nu},Q_a]&=\frac{i}{2}\epsilon_{\mu\nu\rho}(\sigma^\rho)^b_a Q_b,\qquad\quad\quad~~[P_\mu,S_a]=i(\sigma_\mu)^a_b Q_b,
    \notag\\
     [M_{\mu\nu},S_a]&=\frac{i}{2}\epsilon_{\mu\nu\rho}(\sigma^\rho)^b_a S_b,\quad\qquad\quad~~\{Q_a,S_b\}=\epsilon_{ab}D-\frac{i}{2}\epsilon_{\mu\nu\rho}(\sigma^\rho)_{ab}M^{\mu\nu},
\end{align}
with the (anti) commutators not listed above being equal to zero. 
\hint{Also take the commutator of $K$ with $Q$ to be the definition of $S$. The only non-trivial (anti-)commutators one needs to check are $\{S,S\},\{Q,S\}$. The rest pretty much follow by dimensional analysis and index structure. Use the graded Jacobi identity of a Lie super-algebra if required.}
\end{exercise}
Our objects of interest are what are known as super-current multiplets. Before defining them, let us introduce the concept of super-space which shall be the stage for our calculations. Given the fact that $Q$ is kind of a fermionic square root of $P$, it can be interpreted as a fermionic translation operator. This motivates the following construction. In addition to the spacetime coordinates $x^\mu$, lets define two additional fermionic coordinates $\theta^a$. We then interpret $Q_a$ as generating a translation $\theta^a\to \theta^a+\epsilon^a$ for some Grassmann spinor $\epsilon$. However, given the fact that $\{Q,Q\}=P$, two of these fermionic translations need to lead to the usual space-time translation. Therefore, we consider the following representation:
\begin{align}
    Q_a=\frac{\partial}{\partial\theta^a}+\frac{1}{2}\theta^b p_{ab},~~p_{ab}=-i\frac{\partial}{\partial x^{ab}}.
\end{align}
One can easily verify that the anti-commutation relation \eqref{QQisP} holds for this representation. Given the fact that $Q_a$ is spin-$1/2$ and is a symmetry generator, we expect it to relate conformal primaries whose spin is separated by $\frac{1}{2}$. Indeed, consider let us consider an ansatz for a super-current respecting this. First of all, in super-space, objects are functions of $x^\mu$ and $\theta^a$. However, the $\theta$ expansion of any function terminates at second order due to the Grassmann nature of its components. Therefore we write,
\begin{align}
    \mathbf{J}_{s}^{a_1\cdots a_{2s}}(x,\theta)=J_s^{a_1\cdots a_{2s}}(x)+\theta_a J_{s+\frac{1}{2}}^{a a_1\cdots a_{2s}}(x)+\theta^2 F_s^{a_1\cdots a_{2s}}(x).
\end{align}
However, one must still implement the super version of conservation. This is acheived by the super-covariant derivative given by,
\begin{align}
    D_a=\frac{\partial}{\partial \theta^a}-\frac{1}{2}\theta^b p_{ab}.
\end{align}
This implements conservation at the component level for $J_s$ and $J_{s+\frac{1}{2}}$. Further, it also demands that $F_s$ is related to $J_s$.
\begin{exercise}
    Consider the shortening condition (conservation),
    \begin{align}\label{SuperCons}
        D_{a_1}J_s^{a_1\cdots a_{2s}}=0.
    \end{align}
    Show that this implies that the super-current is given by,
    \begin{align}\label{Jsinthetaposspace}
        &\mathbf{J}_{s}^{a_1\cdots a_{2s}}(\theta,\mathbf{x})=J_{s}^{a_1\cdots a_{2s}}(\mathbf{x})+\theta_{m}J_{s+\frac{1}{2}}^{(a_1\cdots a_{2s_1}m)}(\mathbf{x})-i\frac{\theta^2}{4}(\slashed{\partial})^{a_1}_{m}J_{s}^{a_2\cdots a_{2s}m}(\mathbf{x}).
    \end{align}
\end{exercise}
\begin{exercise}
    Lets now pay justice to the name supersymmetry. Apply the supersymmetry generator on \eqref{Jsinthetaposspace}. This results in,
\begin{align}
    Q_a \mathbf{J}_s^{a_1\cdots a_{2s}}=\frac{i}{2}\theta^b \frac{\partial}{\partial x^{ab}}J_s^{a_1\cdots a_{2s}}-J_{s+\frac{1}{2}}^{a_1\cdots a_{2s}a}-\frac{i\theta^2}{2} \frac{\partial}{\partial x^{ab}}J_{s+\frac{1}{2}}^{a_1\cdots a_{2s}m}-\frac{i}{2}\theta_a \slashed{\partial}_m^{a_1}J_s^{a_2\cdots a_{2s}m}.
\end{align}
This shows that supersymmetry mixes operators with different spins differing by half a unit.
\end{exercise}
The correlation functions of the super-currents \eqref{Jsinthetaposspace} are the objects of interest to us. For example take $s=1/2$. This leads to a spin-1/2 super-current that contains a spin-1/2 operator and a spin-1 conserved current. From an AdS bulk perspective, this corresponds to a multiplet containing a Fermion and a photon. One can now go over to momentum space, to spinor helicity variables and develop its supersymmetric version. We do this when we shall develop the super-Witten transform. For now, we shall directly head over to Twistor space.

\section{The Geometry of Super Twistor space}\label{sec:GeometrySuperTwistor}
Twistor space for $\mathbb{R}^{2,1}$ is an open subset of the real projective space $\mathbb{RP}^{3}$. Given the fact that super-space counterpart to $\mathbb{R}^{2,1}$ is $\mathbb{R}^{2,1|2\mathcal{N}}$, which has $3$ bosonic and $2\mathcal{N}$ fermionic directions, a natural candidate for the super-Twistor space is an open subset of $\mathbb{RP}^{3|2\mathcal{N}}$. As we shall see shortly, it will turn out that $\mathcal{N}$ out of the $2\mathcal{N}$ fermionic projective coordinates will not feature atleast in the analysis of conserved super-currents. However, they do play an important role when it comes to super-scalars and other generic operators which see the presence of all $2\mathcal{N}$ coordinates. For now, lets focus on $\mathbb{RP}^{3|\mathcal{N}}$. It is spanned by bosonic coordinates $Z^A$ and fermionic coordinates $\psi^N$ where $A=1,2,3,4$ is the $Sp(4)$ fundamental index and $N=1,\cdots,\mathcal{N}$ is the $O(\mathcal{N})$ fundamental index. Together they form the super twistor space coordinates $\mathcal{Z}^{\mathcal{A}}=(Z^A,\psi^N)=(\lambda^a,\Bar{\mu}_{a'},\psi^N)$ which is in the fundamental representation of the the super-conformal group $OSp(\mathcal{N}|4;\mathbb{R})$.
The coordinates $Z^A$ are real and $\psi^N$ satisfy $(\psi^N)^*=i \psi^N$. Similar to the non-supersymmetric case, there exist dual twistor space coordinates $\mathcal{W}_{\mathcal{A}}=(W_A,\bar{\psi}_N)=(\mu_{a},\bar{\lambda}^{a'}, \bar{\psi}_N)$ with $W^{A}$ being real and $(\bar{\psi}^N)*= i \Bar{\psi}^N$. Indices of super-twistors are raised and lowered using the OSp($\mathcal{N}|4;\mathbb{R})$ invariant graded symplectic form $\Omega_{\mathcal{AB}}$ which is simply the direct sum of the symplectic form $\Omega_{AB}$ in \eqref{OmegaandEpsilon} and a $\mathcal{N}\times \mathcal{N}$ identity matrix that acts on the Grassmann variables. 

\section{The Super Penrose Transform}\label{sec:SuperPenroseTransform}
In this section, we shall generalize the Penrose transform for conserved currents \eqref{PenroseTransform} to the supersymmetric setting. The super-Penrose transform for 3d SCFT was developed by the authors of \cite{Bala:2025qxr} which we now review.

A conserved super-current in position super-space takes the form \eqref{Jsinthetaposspace}:
\begin{align}\label{SuperFieldExpansion}
\mathbf{J}_s^{a_1\cdots a_{2s}}(x,\theta)=J_s^{a_1\cdots a_{2s}}(x)+\theta_a J_{s+\frac{1}{2}}^{aa_1\cdots a_{2s}}(x)-\frac{i\theta^2}{4}\slashed{\partial}_{a}^{a_1}J_s^{a_2\cdots a_{2s}a}(x).
\end{align}
Given the super-field expansion \eqref{SuperFieldExpansion} and the fact that we know the Penrose transform of each component current appearing there \eqref{PenroseTransform}, we can re-write \eqref{SuperFieldExpansion} as,
\small
\begin{align}\label{SuperCurrentcomponent1}
    &\mathbf{J}_s^{a_1\cdots a_{2s}}(x,\theta)=\int \langle \lambda d\lambda\rangle \lambda^{a_1}\cdots \lambda^{a_{2s}}\bigg(\hat{J}_s^{+}(\lambda,\Bar{\mu})+\theta_a\lambda^a \hat{J}_{s+\frac{1}{2}}^{+}(\lambda,\Bar{\mu})\bigg)\bigg|_{X}-\frac{i\theta^2}{4}\slashed{\partial}_a^{a_1}\int \langle \lambda d\lambda \rangle \lambda^{a_2}\cdots \lambda^{a_{2s}}\lambda^a \hat{J}_s^{+}(\lambda,\Bar{\mu})|_{X}\notag\\
    &=\int \langle \lambda d\lambda\rangle \lambda^{a_1}\cdots \lambda^{a_{2s}}\bigg(\hat{J}_s^{+}(\lambda,\Bar{\mu})+\theta_a\lambda^a \hat{J}_{s+\frac{1}{2}}^{+}(\lambda,\Bar{\mu})\bigg)\bigg|_{X}-\frac{i\theta^2}{4}\int \langle\lambda d\lambda\rangle \lambda^{a_2}\cdots \lambda^{a_{2s}}\lambda^a\frac{\partial\Bar{\mu}^b}{\partial x^{a}_{a_1}}\frac{\partial}{\partial \Bar{\mu}^b}\hat{J}_s^{+}(\lambda,\Bar{\mu})|X\notag\\&=\int \langle \lambda d\lambda\rangle \lambda^{a_1}\cdots \lambda^{a_{2s}}\bigg(\hat{J}_s^{+}(\lambda,\Bar{\mu})+\theta_a\lambda^a \hat{J}_{s+\frac{1}{2}}^{+}(\lambda,\Bar{\mu})\bigg)\bigg|_{X}-\frac{i\theta^2}{4}\int \langle\lambda d\lambda\rangle \lambda^{a_2}\cdots \lambda^{a_{2s}}\lambda^a(-2\lambda^{a_1}\delta^{b}_a+\lambda^b \delta^{a_1}_{a})\frac{\partial}{\partial\Bar{\mu}^b}\hat{J}_s^{+}(\lambda,\Bar{\mu})|_X\notag\\
    &=\int \langle \lambda d\lambda\rangle \lambda^{a_1}\cdots \lambda^{a_{2s}}\bigg(\hat{J}_s^{+}(\lambda,\Bar{\mu})+\theta_a\lambda^a \hat{J}_{s+\frac{1}{2}}^{+}(\lambda,\Bar{\mu})+\frac{i\theta^2}{4}\lambda^a\frac{\partial}{\partial\Bar{\mu}^a}\hat{J}_s^{+}(\lambda,\Bar{\mu})\bigg)\bigg|_{X}.
\end{align}
\normalsize
where $X$ as usual denote that the familiar incidence relations \eqref{IncidenceRelation} should be imposed. Lets now re-write this more covariantly as an integral over what we shall call the super-twistor space currents $\mathbf{\hat{J}}_s^{+}(\mathcal{Z})=\mathbf{\hat{J}}_s^{+}(\lambda,\Bar{\mu},\psi)$. Taking a hint from its non-supersymmetric counterpart sitting inside \eqref{PenroseTransform}, lets give this quantity the same homogeneity:
\begin{align}
    \mathbf{\hat{J}}_s^{+}(r \mathcal{Z})=\frac{1}{r^{2s+2}}\mathbf{\hat{J}}_s^{+}(\mathcal{Z}).
\end{align}
Further, lets expand this in the Grassmann parameter $\psi$ as,
\begin{align}
     \mathbf{\hat{J}}_s^{+}(r \mathcal{Z})=\big(\hat{J}_s^{+}(\lambda,\Bar{\mu})+\frac{e^{\frac{i\pi}{4}}\psi}{\sqrt{2}}\hat{J}_{s+\frac{1}{2}}^{+}(\lambda,\Bar{\mu})\big).
\end{align}
This is simply because the $\mathcal{N}=1$ super-current should have as components a spin-s and spin-$s+\frac{1}{2}$ current\footnote{We shall derive this expression later from Witten's half Fourier transform.}. Further, only two terms in the expansion are possible since $\psi^2=0$ as its a Grassmann number. This leads us to the following well defined projective integral involving this current:
\begin{align}\label{ansatzSUSYPenrose}
    \mathbf{J}_s^{a_1\cdots a_{2s}}(x,\theta)=\int \langle \lambda d\lambda\rangle \lambda^{a_1}\cdots \lambda^{a_{2s}} \mathbf{J}_s^{+}(\lambda,\Bar{\mu},\psi)|_{\mathcal{X}}=\int \langle \lambda d\lambda\rangle \lambda^{a_1}\cdots \lambda^{a_{2s}}\big(\hat{J}_s^{+}(\lambda,\Bar{\mu})+\frac{e^{\frac{i\pi}{4}}\psi}{\sqrt{2}}\hat{J}_{s+\frac{1}{2}}^{+}(\lambda,\Bar{\mu})\big)|_{\mathcal{X}}.
\end{align}
The subscript $\mathcal{X}$ indicates the superincidence relations that we need to derive. Let us write down an ansatz for the same using dimension analysis and the degree of homogeneity:
\begin{align}\label{Superincidenceansatz}
    \mathcal{X}=\big\{\Bar{\mu}_a=-x_{ab}\lambda^b+\alpha \theta^2 \lambda_a~,~\psi=\beta\theta^a\lambda_a\big\}.
\end{align}
Substituting \eqref{Superincidenceansatz} in \eqref{ansatzSUSYPenrose} and performing the Grassmann expansion yields,
\begin{align}\label{SUSYPenroseansatz2}
    \mathbf{J}_s^{a_1\cdots a_{2s}}(x,\theta)=\int \langle \lambda d\lambda\rangle\lambda^{a_1}\cdots \lambda^{a_{2s}}\bigg(\hat{J}_s^{+}(\lambda,\Bar{\mu})-\frac{e^{\frac{i\pi}{4}}}{\sqrt{2}}\beta \theta_a\lambda^a~\hat{J}_{s+\frac{1}{2}}^{+}(\lambda,\Bar{\mu})+\alpha\theta^2\lambda^a\frac{\partial}{\partial\Bar{\mu}^a}\hat{J}_s^{+}(\lambda,\Bar{\mu})\bigg)\bigg|_{X},
\end{align}
where $X$ now denotes the usual component level usual incidence relations \eqref{IncidenceRelation}.
Comparing the component expansion \eqref{SuperCurrentcomponent1} to our ansatz \eqref{SUSYPenroseansatz2} yields the values,
\begin{align}
    \alpha=\frac{i}{4},\beta=-\sqrt{2}e^{-\frac{i\pi}{4}}.
\end{align}
Putting everything together, we obtain the Super-Penrose transformation:
\begin{align}\label{SuperPenroseTransform}
    \mathbf{J}_s^{a_1\cdots a_{2s}}(x,\theta)=\int \langle \lambda d\lambda\rangle\lambda^{a_1}\cdots \lambda^{a_{2s}}\hat{\mathbf{J}}_s^{+}(\lambda,\Bar{\mu},\psi)\bigg|_{\mathcal{X}},
\end{align}
where the super-incidence relations are given by,
\begin{align}\label{Superincidence}
    \mathcal{X}=\big\{\Bar{\mu}_a=-x_{ab}\lambda^b+\frac{i}{4} \theta^2 \lambda_a~,~\psi=-\sqrt{2}e^{-\frac{i\pi}{4}}\theta^a\lambda_a\big\}.
\end{align}

\begin{exercise}
    As a consistency check, show that \eqref{SuperPenroseTransform} along with \eqref{Superincidence} satisfies super-conservation \eqref{SuperCons}.
    \bonus{Can you generalize this construction to obtain the supersymmetric version of the derivative based Penrose transform \eqref{negativehelicityPenroseTransform}? How about the dual Twistor space super Penrose transforms that generalize \eqref{dualPenroseTransform}?}
\end{exercise}

Let us now reflect on the super incidence relations \eqref{Superincidence} briefly. Given a point in position superspace $(x_{ab},\theta_c)$ we see that \eqref{Superincidence} defines a projective line $\mathbb{RP}^{1|1}\in \mathbb{RP}^{3|1}$ which is the natural generalization of the non susy setting.
\begin{exercise}
    What does a point in super-Twistor space correspond to in super-space-time?
\end{exercise}
The generalization to extended supersymmetry should be straightforward with one just having to take into account the fact that the Grassmann coordinate $\psi$ becomes a vector of $SO(\mathcal{N})$ which is the R-symmetry group. The $R$ symmetry group is the set of transformations that preserve the $\{Q,Q\}=i P$ anti-commutator. In 3d, it is $SO(\mathcal{N})$ that rotate the supersymmetry generators among themselves. Thus, one needs to impose incidence relations for each component. Further, for $\mathcal{N}\ge 2$, one can form contractions in different ways using the invariants of the $R-$symmetry group. 
\begin{exercise}
    Can you generalize the above construction for a $\mathcal{N}=2$ super-current? The expression for the Grassmann expansion of the current can be found for instance in \cite{Nizami:2013tpa}.
\end{exercise}

\section{Super Spinor Helicity and the Super Witten Transform}\label{sec:SuperWittenTransform}
In the non-supersymmetric case, we had two routes to head over to twistor space. The route laid out by Penrose and the route laid out by Witten. We have generalized the former to the supersymmetric case so it is only fair that we do so far the latter too \cite{Bala:2025qxr}. For this, we need to take the position space super-current \eqref{Jsinthetaposspace} and convert to momentum space and then spinor helicity variables. The Fourier transform is the same as its non susy counterpart \eqref{FourierTrans1} and is given by,
\begin{align}\label{Jsinthetamomspace}
    &\mathbf{J}_{s}^{a_1\cdots a_{2s}}(\theta,p)=\int d^3 p e^{2ip\cdot x}\mathbf{J}_{s}^{a_1\cdots a_{2s}}(\theta,x)=J_{s}^{a_1\cdots a_{2s}}(p)+\theta_{m}J_{s+\frac{1}{2}}^{(a_1\cdots a_{2s_1}m)}(p)-\frac{\theta^2}{2}(\slashed{p})^{a_1}_{m}J_{s}^{a_2\cdots a_{2s}m}(p).
\end{align}
Lets contract this super-current with the polarization spinors \eqref{3dpolarizationspinors}. Of course, there are only two non-zero components: The ones with helicity $\pm s$. Focusing on the positive helicity case we have,
\begin{align}
    \mathbf{J}_s^{+}(\lambda,\Bar{\lambda},\theta)=J_s^{+}(\lambda,\Bar{\lambda})+\theta_m J_{s+\frac{1}{2}}^{m +}(\lambda,\Bar{\lambda})-\frac{p \theta^2}{2}J_s^{+}(\lambda,\Bar{\lambda}).
\end{align}
To be democratic, we better convert $\theta$ into spinor helicity variables. The fact that it is a spinor means that we can express it in the basis of $\lambda$ and $\Bar{\lambda}$. In particular we take,
\begin{align}\label{etavariables}
    \theta^a=\frac{\Bar{\eta}\lambda^a+\eta\Bar{\lambda}^a}{2p}.
\end{align}
In Lorentzian signature and working with space-like momenta, $\eta$ and $\Bar{\eta}$ are real and independent Grassmann numbers.
Recall that the little group transformation $\lambda\to r^{-1}\lambda$ and $\Bar{\lambda}\to  r \Bar{\lambda}$ leaves the momentum vector invariant. Obviously, it should leave the spinor $\theta^a$ invariant as well. This implies that under little group scaling we must have,
\begin{align}
    \lambda\to r^{-1}\lambda, \Bar{\lambda}\to  r \Bar{\lambda}~\text{and}~\Bar{\eta}\to r \Bar{\eta},\eta\to r^{-1}\eta.
\end{align}
 Lets now express the Grassmann spinor $\theta^a$ in terms of the $\eta,\,\Bar{\eta}$ variables defined in \eqref{etavariables}. We obtain the much simpler super-current in the $\pm s$ helicities,
\begin{align}\label{Jsineta}
    &\mathbf{J}_s^{-}(\lambda,\Bar{\lambda},\eta,\Bar{\eta})=e^{-\frac{\eta \,\Bar{\eta}}{2}}J_s^{-}+\frac{\,\Bar{\eta}}{2\sqrt{p}}J_{s+\frac{1}{2}}^{-},\notag\\
    &\mathbf{J}_s^{+}(\lambda,\Bar{\lambda},\eta,\Bar{\eta})=e^{\frac{\eta \,\Bar{\eta}}{2}}J_s^{+}+\frac{\eta}{2\sqrt{p}}J_{s+\frac{1}{2}}^{+},
\end{align}
The super-symmetric half-Fourier transform is then defined as,
\begin{align}\label{superWittentransform}
    \mathbf{J}_s^{+}(\mathcal{Z})=\mathbf{J}_s^{+}(\lambda,\Bar{\mu},\psi)=\int \frac{d^2\Bar{\lambda}}{(2\pi)^2}\int d \xi_{-}\int d \Bar{\eta}~ e^{i\Bar{\lambda}\cdot \Bar{\mu}+\frac{(\xi_-+\sqrt{2} e^{\frac{i\pi}{4}}\psi) \Bar{\eta}}{2}}\hat{\mathbf{J}}_s^{+}(\lambda,\Bar{\lambda},\sqrt{2}e^{\frac{i\pi}{4}}\psi-\xi_-,\Bar{\eta}),
\end{align}
and similarly for its negative helicity counterpart. We have also rescaled the super-current by dividing by $p^{s-1}$ as in \eqref{WittenTransform}.  The final results are,
\begin{align}\label{supertwistorcurrent1}
\notag\mathbf{\hat{J}_s^+}(\lambda, \bar{\mu},\psi)=& \left( \hat{J}_s^+ (\lambda, \bar{\mu}) + e^{\frac{\textbf{i}\pi}{4}}\psi\;\hat{J}_{s+\frac{1}{2}}^+(\lambda, \bar{\mu})\right), \\
\mathbf{\hat{J}_s^-}(\lambda, \bar{\mu},\psi)=& \left(e^{\frac{\textbf{i}\pi}{4}}\psi\; \hat{J}_s^- (\lambda, \bar{\mu}) + \hat{J}_{s+\frac{1}{2}}^-(\lambda, \bar{\mu})\right).
\end{align}
For convenience we present the supersymmetry generator in these variables since one of the exercises require it:
\begin{align}\label{QTwistor}
[Q_{a},\mathbf{\hat{J}_s^\pm}(\lambda, \bar{\mu},\psi)]   &= \frac{e^{-\frac{i\pi}{4}}}{2\sqrt{2}}\Big( \lambda_a \frac{\partial}{\partial \psi} + \psi \frac{\partial}{\partial \bar{\mu}^a}\Big)\mathbf{\hat{J}_s^\pm}(\lambda, \bar{\mu},\psi).
\end{align}
In dual-twistor space we have,
\begin{align}\label{superdualtwistorcurrent1}
\notag\mathbf{\hat{J}_s^+}(\mu,\bar{\lambda},\bar{\psi})=& \left( e^{\frac{\textbf{i}\pi}{4}}\bar{\psi}\;\hat{J}_s^+ (\mu,\bar{\lambda}) + \hat{J}_{s+\frac{1}{2}}^+(\mu,\bar{\lambda}) \right),\\
\mathbf{\hat{J}_s^-}(\mu,\bar{\lambda},\bar{\psi})=& \left( \hat{J}_s^- (\mu,\bar{\lambda}) + e^{\frac{\textbf{i}\pi}{4}}\bar{\psi}\;\hat{J}_{s+\frac{1}{2}}^-(\mu,\bar{\lambda}) \right).
\end{align}
The action of supersymmetry generator is\footnote{The dual-twistor representation of the generators can be obtained by a super-Fourier transform of the corresponding twistor results using.},
\begin{align}\label{QDualTwistor}
[Q_{a},\mathbf{\hat{J}_s^\pm}(\mu,\bar{\lambda},\bar{\psi})]   &= \frac{e^{-\frac{i\pi}{4}}}{2\sqrt{2}}\Big(\bar{\lambda}_a \frac{\partial}{\partial \bar{\psi}} - \bar{\psi} \frac{\partial}{\partial \mu^a}\Big) \mathbf{\hat{J}_s^\pm}(\mu, \bar{\lambda},\bar{\psi}).
\end{align}
\section{A Tale of a Transform and two super-Transforms}\label{sec:SuperPenroseandWitten}
In an earlier section, we showed that the Penrose transform can be derived by stacking the Witten transform on top of the Fourier transform while working in the helicity basis. It is natural to ask whether such a derivation is possible in the supersymmetric case. The answer is affirmative. We do not do it in detail here since the calculation is quite involved but the interested reader can either refer to section 5.3 of \cite{Bala:2025qxr} where the authors explicitly work this out or work it out themselves following the steps outlined below.
\begin{exercise}
   Prove that,
    \begin{equation}\label{GrassmannId}
(\lambda \cdot \bar{\lambda}) \int d\eta\, d\bar{\eta}~
\delta^2\left( \theta^a + \frac{\bar{\eta} \lambda^a + \eta \bar{\lambda}^a}{\lambda \cdot \bar{\lambda}} \right) = 1.
\end{equation}
By following the steps in the non-supersymmetric case of section \ref{sec:PenroseFromWittenandFourier} and inserting \eqref{GrassmannId} appropriately to accommodate the super-spinor helicity description (This is basically sort of a Grassmann version of the Fadeev-Popov trick), show that the ordinary Fourier transform in conjunction with the super-Witten transform \eqref{superWittentransform} leads to the super-Penrose transform \eqref{SuperPenroseTransform}.  
\end{exercise}

\section{The Super-Conformal Generators and Ward Identities}\label{sec:SuperConformalWardId}
Having established the construction of $\mathcal{N}=1$ super twistor space, the super-Penrose transform and the super-Witten transform (with the generalization to higher $\mathcal{N}$ a natural extension), we now move on to the task of setting up the super-conformal Ward identities in super-Twistor variables and solving for two and three point parity even functions of super-currents. To this end, we first write down the super-conformal generators in these variables and solve the Ward identities.

\subsection{The generators of OSp$(\mathcal{N}|4)$}
Our supertwistor space carries a natural action of the super conformal group OSp($\mathcal{N}|4;\mathbb{R}$) with generators acting on super-currents as follows \cite{Bala:2025jbh},
\begin{align}\label{supertwistorgeneratorZ/W}  &[\mathcal{T}^{\mathcal{A}\mathcal{B}},\mathbf{\hat{J}}_s^{\pm}(\mathcal{Z})]= \mathcal{Z}^{\mathcal{(A}}\frac{\partial}{\partial\mathcal{Z
}_{B]}}\mathbf{\hat{J}}_s^{\pm}(\mathcal{Z})=\bigg(\mathcal{Z}^{\mathcal{A}} \frac{\partial}{\partial\mathcal{Z}_{\mathcal{B}}} + (-1)^{\eta_{\mathcal{A}}\eta_{\mathcal{B}}} \mathcal{Z}^{\mathcal{B}} \frac{\partial}{\partial\mathcal{Z}_{\mathcal{A}}}\bigg)\mathbf{\hat{J}}_s^{+}(\mathcal{Z}),\notag\\&[\mathcal{T}^{\mathcal{A}\mathcal{B}},\mathbf{\hat{J}}_s^{\pm}(\mathcal{W})]= \mathcal{W}^{\mathcal{(A}}\frac{\partial}{\partial\mathcal{W}_{B]}}\mathbf{\hat{J}}_s^{\pm}(\mathcal{W})=\bigg(\mathcal{W}^{\mathcal{A}} \frac{\partial}{\partial\mathcal{W}_{\mathcal{B}}} + (-1)^{\eta_{\mathcal{A}}\eta_{\mathcal{B}}} \mathcal{W}^{\mathcal{B}} \frac{\partial}{\partial\mathcal{W}_{\mathcal{A}}}\bigg)\mathbf{\hat{J}}_s^{\pm}(\mathcal{W}),
\end{align}
where we have defined a graded symmetrization operation. Indices are raised and lowered now using the graded-symplectic form,
\begin{align}\label{gradedOmega}
\Omega_{\mathcal{A}\mathcal{B}}=\begin{pmatrix}
        0&\delta_a^{b}&0\\
        -\delta^b_{a} & 0&0\\
        0&0&\mathbb{I}_{\mathcal{N}\cross \mathcal{N}}
    \end{pmatrix}=\Omega^{\mathcal{A}\mathcal{B}},
\end{align}
where $\Omega_{\mathcal{A}\mathcal{B}}\Omega^{\mathcal{AC}}=\delta_\mathcal{B}^\mathcal{C}$.

These generators are identical to their non-supersymmetric counterparts \eqref{TABtwistoranddual} with Twistors replaced by super-Twistors and symmetrization replaced by graded-symmetrization. The reason for the latter point is simple: The symplectic group has symmetric generators whereas the $R$ symmetry group for the Grassmann coordinates is $SO(\mathcal{N})$ which has anti-symmetric generators. The grade $\eta$ is zero for the bosonic generators and is one for the fermionic ones. The former however, is a consequence of choosing the right set of variables and also the fact that the Super-Penrose transform \eqref{SuperPenroseTransform} and Super-Witten transforms \eqref{superWittentransform} are natural generalizations of their non susy counterparts \eqref{PenroseTransform}, \eqref{WittenTransform}.
\begin{exercise}
    Show that the super-conformal generators obey the $\mathfrak{osp(\mathcal{N}|4)}$ Lie super-algebra,
    \begin{align}\label{OSpN4algebra}
[\mathcal{T}^{\mathcal{AB}},\mathcal{T}^{\mathcal{CD}})=\Omega^{\mathcal{CB}}\mathcal{T}^{\mathcal{AD}}+(-1)^{\eta_{\mathcal{C}}\eta_{\mathcal{D}}}\Omega^{\mathcal{DB}}\mathcal{T}^{\mathcal{AC}}+ (-1)^{\eta_\mathcal{A}\eta_\mathcal{B}}\Omega^{\mathcal{CA}} \mathcal{T}^{\mathcal{BD}}+(-1)^{\eta_\mathcal{A}\eta_\mathcal{B}+\eta_\mathcal{C}\eta_\mathcal{D}}\Omega^{\mathcal{DA}}\mathcal{T}^{\mathcal{BC}},
\end{align}
where $[\mathcal{T}^{\mathcal{AB}},\mathcal{T}^{\mathcal{CD}})\equiv \mathcal{T}^{\mathcal{AB}}\mathcal{T}^{\mathcal{CD}}-(-1)^{(\eta_{\mathcal{A}}+\eta_{\mathcal{B}})(\eta_{\mathcal{C}}+\eta_{\mathcal{D}})}\mathcal{T}^{\mathcal{CD}}\mathcal{T}^{\mathcal{AB}}$.
\end{exercise}
\begin{exercise}
    Using the fact that the super-Twistor $\mathcal{Z}^A=(\lambda^a,\Bar{\mu}_{a'},\psi^N)$, find all component generators by directly computing each component of the generators \eqref{supertwistorgeneratorZ/W}. You should find,
    \begin{align}\label{supertwistormatrix}
 \mathcal{T}^{\mathcal{AB}} & =
    \begin{pmatrix}
    -i P^{ab} &\quad  -iM^a_{b} -2i\delta_{b}^{a} D &  \quad\sqrt{2}e^{\frac{i\pi}{4}}Q^{a\, N}   \\
   -iM^a_{b} -2i\delta_{b}^{a} D & -iK^{ab} &  \frac{e^{\frac{i\pi}{4}}}{\sqrt{2}}S_{a}^{\,N} \\ 
    \sqrt{2}e^{\frac{i\pi}{4}}Q^{a\,M}   & \frac{e^{\frac{i\pi}{4}}}{\sqrt{2}}S_{a}^{M}  &  R^{MN} 
    \end{pmatrix}.
\end{align}
\end{exercise}

\subsection{The super-conformal Ward identities}
The superconformal Ward identities for supercorrelators of conserved supercurrents $\mathbf{J}_s$ is thus,
\begin{align}\label{manifesttwistorsuperconformalWard}
\langle 0|\cdots [\mathcal{T}^{\mathcal{AB}},\mathbf{J}_s]\cdots|0\rangle=0.
\end{align}
We need to supplement this with the helicity counting identity (invariance under projective rescalings of the super-Penrose transform \eqref{SuperPenroseTransform}) which reads,
\begin{align}\label{superhelicity1}
    &\mathbf{h}_j\langle\cdots \hat{\mathbf{J}}_{s_j}^{+}(\mathcal{Z}_j)\cdots\rangle=-\frac{1}{2} \big(\mathcal{Z}_{j\mathcal{A}}\frac{\partial}{\partial \mathcal{Z}_{j\mathcal{A}}}+2\big)\langle\cdots \hat{\mathbf{J}}_{s_j}^{+}(\mathcal{Z}_j)\cdots\rangle = +s_{j} \langle\cdots \hat{\mathbf{J}}_{s_j}^{+}(\mathcal{Z}_j)\cdots\rangle, \notag \\
    & \mathbf{h}_j\langle\cdots\hat{\mathbf{J}}_{s_j}^{-}(\mathcal{W}_j)\cdots\rangle=\frac{1}{2} \big(\mathcal{W}_{j\mathcal{A}}\frac{\partial}{\partial \mathcal{W}_{j\mathcal{A}}}+2 \big)\langle\cdots \hat{\mathbf{J}}_{s_j}^{-}(\mathcal{W}_j)\cdots\rangle = -s_{j} \langle\cdots \hat{\mathbf{J}}_{s_j}^{-}(\mathcal{W}_j)\cdots\rangle.
\end{align}
The super-twistor and dual super-twistor variables are related by a super-Fourier transform analogous to the non susy case \eqref{ZtoWfullFourier},
\begin{align}\label{superfouriertransform}
    f(\mathcal{W}) = \int \frac{d^{4|\mathcal{N}}\mathcal{Z}}{(2\pi)^{2+\frac{\mathcal{N}}{2}}} \, e^{-i \mathcal{Z}\cdot\mathcal{W}} \, \tilde{f}(\mathcal{Z}), \qquad \tilde{f}(\mathcal{Z}) = \int \frac{d^{4|\mathcal{N}}\mathcal{W}}{(2\pi)^{2+\frac{\mathcal{N}}{2}}} \, e^{i \mathcal{Z}\cdot\mathcal{W}} \, f(\mathcal{W}).
\end{align}
\begin{exercise}
Using \eqref{superfouriertransform} and \eqref{superhelicity1}, show that,
    \begin{align}\label{superhelicity2}
&\mathbf{h}_j\langle\cdots\hat{\mathbf{J}}_{s_j}^{+}(\mathcal{W}_j)\cdots\rangle=\frac{1}{2} \big(\mathcal{W}_{j\mathcal{A}}\frac{\partial}{\partial \mathcal{W}_{j\mathcal{A}}}+2 - \mathcal{N} \big)\langle\cdots \hat{\mathbf{J}}_{s_j}^{+}(\mathcal{W}_j)\cdots\rangle = +s_{j} \langle\cdots \hat{\mathbf{J}}_{s_j}^{+}(\mathcal{W}_j)\cdots\rangle \notag, \\
&\mathbf{h}_j\langle\cdots\hat{\mathbf{J}}_{s_j}^{-}(\mathcal{Z}_j)\cdots\rangle=-\frac{1}{2} \big(\mathcal{Z}_{j\mathcal{A}}\frac{\partial}{\partial \mathcal{Z}_{j\mathcal{A}}}+2 - \mathcal{N} \big)\langle\cdots \hat{\mathbf{J}}_{s_j}^{-}(\mathcal{Z}_j)\cdots\rangle = -s_{j} \langle\cdots \hat{\mathbf{J}}_{s_j}^{-}(\mathcal{W}_j)\cdots\rangle. 
\end{align}
Notice the extra factor of $\mathcal{N}$ that explicitly features in the helicity identities. Therefore, it is ``natural" to express positive helicity quantities in super-twistor variables and negative helicity quanties in terms of dual-super twistors.
\end{exercise}

The solutions to \eqref{manifesttwistorsuperconformalWard} include arbitrary functions of the invariants constructed out of the supertwistors $\mathcal{Z}_i$ and $\mathcal{W}_j$. While $\mathcal{Z}$ and $\mathcal{W}$ can be directly contracted, we must make use of the graded-symplectic form $\Omega$ \eqref{gradedOmega} to contract a (dual)twistor with another (dual)twistor. These are given by,
\begin{align}\label{supertwistorsdotprods}
\notag \mathcal{Z}_i \cdot \mathcal{Z}_j&=-\mathcal{Z}_i^{\mathcal{A}}\Omega_{\mathcal{A}\mathcal{B}}\mathcal{Z}_j^{\mathcal{B}}= Z_i \cdot Z_j-\delta_{MN}\psi_i^M\psi_{j}^N,\\ 
\notag\mathcal{W}_i \cdot \mathcal{W}_j&=\mathcal{W}_{i\mathcal{A}}\Omega^{\mathcal{A}\mathcal{B}}\mathcal{W}_{j\mathcal{B}}= W_i \cdot W_j+\delta_{MN}\bar{\psi}_i^M\bar{\psi}_{j}^N,\\
\notag\mathcal{Z}_i \cdot \mathcal{W}_j&=-\mathcal{Z}_i^{\mathcal{A}}\mathcal{W}_{j \mathcal{A}}= Z_i \cdot W_j+\delta_{MN}\psi_i^M\bar{\psi}_{j}^N,\\
\mathcal{W}_i\cdot \mathcal{Z}_j&=\mathcal{W}_{i \mathcal{A}}\mathcal{Z}_j^{\mathcal{A}}= W_i \cdot Z_j+\delta_{MN}\bar{\psi}_i^M\psi_{j}^N.
\end{align}
There are also projective super delta function invariants much like in the non susy case. \begin{align}\label{nptansatzdelta4susy}
    \mathcal{F}(\mathcal{Z}_1,\cdots \mathcal{Z}_n)= \int dc_1\cdots dc_{n-1}~ f(c_1,\cdots,c_{n-1})\delta^{4|\mathcal{N}}(c_1 \mathcal{Z}_1+\cdots +\mathcal{Z}_n).
\end{align}
To erase arbitrariness in the parameters $c_i$, we have integrated them on the support of a function $f(c_1,c_2,\cdots c_{n-1})$. For the projectiveness of this quantity and to be well defined in the super-Penrose transform, this function will be constrained by the super-helicity identity at each point viz \eqref{superhelicity1}, \eqref{superhelicity2}. 
\begin{exercise}
    At the level of three points show that the unique invariant we can form of this kind with three super-twistors is,
     \begin{align}\label{susydelta3}
    &\delta^{3|\mathcal{N}}(\mathcal{Z}_1,\mathcal{Z}_2,\mathcal{Z}_3;\alpha_{12}, \alpha_{23},\alpha_{31})= (-i)^{\alpha}\delta^{[\alpha_{12}]}(\mathcal{Z}_1\cdot \mathcal{Z}_2) \int dc_{23}dc_{31} c_{23}^{\alpha_{23}}c_{31}^{\alpha_{31}} \delta^{4|\mathcal{N}}(\mathcal{Z}_3+c_{23} \mathcal{Z}_1 + c_{31} \mathcal{Z}_2).
\end{align}
\bonus{Can you construct invariants of this form with four twistors? How many undetermined parameters do such solutions in general have?}
\end{exercise}

\subsection{Replace Twistors by Super-Twistors! The quickest route towards SUSY}
The first thing to notice from the form of the invariants \eqref{supertwistorsdotprods} and \eqref{nptansatzdelta4susy} is that is that tey are simply the same as their non supersymmetric counterparts \eqref{twistordotprods} and \eqref{nptansatz}! Supplementing this by the super-helicity identities \eqref{superhelicity1} and \eqref{superhelicity2}, we see that super-correlators can be obtained from their non-susy just by replacing twistors by super-twistors. The hard work was in finding the correct set of variables. The rest is all fun and games.

\subsection{Examples of Two and Three point functions}
Starting at two points we pretty much have a unique solution for a spin-s super-current,
\begin{align}\label{susyJsJsTwopoint}
    \langle0|\mathbf{\hat{J}_{s}^{+}}(\mathcal{Z}_1)\mathbf{\hat{J}_{s}^{+}}(\mathcal{Z}_2)|0\rangle = \frac{1}{(\mathcal{Z}_1\cdot\mathcal{Z}_2)^{2s+2}}, ~~ \langle0|\mathbf{\hat{J}_{s}^{-}}(\mathcal{Z}_1)\mathbf{\hat{J}_{s}^{-}}(\mathcal{Z}_2)|0\rangle = \frac{1}{(\mathcal{Z}_1\cdot\mathcal{Z}_2)^{2(-s+\frac{\mathcal{N}}{2})+2}}.
\end{align}
Note the extra factor of $\frac{\mathcal{N}}{2}$ for the negative helicity case which is a consequence of \eqref{superhelicity2}. This will be reversed when working in dual super-twistor space. 
\begin{exercise}
    Using the super-field expansion \eqref{supertwistorcurrent1} and performing the Grassmann expansion, show that we recover the correct component level results viz \eqref{JsJsevenTwopointTwistor}, \eqref{JsJsevenTwopointTwistormm}.
\end{exercise}
\begin{exercise}
    Lets now pretend for a moment that we do not know the manifest super-conformal generators and that we only know the supersymmetry generator \eqref{QTwistor}. Lets also assume that we know all the component correlators by solving the ordinary twistor space conformal Ward identities for each component correlator. By using the super field expansion \eqref{supertwistorcurrent1} and solving the supersymmetry Ward identity due to \eqref{QTwistor}, show that we recover the results \eqref{susyJsJsTwopoint}.
\end{exercise}
Lets move on to three point functions focusing on $\mathcal{N}=1$ theories. We have for three integer spins,
\begin{align}\label{3deltaSusydot}
    \langle 0|\mathbf{\hat{J}_{s_1}^+ }(\mathcal{Z}_1)\mathbf{\hat{J}_{s_2}^+ }(\mathcal{Z}_2)\mathbf{\hat{J}_{s_3}^+ }(\mathcal{Z}_3)|0\rangle_h=  (-i)^{s_T} \delta^{[s_1+s_2-s_3]}(\mathcal{Z}_1\cdot \mathcal{Z}_2)\delta^{[s_2+s_3-s_1]}(\mathcal{Z}_2\cdot \mathcal{Z}_3)\delta^{[s_3+s_1-s_2]}(\mathcal{Z}_3 \cdot \mathcal{Z}_1).
\end{align}
The helicity identities of course allow solutions that are of a polynomial type just as in the non-susy case. However, they are ruled out here for the same reason they were ruled out there thus leaving us with \eqref{3deltaSusydot}. What about the other solution involving projective delta functions? Well first of all, let us note that for three half integer spins we have,
\begin{align}\label{delta4+++}
    &\langle0|\mathbf{\hat{J}_{s_1}^+ }(\mathcal{Z}_1)\mathbf{\hat{J}_{s_2}^+ }(\mathcal{Z}_2)\mathbf{\hat{J}_{s_3}^+ }(\mathcal{Z}_3)|0\rangle_{nh}\notag\\&= (-i)^{s_T}\delta^{[s_1+s_2+s_3]}(\mathcal{Z}_1\cdot\mathcal{Z}_2) \int dc_{23} dc_{31} c_{23}^{-s_2-s_3+s_1} c_{31}^{-s_3-s_1+s_2} \delta^{4|1}(\mathcal{Z}_3^{\mathcal{A}}+ c_{31} \mathcal{Z}_2^{\mathcal{A}}+c_{23}\mathcal{Z}_1^{\mathcal{A}})
\end{align}
Notice that there only exists a homogeneous solution for integer spins and a non-homogeneous solution for half-integer spins. Super-conformal invariance in the $(+++)$ helicity configuration allows for two solutions which are \eqref{3deltaSusydot} and \eqref{susydelta3}. However, the latter of these solutions is Grassmann odd due to the fermionic part of the delta function! Lets look at the super-current expansion \eqref{supertwistorcurrent1}. For integer spins, expanding into components we obtain,
\small
\begin{align}\label{susys1s2s3intcomp}
    &\langle0|\mathbf{\hat{J}_{s_1}^+ }(\mathcal{Z}_1)\mathbf{\hat{J}_{s_2}^+ }(\mathcal{Z}_2)\mathbf{\hat{J}_{s_3}^+ }(\mathcal{Z}_3)|0\rangle=\langle0| \hat{J}_{s_{1}}^+ \hat{J}_{s_{2}}^+ \hat{J}_{s_{3}}^+|0\rangle - i\psi_{1}\psi_{2} \langle0| \hat{J}_{s_{1}+\frac{1}{2}}^+ \hat{J}_{s_{2}+\frac{1}{2}}^+ \hat{J}_{s_{3}}^+|0\rangle-i\psi_{2}\psi_{3} \langle0| \hat{J}_{s_{1}}^+ \hat{J}_{s_{2}+\frac{1}{2}}^+ \hat{J}_{s_{3}+\frac{1}{2}}^+|0\rangle\notag\\& +i\psi_{3}\psi_{1} \langle0| \hat{J}_{s_{1}+\frac{1}{2}}^+ \hat{J}_{s_{2}}^+ \hat{J}_{s_{3}+\frac{1}{2}}^+|0\rangle.
\end{align}
\normalsize
The remaining component correlators have net half-integer spin which vanish by Lorentz invariance. Similarly, for three half-integer spins we obtain,
\begin{align}\label{susys1s2s3halfintcomp}
    &\langle0|\mathbf{\hat{J}_{s_1}^+ }(\mathcal{Z}_1)\mathbf{\hat{J}_{s_2}^+ }(\mathcal{Z}_2)\mathbf{\hat{J}_{s_3}^+ }(\mathcal{Z}_3)|0\rangle=e^{\frac{i\pi}{4}}\psi_3\langle 0|\hat{J}_{s_1}^{+}\hat{J}_{s_2}^{+}\hat{J}_{s_3+\frac{1}{2}}^{+}|0\rangle-e^{\frac{i\pi}{4}}\psi_2\langle 0|\hat{J}_{s_1}^{+}\hat{J}_{s_2+\frac{1}{2}}^{+}\hat{J}_{s_3}^{+}|0\rangle+e^{\frac{i\pi}{4}}\psi_1\langle 0|\hat{J}_{s_1+\frac{1}{2}}^{+}\hat{J}_{s_2}^{+}\hat{J}_{s_3}^{+}|0\rangle\notag\\&-e^{\frac{3i\pi}{4}}\psi_1\psi_2\psi_3\langle 0|\hat{J}_{s_1+\frac{1}{2}}^{+}\hat{J}_{s_2+\frac{1}{2}}^{+}\hat{J}_{s_3+\frac{1}{2}}^{+}|0\rangle,
\end{align}
with the remaining ones vanishing by Lorentz invariance. Component level conformal invariance allows each of the correlators in \eqref{susys1s2s3intcomp} and \eqref{susys1s2s3halfintcomp} to be either a product of three delta function derivatives (homogeneous correlator) or a projective delta function (non-homogeneous correlator) \eqref{3pointparityevenTwistorppp}. Lets focus on \eqref{susys1s2s3intcomp}. If each component is a projective delta function type solution, then there is no way it can combine to a super-projective delta function since the latter is Grassmann odd! Therefore, it would lead to an answer that is not super-covariant and consistent with the super-Twistor Ward identities (in particular the supersymmetry Ward identity). Thus, the component correlators arrange themselves to be purely of the delta function product type in order to combine to \eqref{3deltaSusydot}. Similarly, the half-integer correlators conspire together to lead to \eqref{susys1s2s3halfintcomp}.
\begin{exercise}
    Determine the three point super-correlators with two integer and one half integer spin currents and vice versa. There is no need to do a calculation if you use the above arguments.
\end{exercise}
\section{Parity odd super-sector: The rise of the super Infinity Twistor}\label{sec:SuperParityOdd}
Just as we did in the non susy case, lets go beyond the parity even sector and tackle the parity odd cases following \cite{Bala:2025qxr}.
\subsection{The epsilon transform of the super-Penrose transform}
Lets generalize the epsilon transform \eqref{EPTspinornotation} to the supersymmetric setting. For super-currents the epsilon transform is a simple generalization of the same and reads,
\begin{align}\label{superspacecurrentEPT}
    (\epsilon\cdot \mathbf{J}_s(x,\theta))^{a_1\cdots a_{2s}}=-i\int \frac{d^3 y}{|y-x|^2}\frac{\partial}{\partial y^{b}_{(a_1}}\mathbf{J}^{a_2\cdots a_{2s})b}_{s}(y,\theta)
\end{align}
The super-twistor space epsilon transform can then therefore be derived from \eqref{superspacecurrentEPT} and the super-Penrose transform \eqref{SuperPenroseTransform}. We find,
\begin{align}\label{supertwistorEPTspins}
    (\epsilon \cdot \hat{\mathbf{J}}_s)^+(\mathcal{Z})= 
    -i \int \frac{d^3 y}{y^2} \,
    \mathcal{Z}^A \mathcal{I}_{AB} \frac{\partial}{\partial \mathcal{Z}_B}
    \hat{\mathbf{J}}_s^+(\lambda, \bar{\mu},\psi) \bigg|_{\bar{\mu}^a \to \bar{\mu}^a + y^{ab} \lambda_b},
\end{align}
where $\mathcal{I}_{AB}$ is the super-infinity twistor \eqref{superinfinitytwistor} to be defined shortly .
The parity odd super-Penrose transform is thus given by,
\begin{align}\label{parityoddSuperPenrose}
    (\epsilon\cdot \mathbf{J}_s)^{a_1\cdots a_{2s}}(x,\theta)=\int \langle \lambda d\lambda\rangle \lambda^{a_1}\cdots \lambda^{a_{2s}}(\epsilon\cdot \hat{\mathbf{J}}_s)^{+}(\mathcal{Z})|_{\mathcal{X}},
\end{align}
where $\mathcal{X}$ denote the super-incidence relations. We have also defined the super-infinity twistor that is equal to,
\begin{align}\label{superinfinitytwistor}
    \mathcal{I}_{\mathcal{A}\mathcal{B}}=I_{AB}\oplus 0_{\mathcal{N}\times \mathcal{N}}.
\end{align}
Similarly, the parity odd super-Witten transform obtained using the momentum space version of the super-epsilon transform \eqref{superspacecurrentEPT} and performing a super half-Fourier transform like in \eqref{superWittentransform} leads to,
\begin{align}\label{superWittentransformparityodd}
    \mathbf{J}_s^{+}(\mathcal{Z})=\mathbf{J}_s^{+}(\lambda,\Bar{\mu},\psi)=\int \frac{d^2\Bar{\lambda}}{(2\pi)^2}\int d \xi_{-}\int d \Bar{\eta}~ e^{i\Bar{\lambda}\cdot \Bar{\mu}+\frac{(\xi_-+\sqrt{2} e^{\frac{i\pi}{4}}\psi) \Bar{\eta}}{2}}i\text{Sign}(\frac{(\lambda\cdot \Bar{\lambda})^2}{4})\hat{\mathbf{J}}_s^{+}(\lambda,\Bar{\lambda},\sqrt{2}e^{\frac{i\pi}{4}}\psi-\xi_-,\Bar{\eta}).
\end{align}
\subsection{Two and three point functions}
Using the parity odd super-Penrose transform \eqref{parityoddSuperPenrose}, we can easily determine the parity odd two and three point functions given their parity even counterparts. For two point functions, we obtain using \eqref{susyJsJsTwopoint} and \eqref{supertwistorEPTspins} to perform an epsilon transform with one of the two currents we get,
\begin{align}\label{twopointoddSUSY}
    \langle 0|\mathbf{\hat{J}}_{s_1}^{+}(\mathcal{Z}_1)\mathbf{\hat{J}}_{s_2}^{+}(\mathcal{Z}_2)|0\rangle=\text{Sgn}(\mathcal{Z}_1\mathcal{I}\mathcal{Z}_2)\delta^{[2s+1]}(\mathcal{Z}_1\cdot \mathcal{Z}_2).
\end{align}
\begin{exercise}
    Read off the component correlators in \eqref{twopointoddSUSY} by using the super-current expansion \eqref{SuperCurrentcomponent1}. Use the super-Penrose transform \eqref{parityoddSuperPenrose}, \eqref{SuperPenroseTransform} to go to position space (assume that the epsilon transform has been done with respect to the first current). What happens to the position space result when $x_1^\mu\ne x_2^\mu$?
\end{exercise}
Moving on, lets epsilon transform the parity even three point result for integer spins viz \eqref{3deltaSusydot} with respect to the first operator. The result is,
\small
\begin{align}\label{SUSY3pointODD}
  &\langle 0|\mathbf{\hat{J}}_{s_1}^{+}(\mathcal{Z}_1)\mathbf{\hat{J}}_{s_2}^{+}(\mathcal{Z}_2)\mathbf{\hat{J}}_{s_3}^{+}(\mathcal{Z}_3)|0\rangle\notag\\&=\frac{i}{2}\int dc_{12}dc_{23}dc_{31}c_{12}^{s_1+s_2-s_3}c_{23}^{s_2+s_3-s_1}c_{31}^{s_3+s_1-s_2}\text{Sgn}\big(c_{12}\langle \mathcal{Z}_1\mathcal{I}\mathcal{Z}_2\rangle+c_{31}\langle \mathcal{Z}_3 \mathcal{I}\mathcal{Z}_2\rangle\big)e^{-ic_{12}\mathcal{Z}_1\cdot \mathcal{Z}_2-ic_{23}\mathcal{Z}_2\cdot \mathcal{Z}_3-ic_{31}\mathcal{Z}_3\cdot \mathcal{Z}_1}\bigg|_{\mathcal{X}}.
\end{align}
\normalsize
Note that the super-infinity twistor features in parity odd super-correlators just as its non-susy featured earlier. The dependence, just as before is only through sign factors again.
\section{Extensions to super-scalars and generic operators}\label{sec:superscalarsgeneric}
Lets generalize the construction to super-scalars now \cite{Bala:2025qxr}. We do so for scaling dimension one and leave the generic case as an exercise to the reader. In position space, the scalar super-field is given by,
\begin{align}\label{J0posspace}
    \mathbf{J}_0(x,\theta)=O_1(x)+\theta_a O_{1/2}^{a}(x)+\theta^2 O_2(x).
\end{align}
We have seen that the coordinates for super-twistor space adapted to conserved super-currents are $\mathcal{Z}^{\mathcal{A}}=(Z^A,\psi)$. For super-scalar operators to be incorporated into this framework however, we need an additional Grassmann coordinate, $\psi_-$. The reason for the following is that the super-scalar in $\mathcal{N}=1$ theories consists of scalars with $\Delta=1$ and $\Delta=2$ as well as the positive and negative helicity components of a spin-1/2 operator. The explicit expression is given by \cite{Bala:2025qxr}:
\begin{align}\label{scalarmultiplet}
    \mathbf{\hat{J}}_0(\mathcal{Z},\psi_-)=\frac{e^{\frac{i\pi}{4}}}{2\sqrt{2}}(\psi+\psi_-)\hat{O}_1(Z)+\frac{1}{2}\hat{O}_{1/2}^{-}(Z)+\frac{i}{2}\psi_- \psi \hat{O}_{1/2}^{+}(Z)-\sqrt{2}e^{\frac{i\pi}{4}}(\psi-\psi_-)\hat{O}_2(Z).
\end{align}

For conserved super-currents, the Penrose transform was given by \eqref{SuperPenroseTransform}. For the super-scalar \eqref{J0posspace}, we propose that it is related to \eqref{scalarmultiplet} as follows:
\begin{align}\label{SuperPenroseTransformScalar}
    \mathbf{J}_0(x,\theta)=\int \langle \lambda d\lambda\rangle\int e^{-\frac{i\pi}{4}}d\psi_-~\mathbf{\hat{J}}_0(\mathcal{Z},\psi_-)|_{\mathcal{X}},
\end{align}
where $\mathcal{X}$ are the super-incidence relations \eqref{Superincidence}.
A few words are in order. From \eqref{scalarmultiplet}, we can see that $\psi,\psi_-$ both have helicity $-\frac{1}{2}$. Therefore, $\mathbf{\hat{J}}_0(\mathcal{Z})$ behaves like a $s=-\frac{1}{2}$ current (and not like a scalar with  $s=0$!) and satisfies,
\begin{align}\label{J0projective}
    \mathbf{\hat{J}}_0(r\mathcal{Z},r\psi_-)=\frac{1}{r}\mathbf{\hat{J}}_0(\mathcal{Z},\psi_-).
\end{align}
Then under $\lambda\to r\lambda,\psi_-\to r\psi_-$, \eqref{SuperPenroseTransformScalar}, the measure transforms as $\langle\lambda d\lambda\rangle d\psi_-\to r^2 \langle \lambda d\lambda\rangle d(r\psi_-)=r^2 \langle \lambda d\lambda\rangle \frac{1}{r}d(\psi_-)=r \langle \lambda d\lambda\rangle d(\psi_-)$. This is precisely canceled out by the transformation of $J_0$ \eqref{J0projective} and thus the integral is invariant under projective rescalings.
Next, the super-Witten transform for scalars is identical to the spinning case \eqref{superWittentransform} with $s=0$ with one major difference that the $\xi_-$ coordinate is not integrated over. 
The main difference from the spinning case \eqref{superWittentransform} is that the scalar super-field \eqref{scalarmultiplet} depends on the difference $\chi-\eta\sim \psi_-$ in a non-trivial way unlike the spinning case \eqref{superWittentransform} where it appears as an overall multiplicative quantity.
\subsection{Two point functions and supersymmetric contact terms}
Lets start with two point functions. An ansatz consistent with projective rescaling and dimensional analysis is given by,
\begin{align}
    \langle 0|\mathbf{\hat{J}}_0(\mathcal{Z}_1,\psi_{1-})\mathbf{\hat{J}}_0(\mathcal{Z}_2,\psi_{2-}|0\rangle=\frac{c_{\frac{1}{2}}}{(\mathcal{Z}_1\cdot \mathcal{Z}_2)}+\frac{c_{\Delta=1}\psi_{1-}\psi_{2-}}{(\mathcal{Z}_1\cdot \mathcal{Z}_2)^2}.
\end{align}
However, the supersymmetric Penrose transform \eqref{SuperPenroseTransformScalar} for scalars instructs us to integrate over the $\psi_-$ coordinates thus yielding the simpler result,
\begin{align}\label{scalarsuperpenrose2point}
    \langle 0|\mathbf{J}_0(x_1,\theta_1)\mathbf{J}_0(x_2,\theta_2)|0\rangle&=\int \langle \lambda_1d\lambda_1\rangle\langle\lambda_2d\lambda_2\rangle\int d\psi_{1-}d\psi_{2-}\bigg(\frac{c_{\frac{1}{2}}}{(\mathcal{Z}_1\cdot \mathcal{Z}_2)}+\frac{c_{\Delta=1}\psi_{1-}\psi_{2-}}{(\mathcal{Z}_1\cdot \mathcal{Z}_2)^2}\bigg)\notag\\
    &=\int \langle \lambda_1d\lambda_1\rangle\langle\lambda_2d\lambda_2\rangle\frac{c_{\Delta=1}}{(\mathcal{Z}_1\cdot \mathcal{Z}_2)^{2}}.
\end{align}
which is simply obtained from the spinning result \eqref{susyJsJsTwopoint} by setting $s=0$.
Using the super incidence relations \eqref{Superincidence} and performing a Grassmann expansion results in,
\begin{align}\label{supertwistorexpansion2pointscalar}
    &\langle 0|\mathbf{J}_0(x_1,\theta_1)\mathbf{J}_0(x_2,\theta_2)|0\rangle\notag\\
    &=\int \langle \lambda_1d\lambda_1\rangle\langle\lambda_2d\lambda_2\rangle \frac{i^{2s+2}}{(Z_1\cdot Z_2)_0^{2}}-4i\theta_{1a}\theta_{2b}\int \langle \lambda_1d\lambda_1\rangle\langle\lambda_2d\lambda_2\rangle \frac{i^{2}\lambda_1^a\lambda_2^b}{(Z_1\cdot Z_2)_0^{3}}\notag\\
    &-\frac{i}{2}(\theta_1^2+\theta_2^2)\int \langle \lambda_1d\lambda_1\rangle\langle\lambda_2d\lambda_2\rangle \frac{\langle 1 2\rangle}{(Z_1\cdot Z_2)_0^{3}}-\frac{3}{8}\theta_1^2\theta_2^2\int \langle \lambda_1d\lambda_1\rangle\langle\lambda_2d\lambda_2\rangle \frac{\langle 1 2\rangle^2}{(Z_1\cdot Z_2)_0^{4}}.
\end{align}

The first integral is the $\langle 0|O_1(x_1)O_1(x_2)|0\rangle$, the second one is $\langle O_{1/2}^{a}(x_1)O_{1/2}^{b}(x_2)\rangle$ whereas the fourth integral is $\langle 0|O_2O_2|0\rangle$. As desired, the Supersymmetric Penrose transform has reproduced for us twistor space expression for $\langle O_2(x_1)O_2(x_2)\rangle$ which is obtainable by a Legendre transform \cite{Jain:2024bza}. Thus, we see that the dependence on the infinity twistor \eqref{infinitytwistor} naturally emerges via the super-incidence relations \eqref{Superincidence}. The third integral in \eqref{supertwistorexpansion2pointscalar} is zero since it is odd under the exchange of $\lambda_1$ and $\lambda_2$. Thus, we have obtained a result that is perfectly consistent with the superfield expansion \eqref{J0posspace}. 
\begin{exercise}
    In a conformal field theory, is it true that vacuum two point functions of non-identical operators is always zero? Lets work in d dimensions for this exercise. Can you come up with a situation where a scalar with dimension $\Delta_1$ and one with dimension $\Delta_2$ can have a non-zero two point function? Attempt to write down all possible solutions in position space consistent with scaling and check invariance under SCTs.
    \hint{Do not neglect Dirac delta functions.}
    \bonus{If you agree that non-identical operators can have non-zero two point functions, show that,
    \begin{align}
         \langle 0|\mathbf{\hat{J}}_0(\mathcal{Z}_1,\psi_{1-})\mathbf{\hat{J}}_0(\mathcal{Z}_2,\psi_{2-}|0\rangle= \psi_{1-}\psi_{2-} \text{Sgn}(\langle \mathcal{Z}_1 \mathcal{I}\mathcal{Z}_2\rangle)\delta^{[1]}(\mathcal{Z}_1\cdot\mathcal{Z}_2),
    \end{align}}
    and the super-field expansion \eqref{scalarmultiplet} lead to a non-zero $\langle O_1 O_2\rangle$ two point function and a parity odd spin-1/2 correlator. Convert this to position space using the super-Penrose transform and show that it reproduces a beautiful supersymmetric contact term \cite{Bala:2025qxr}. 
\end{exercise}

The general case as well as the extension to generic non-conserved super operators have not yet appeared in the literature and are left as an exercise to the interested reader.

\part{The Road so Far}\label{part:Conclusion}
In these lecture notes, we have discussed the formulation of 3d CFT in the language of spinors and twistors. Let us summarize each part now.
\section{Summary of part \ref{part:SH}}
We started by developing off-shell spinor helicity variables for three dimensional QFT in section \ref{sec:SH3d}. We discussed their construction in the distinct cases of Lorentzian (subsection \ref{subsec:LorentzSH}) and Euclidean (subsection \ref{subsec:EuclidSH}) signature space-times and how they naturally arise from a dimensional reduction of the usual on-shell massless spinor helicity variables in Klein and Minkowski space-times respectively. We then turned our interest to correlators of conserved currents and recast them in the helicity basis in subsection \ref{subsec:HelicityBasis}. In contrast to the usual representation of currents as Lorentz tensors, the helicity basis trades them with just two simple scalar like quantities: The positive and negative helicity components. In the AdS/CFT context, these are dual to the two helicity components of the dual bulk gauge fields. We then set up the conformal Ward identities in subsection \ref{subsec:ConformalWardId} and discussed the general form of their two and three point correlators in subsection \ref{subsec:TwoThreePointSH}. We then discussed the construction of conformal partial waves which are the building blocks for higher point correlators in subsection \ref{subsec:FourHigherPointSH}.

We then turned to some applications of this formalism. In particular, we discussed double copy relations in section \ref{sec:DoubleCopy} and a flat space limit of AdS$_4$ correlators in section \ref{sec:flatlimit}. In section \ref{sec:CSmattercorrelators} we specialized to a particular class of theories (Chern-Simons matter theories) and discussed how the spinor helicity formalism makes certain aspects manifest such as the anyonic nature of correlation functions. Finally, in section \ref{sec:chiralhigherspin} , we saw how this formalism enables us to identify the holographic dual of Ads$_4$ chiral higher spin theory.

Before proceeding to the next part on Twistors, we required one more ingridient: The formalism of Wightman functions in Lorentzian CFT. We discussed just that in section \ref{sec:LorentzianCFT}. We discussed how CPT acts on spinor helicity variables in subsection \ref{subsec:CPT}. In subsection \ref{subsec:Definitions}, we discussed the general non-perturbative structure of Wightman functions in field theories. We then proceeded to discuss how Wightman functions can be obtained by a suitable Wick rotation of Euclidean correlators in momentum space in subsection \ref{subsec:EuclidtoWightmangeneral}. We concluded this section with a simple example of a Wightman function of a stress tensor with two $U(1)$ conserved currents in subsection \ref{subsec:WightmanExample}.

\section{Summary of part \ref{part:Twistors}}
We began with a discussion of the concept and geometry of Twistor space in section \ref{sec:TwistorGeometry}.
We then derived and developed the Penrose transform for conserved currents in section \ref{sec:PenroseTransform} and the Witten transform in \ref{sec:WittenTransform}. We discussed how the two are related in section \ref{sec:PenroseFromWittenandFourier}. In section \ref{sec:conformalWardIdTwistor}, we setup the conformal Ward identities in Twistor variables and discussed the two and three point functions of conserved currents. Realizing that we were missing parity odd contributions, we saw in section \ref{sec:TwistorParityodd} that the Infinity Twistor of $\mathbb{R}^{2,1}$ allows us to incorporate them. Finally, in section \ref{sec:ScalarsGenericTwistor}, we derived the Penrose transform for scalars and generic primary operators. We saw the interesting feature that the representation of the conformal generators acting on general operators (not conserved currents) involve non-local terms in Twistor space.

\section{Summary of part \ref{part:SuperTwistors}}
We began with a brief recap of superconformal algebras in three dimensions in section \ref{sec:susyalgebra}. In section \ref{sec:GeometrySuperTwistor}, we discussed the construction and geometry of the super-Twistor space. In section \ref{sec:SuperPenroseTransform}, we derived the super-Penrose transform for conserved super-currents. We proceeded to discuss the super-Witten transform in section \ref{sec:SuperWittenTransform} and how it is related to the super-Penrose transform in section \ref{sec:SuperPenroseandWitten}. In section \ref{sec:SuperConformalWardId}, we setup the super-conformal Ward identities that constrain the form of super-correlators of conserved super-currents. We derived the general form of the super-conformal invariants for $n-$point functions and explicitly solved for two and three point super-current correlators. We discussed the parity odd case separately in section \ref{sec:SuperParityOdd} as it required the use of the super-Infinity twistor. Finally, we briefly discussed the construction of super-scalars in section \ref{sec:superscalarsgeneric}.

\subsection*{Acknowledgments}
I would like to thank the organizers of $\text{ST}^4$ $2025$ for putting together an intellectuality stimulating workshop and the opportunity to lecture there. I also convey my thanks to the participants for their excellent questions and comments.

\part{Appendix}
\appendix
\section{A Brief guide to 3d CFT in position space}\label{app:CFTreview}

The theories of interest to us in these lectures are conformal field theories. In addition to the usual Poincare invariance, these theories possess \textit{scale invariance} as well as invariance under what are called \textit{special conformal transformations}. 
\begin{exercise}
    Consider an infinitesimal diffeomorphism $x^\mu\to x^\mu+\epsilon^\mu(x)$. If we demand that the metric\footnote{This is short for $\delta_{\mu\nu}=\text{diag}(1,1,1)$ in the Euclidean context or $\eta_{\mu\nu}=\text{diag}(-1,1,1)$ for its Lorentzian counterpart.} $\delta_{\mu\nu}$ should only change up to an overall factor $\Omega^2(x)$, show that $\epsilon^\mu(x)$ obeys the conformal Killing equation,
    \begin{align}\label{conformalKillingequation}
        \partial_\mu \epsilon_\nu+\partial_\nu \epsilon_\mu=(1-\Omega^2(x))\delta_{\mu\nu}=\frac{2}{d}\partial\cdot\epsilon~\delta_{\mu\nu}.
    \end{align}
    Further, show that the solutions to these equations are at most quadratic in $x$ and explicitly given by,
\begin{align}\label{conformalkillingvectors}
        \epsilon_{\mu}(x)=a_\mu+\omega_{[\mu\nu]}x^\nu+\lambda x_{\mu}-2(b\cdot x)x_\mu+x^2 b_\mu.
    \end{align}
$a_\mu$ and $\omega_{[\mu\nu]}$ correspond to translations and Lorentz transformations (Rotations in the Euclidean context) which leave the metric invariant. $\lambda$ represents a scale transformation whereas the vector parameter $b$ generates a special conformal transformation. The former changes the metric up to an overall constant whereas the latter leads to a position dependent conformal factor. Therefore, there a total of $3+\frac{3(3-1)}{2}+1+3=10$ conformal generators in $d=3$.
\end{exercise}
Continuing on from the previous exercise, lets now find out how a function $f(x)$ transforms under infinitesimal conformal transformations. Given the form of the general conformal transformations \eqref{conformalkillingvectors}, we have,
\begin{align}
    f(x'=x+\epsilon)=f(x)+\epsilon^\mu\partial_\mu f(x)+\order{\epsilon^2}.
\end{align}
Performing this using \eqref{conformalkillingvectors} show that,
\begin{align}
   f(x')-f(x)=(a^\mu \partial_\mu+\omega_{\mu\nu}x^{[\mu}\partial^{\nu]}+\lambda x^\mu \partial_\mu+b^\mu(-2 x_\mu x^\nu \partial_\nu+x^2\partial_\mu))f(x)+\order{\epsilon^2}.
\end{align}
This allows us to identify the translation, Lorentz transformation/Rotation, Dilatation and special conformal generators,
\begin{align}\label{confgen1}
    &P_\mu=-i\partial_\mu~,M_{\mu\nu}=-i(x_\mu \partial_\nu-x_\nu\partial_\mu),\notag\\
    &D=-i(x^\mu\partial_\mu), K_\mu=-i(-2x_\mu x^\nu\partial_\nu+x^2\partial_\mu).
\end{align}
So far, we have focused on how ordinary functions transform under conformal transformations. We now generalize this construction to include arbitrary representations of the conformal group. First of all, we see that the Cartan sub-algebra of \eqref{confalgebra} is spanned by the dilatation operator $D$ and the Lorentz/Rotation generators $M_{\mu\nu}$. Therefore, we can label an irreducible representation by their eigenvalues. The representations that we are interested in are local operators such as quantum fields and composite operators constructed out of them. Lets first consider a translation parametrized by $a^\mu$. A operator $\mathcal{O}(x)$ transforms as, 
\begin{align}
    e^{i P\cdot a}\mathcal{O}(x)e^{-i P\cdot a}=\mathcal{O}(x').
\end{align}
Expanding both sides for infinitesimal translations we obtain,
\begin{align}\label{OperatorPGen}
    [P_\mu,\mathcal{O}(x)]=-i\partial_\mu \mathcal{O}(x).
\end{align}
To accommodate non-trivial representations we take the operator placed at the origin to be a simultaneous eigenstate of the dilatation and the Lorentz generators. This makes sense as they generate the Cartan sub-algebra as mentioned above. Therefore we have,
\begin{align}
    [D,\mathcal{O}(0)]=-i\Delta\mathcal{O}(0),~~[M_{\mu\nu},\mathcal{O}(0)]=\mathcal{M}_{\mu\nu}\mathcal{O}(0).
\end{align}
The first of these equations coincides with the notation of the scaling dimension of an operator at a classical level and the second accomodates the fact that the operator can have a non-zero intrinsic spin. Using translations to obtain the action at non-zero positions yields,
\begin{align}\label{OperatorDandMgen}
    &[D,\mathcal{O}(x)]=-i\big(x^\mu\partial_\mu+\Delta)\mathcal{O}(x),~~[M_{\mu\nu},\mathcal{O}(x)]=-i\big(x_\mu\partial_\nu-x_\nu\partial_\mu+i\mathcal{M}_{\mu\nu}\big)\mathcal{O}(x).
\end{align}
Finally, note that the special conformal generator acts like  a lowering operator for $D$ as can be seen from the algebra \eqref{confalgebra}. It is thus natural to label an irreducible representation called a \textit{primary} operator as one that is annihilated by $K_\mu$ at the origin.
\begin{align}\label{KonOof0}
    [K_\mu,\mathcal{O}(0)]=0\implies \mathcal{O} ~\text{is primary}.
\end{align}
To obtain the action at finite $x$, we can once again make use of the translation generator, the Baker-Cambell Hausdorff formula and the algebra \eqref{confalgebra}. The result is,
\begin{align}\label{OperatorKgen}
    [K_\mu,\mathcal{O}(x)]=-i\big(-2x^\mu x^\nu \partial_\nu+x^2 \partial_\mu-2\Delta x_\mu-2i x^\nu\mathcal{M}_{\mu\nu})\mathcal{O}(x).
\end{align}
\begin{exercise}
    Show that the conformal algebra \eqref{confalgebra} is obeyed by the representations \eqref{OperatorPGen},\eqref{OperatorDandMgen} and \eqref{OperatorKgen}. Find the Quadratic and Quartic Casimirs of the conformal algebra. Show that their eigenvalues are functions of the dilatation eigen-value $\Delta$ and the spin $s$ of the operator.
    \bonus{Find the set of transformations $(\Delta,s)\to (\Delta'(\Delta),s'(s))$ that leave these eigenvalues invariant.}
\end{exercise}
We have suppressed possible ($SU(2)$ or $SL(2,\mathbb{R})$ Lorentz spinor) indices for the primary operator $\mathcal{O}(x)$. We shall also henceforth denote it with subscripts $\Delta,s$ for clarity. Therefore, a generic primary operator is denoted by,
\begin{align}\label{primaryop}
    \mathcal{O}_{\Delta,s}^{a_1\cdots a_{2s}}(x),
\end{align}
and transforms under conformal transformations like \eqref{OperatorPGen},\eqref{OperatorDandMgen} and \eqref{OperatorKgen}. Given a primary operator, one can form its descendants by acting with $P_\mu$ which acts as a raising operator for the dilatation operator as can be seen from the algebra \eqref{confalgebra}. A general descendant takes the form,
\begin{align}
    P^{\nu_1}\cdots P^{\nu_k}(P_\mu P^\mu)^l \mathcal{O}_{\Delta,s}^{\mu_1\cdots \mu_s}(x),~k,l\in\mathbb{R},
\end{align}
and is not annihilated by $K_\mu$ at the origin unlike the primary \eqref{KonOof0}.
Therefore, knowledge of the primary operator entails knowledge of the descendants and thus the former will be our primary focus\footnote{Pun intended.}.

There is one operator that is very important in the conformal theories we study: The stress tensor $T_{\mu\nu}$. To illustrate its fundamental importance, consider the action for the theory of interest to us (we do not need the explicit form of the action, we just assume that it is made out of a collection of some local fields $\Phi(x)$) in a general background with metric $g_{\mu\nu}$.
\begin{align}
    S=\int d^3 x\sqrt{g}~\mathcal{L}(\Phi(x),\partial \Phi(x),g_{\mu\nu},\cdots).
\end{align}
We have seen that conformal transformations change the metric up to an overall conformal factor $\Omega^2(x)$. The change in the action due to this is given by,
\begin{align}
    \delta S=\int d^3 x ~T^{\mu\nu}(x)\delta g_{\mu\nu},
\end{align}
This defines the stress tensor which is a local function of the fields and satisfies $\partial_\mu T^{\mu\nu}=0$ in a theory which is translation and Lorentz invariant. For conformal transformations we have seen that $\delta g_{\mu\nu}=\partial_{(\mu}\epsilon_{\nu)}=(1-\Omega^2(x))g_{\mu\nu}$ \eqref{conformalKillingequation}. Therefore, we have (we also set the background metric to the flat one $\delta_{\mu\nu}$ now) ,
\begin{align}
    \delta S=\int d^d x (1-\Omega^2(x))\delta_{\mu\nu}T^{\mu\nu}(x).
\end{align}
For dilatations, one can check using \eqref{conformalKillingequation} and \eqref{conformalkillingvectors} that $1-\Omega^2(x)=2\lambda$. Therefore, as long as $g_{\mu\nu}T^{\mu\nu}=\partial_\mu V^\mu$, the action is invariant. However, for special conformal transformations, $1-\Omega^2(x)$ is a position dependent function and therefore, the stress tensor has to be traceless for the action to remain invariant: $g_{\mu\nu}T^{\mu\nu}=T^\mu_\mu=0$. We will also study conformal field theories with larger symmetries in this note. For instance, the theory under question could have a $U(1)$, $SU(N)$ or even a higher spin symmetry. By Noether's theorem all of these continuous symmetries will give rise to conserved currents which we shall generally denote by $J_s^{\mu_1\cdots \mu_s;A}$. The $\mu_i$ indices denote spacetime Lorentz indices whereas $A$ represents a possible internal symmetry group index. With this discussion on operators let us now turn to the implications of conformal symmetry on their correlation functions.

The objects of interest to us our quantum mechanical correlation functions of the primary operators \eqref{primaryop},
\begin{align}
    \Gamma_n(x_1,\cdots,x_n)=\langle \mathcal{O}_1(x_1)\cdots \mathcal{O}_n(x_n)\rangle,
\end{align}
where we have suppressed the indices and labels of the operators and the expectation value is taken in the Vacuum state $|0\rangle$ which is assumed to be unique and invariant under the full conformal group. The correlator above could be a radially ordered correlator or a ``Euclidean" time ordered correlator where we choose one coordinate as time. For the purposes of the Ward-Takahashi identities, one can treat all types of correlators on equal footing.
Given an infinitesimal symmetry generator $G$, the conformal Ward identities takes the form,
\begin{align}\label{WardIdentity1}
    \sum_{i=1}^{n}\langle \cdots [G,\mathcal{O}_i(x_i)]\cdots \rangle=0,
\end{align}
where $G$ can be \eqref{OperatorPGen},\eqref{OperatorDandMgen} and \eqref{OperatorKgen}.
\begin{exercise}
    Prove \eqref{WardIdentity1}.
    \hint{Expand all terms in \eqref{WardIdentity1} and show that the terms cancel out pairwise provided the vacuum is invariant i.e., $G|0\rangle=\langle 0|G=0$. One can also prove this statement from the path integral.}.
\end{exercise}
Given the fact that we know the transformation of the primary operators under conformal transformations \eqref{OperatorPGen},\eqref{OperatorDandMgen} and \eqref{OperatorKgen}, let us see the restrictions imposed on their correlators by \eqref{WardIdentity1}. For example, consider identical scalar operators. 
\begin{exercise}
    Show that the conformal Ward identities fix the form of two, three and four point correlators of identical scalar operators as follows:
    \begin{align}\label{posspaceresultsscalar}
    &\langle O_\Delta(x_1)O_\Delta(x_2)\rangle=\frac{c_{\Delta}}{|x_1-x_2|^{2\Delta}},\notag\\
    &\langle O_{\Delta}(x_1)O_{\Delta}(x_2)O_{\Delta}(x_3)\rangle=\frac{f_{\Delta\Delta\Delta}}{|x_1-x_2|^{\Delta}|x_2-x_3|^{\Delta}|x_3-x_1|^{\Delta}},\notag\\
    &\langle O_{\Delta}(x_1)O_{\Delta}(x_2)O_{\Delta}(x_3)O_{\Delta}(x_4)\rangle=\frac{1}{|x_1-x_2|^{2\Delta}|x_3-x_4|^{2\Delta}}f(u,v),
\end{align}
where $u=\frac{x_{12}^2x_{34}^2}{x_{13}^2x_{24}^2}$ and $v=\frac{x_{14}^2 x_{23}^2}{x_{13}^2 x_{24}^2}$ are called the conformal cross ratios (we use the shorthand $x_{ij}=|x_i-x_j|$).
\bonus{Find the general form of the n-point function. How many cross ratios does it depend on? Is the result a function of the space-time dimension $d$?}
\end{exercise}
One can proceed similarly for non-identical operators and operators with non-zero spin. The result is that two and three point functions are fixed up to several constants by conformal invariance. For details please see a reference to be provided soon. We will have much more to say about this when we work in spinor helicity variables where we shall classify all two and three point functions.

We shall now proceed to a spinor helicity construction of the conformal generators. Naturally, this requires going through momentum space which we now turn to.

\subsection{The generators in the two spinor formalism}
Lets start with the action of the generators on momentum space primary operators which are defined via a Fourier transform:
\begin{align}
    \mathcal{O}_{\Delta,s}^{a_1\cdots a_{2s}}(p)=\int d^3 x~e^{-ip\cdot x}\mathcal{O}_{\Delta,s}^{a_1\cdots a_{2s}}(x)\iff \mathcal{O}_{\Delta,s}^{a_1\cdots a_{2s}}(x)=\int \frac{d^3 p}{(2\pi)^3}e^{ip\cdot x} \mathcal{O}_{\Delta,s}^{a_1\cdots a_{2s}}(p).
\end{align}
Given these formula, one can use the known position space actions of the generators \eqref{OperatorPGen},\eqref{OperatorDandMgen} and \eqref{OperatorKgen} to derive the corresponding momentum space expressions (suppressing operators labels and indices for notational clarity):
\begin{align}\label{momentumspacegenerators}
    &[P_\mu,\mathcal{O}(p)]=p_\mu \mathcal{O}(p),\notag\\
    &[D,\mathcal{O}(p)]=i\big(p^\mu\frac{\partial}{\partial p^\mu}+(3-\Delta)\big)\mathcal{O}(p)~,[M_{\mu\nu},\mathcal{O}(p)]=-i\big(p_\mu\frac{\partial}{\partial p^\nu}-p_\nu\frac{\partial}{\partial p^\mu}+i\mathcal{M}_{\mu\nu}\big),\notag\\
    &[K_\mu,\mathcal{O}(p)]=-\big(p_\mu\frac{\partial}{\partial p^\nu \partial p_\nu}+2(\Delta-3)\frac{\partial}{\partial p^\mu}-2p_\nu\frac{\partial}{\partial p_\nu \partial p^\mu}+2i\frac{\partial}{\partial p_\nu}\mathcal{M}_{\mu\nu}\big)\mathcal{O}(p).
\end{align}
One can now attempt to solve the conformal Ward identities \eqref{WardIdentity1} using the momentum space generators \eqref{momentumspacegenerators} or simply attempt a Fourier transform of the position space results \eqref{posspaceresultsscalar}. The latter approach of performing a direct Fourier transform of \eqref{posspaceresultsscalar} is much more difficult but can also be done. Our interest in these lectures however, are spinor helicity variables and not momentum space directly. We can and will often relate the two languages using \eqref{3dEuclidSH} and \eqref{3dpolarizationspinors}.
\begin{exercise}
    Using \eqref{3dEuclidSH} show that,
    \begin{align}
        \frac{\partial}{\partial p^\mu}=-\frac{(\sigma_\mu)^{b}_a}{(\lambda\cdot\Bar{\lambda})}\bigg(\Bar{\lambda}^a\frac{\partial}{\partial\Bar{\lambda}^b}-\lambda^a\frac{\partial}{\partial \lambda^b}\bigg).
    \end{align}
    Using this formula show that,
    \begin{align}
        &P_\mu=\frac{1}{2}(\sigma^\mu)^a_b\lambda_a\Bar{\lambda}^b,\notag\\
    &M_{\mu\nu}=\frac{1}{2}\epsilon_{\mu\nu\rho}(\sigma^\rho)^{a}_{b}\bigg(\Bar{\lambda}^b\frac{\partial}{\partial\Bar{\lambda}^a}+\lambda^b\frac{\partial}{\partial\lambda^a}\bigg)+\mathcal{M}_{\mu\nu},\notag\\
    &D=\frac{i}{2}\bigg(\Bar{\lambda}^a\frac{\partial}{\partial\Bar{\lambda}^a}+\lambda^a\frac{\partial}{\partial\lambda^a}\bigg)+i\bigg((3-\Delta)-\frac{1}{2}\theta^a\frac{\partial}{\partial\theta^a}\bigg),\notag\\&K^\mu=2(\sigma^\mu)^{ab}\frac{\partial^2}{\partial\lambda^a\partial\Bar{\lambda}^b}+\frac{(\Delta-2)}{p}(\sigma^\mu)^b_a\bigg(\Bar{\lambda}^a\frac{\partial}{\partial\Bar{\lambda}^b}-\lambda^a\frac{\partial}{\partial\lambda^b}\bigg)+\frac{i(\sigma^\nu)^a_b}{p}\bigg(\Bar{\lambda}^a\frac{\partial}{\partial\Bar{\lambda}^b}-\lambda^a\frac{\partial}{\partial\lambda^b}\bigg)\mathcal{M}_{\mu\nu},
    \end{align}
    where as usual $p=-\frac{1}{2}\lambda\cdot\Bar{\lambda}$.
\end{exercise}
However, the objects that we shall mostly work with are conserved currents whose indices have been contracted such as in \eqref{Jspm} and are rescaled as well \eqref{Jsrescaled}.  Lets also move on to a completely spinorial description of the conformal generators by contracting with the Pauli matrices.\\
\underline{\textbf{Constructing the generators acting on currents}}\\
Since the rescaled currents have dimension $2$, we can set $\Delta=2$ in the above generators. They also have no free indices so $\mathcal{M}_{\mu\nu}$ acts trivially. This leads to,
\begin{align}
    &[P_{ab},\hat{J}_s^{\pm}]=\lambda_{(a}\Bar{\lambda}_{b)}\hat{J}_s^{\pm}, [K_{ab},\hat{J}_s^{\pm}]=2\frac{\partial^2}{\partial\lambda^{(a}\partial\Bar{\lambda}^{b)}}\hat{J}_s^{\pm},\notag\\
    &[D,\hat{J}_s^{\pm}]=\frac{i}{2}\big(\lambda^a\frac{\partial}{\partial\lambda^a}+\Bar{\lambda}^a\frac{\partial}{\partial\Bar{\lambda}^a}+2\big)\hat{J}_s^{\pm},[M_{ab},\hat{J}_s^{\pm}]=\frac{i}{2}\big(\lambda_{(a}\frac{\partial}{\partial\lambda^{b)}}+\Bar{\lambda}_{(a}\frac{\partial}{\partial\Bar{\lambda}^{b)}}+2\big)\hat{J}_s^{\pm}.
\end{align}
In the above formula we have defined the rescaled currents $\hat{J}_s$ which effectively remove an amount $s-1$ of dilatation weight from the current and are given by,
\begin{align}
    \hat{J}^{\pm}_s(\lambda,\Bar{\lambda})=\frac{J_s^{\pm}(\lambda,\Bar{\lambda})}{p^{s-1}}.
\end{align}
One can now show that with these generators, we obtain the following Ward-Takahashi identities,
\begin{align}\label{SHgenonJshat1}
    &\bigg(\sum_{i=1}^{n}\lambda_{i(a}\Bar{\lambda}_{ib)}\bigg)\langle \hat{J}_{s_1}^{\pm}(\lambda_1,\Bar{\lambda}_1)\cdots \hat{J}_{s_n}^{\pm}(\lambda_n,\Bar{\lambda}_n)\rangle=0,\notag\\&\frac{i}{2}\bigg(\sum_{i=1}^{n}\lambda_i^a\frac{\partial}{\partial\lambda_i^a}+\Bar{\lambda}_i^a\frac{\partial}{\partial\Bar{\lambda}_i^a}+2\bigg)\langle \hat{J}_{s_1}^{\pm}(\lambda_1,\Bar{\lambda}_1)\cdots \hat{J}_{s_n}^{\pm}(\lambda_n,\Bar{\lambda}_n)\rangle=0,\notag\\&\frac{i}{2}\bigg(\sum_{i=1}^{n}\lambda_{i(a}\frac{\partial}{\partial\lambda_i^{b)}}+\Bar{\lambda}_{i(a}\frac{\partial}{\partial\Bar{\lambda}_i^{b)}}+2\bigg)\langle \hat{J}_{s_1}^{\pm}(\lambda_1,\Bar{\lambda}_1)\cdots \hat{J}_{s_n}^{\pm}(\lambda_n,\Bar{\lambda}_n)\rangle=0,\notag\\&2\bigg(\sum_{i=1}^{n}\frac{\partial^2}{\partial\lambda_{i}^{(a}\partial\lambda_i^{b)}}\bigg)\langle \hat{J}_{s_1}^{\pm}(\lambda_1,\Bar{\lambda}_1)\cdots \hat{J}_{s_n}^{\pm}(\lambda_n,\Bar{\lambda}_n)\rangle=\bigg(\sum_{i=1}^{n}\frac{s_i(s_i+1)}{p_i^{s_i+1}}\zeta_{i\pm a}\zeta_{i \pm b}p_i^{a_1 a_2}\zeta_{i\pm}^{a_3}\cdots \zeta_{i\pm}^{a_{2s}}\bigg)\langle \cdots \hat{J}_{s_i~a_1\cdots a_{2s}}(\lambda_i,\Bar{\lambda}_i)\cdots\rangle.
\end{align}
Note in particular that although the translation, dilatation and rotation generators act as usual, the SCT generator leads to a non-zero RHS which is in particular proportional to the Ward-Takahashi identity of the correlator.
\begin{exercise}
    Prove \eqref{SHgenonJshat1}.
    \hint{Show that the action of this SCT generator on the helicity basis current leads to the usual conformal generator but also generates the term proportional to the Ward Takahashi identity due to current conservation. See appendix $D$ of \cite{Baumann:2020dch} for details.}
\end{exercise}
One final point to note about the correlators of the helicity basis currents is that they obey the helicity identities by virtue of their construction \eqref{Jspm}:
\begin{align}
    h_i\langle \hat{J}_{s_1}^{\pm}(\lambda_1,\Bar{\lambda}_1)\cdots \hat{J}_{s_n}^{\pm}(\lambda_n,\Bar{\lambda}_n)\rangle=-\frac{1}{2}\bigg(\lambda_i^a\frac{\partial}{\partial \lambda_i^a}-\Bar{\lambda}_i^a\frac{\partial}{\partial\Bar{\lambda}_i^a}\bigg)\langle \hat{J}_{s_1}^{\pm}(\lambda_1,\Bar{\lambda}_1)\cdots \hat{J}_{s_n}^{\pm}(\lambda_n,\Bar{\lambda}_n)\rangle=\pm s_i \langle \hat{J}_{s_1}^{\pm}(\lambda_1,\Bar{\lambda}_1)\cdots \hat{J}_{s_n}^{\pm}(\lambda_n,\Bar{\lambda}_n)\rangle.
\end{align}
The finite version of this transformation is what we wrote in the main text in equation \eqref{SHHelicityID}.

\bibliographystyle{JHEP}
\bibliography{biblio}

\end{document}